\documentclass[fleqn,usenatbib]{mnras}

\usepackage{newtxtext,newtxmath}

\usepackage[T1]{fontenc}

\DeclareRobustCommand{\VAN}[3]{#2}
\let\VANthebibliography\thebibliography
\def\thebibliography{\DeclareRobustCommand{\VAN}[3]{##3}\VANthebibliography}

\usepackage{graphicx}	
\usepackage{amsmath}	
\usepackage{xspace}
\usepackage{xcolor}
\usepackage{tabularx}
\makeatletter 
  \patchcmd{\NAT@citex}
    {\@citea\NAT@hyper@{%
      \NAT@nmfmt{\NAT@nm}%
      \hyper@natlinkbreak{\NAT@aysep\NAT@spacechar}{\@citeb\@extra@b@citeb}%
      \NAT@date}}
    {\@citea\NAT@nmfmt{\NAT@nm}%
    \NAT@aysep\NAT@spacechar\NAT@hyper@{\NAT@date}}{}{}

  \patchcmd{\NAT@citex}
    {\@citea\NAT@hyper@{%
      \NAT@nmfmt{\NAT@nm}%
      \hyper@natlinkbreak{\NAT@spacechar\NAT@@open\if*#1*\else#1\NAT@spacechar\fi}%
        {\@citeb\@extra@b@citeb}%
      \NAT@date}}
    {\@citea\NAT@nmfmt{\NAT@nm}%
    \NAT@spacechar\NAT@@open\if*#1*\else#1\NAT@spacechar\fi\NAT@hyper@{\NAT@date}}
    {}{}
\makeatother

\newcommand{\msun}{{\,\rm M_\odot}}

\newcommand{\kms}{\,{\rm km}\,{\rm s}^{-1}}
\newcommand{\cm}{\,{\rm cm}}
\newcommand{\erg}{\,{\rm erg}}
\newcommand{\Gyr}{\,{\rm Gyr}}

\newcommand{\Myr}{\,{\rm Myr}}
\newcommand{\pc}{\,{\rm pc}}
\newcommand{\kpc}{\,{\rm kpc}}

\newcommand{\Mpc}{\,{\rm Mpc}}

\newcommand{\mmag}{\,{\rm mag}}

\newcommand{\thesan}{\textsc{thesan}\xspace}
\newcommand{\thesanzoom}{\textsc{thesan-zoom}\xspace}

\newcommand{\thesanone}{\textsc{thesan-1}\xspace}

\newcommand{\HI}{\ion{H}{I}\xspace}
\newcommand{\HII}{\ion{H}{II}\xspace}
\newcommand{\HeI}{\ion{He}{I}\xspace}
\newcommand{\HeII}{\ion{He}{II}\xspace}
\newcommand{\HeIII}{\ion{He}{III}\xspace}

\def\aap{A\&A}
\def\apj{ApJ}

\def\apjl{ApJ}
\def\mnras{MNRAS}
\def\araa{ARA\&A}
\def\aj{AJ}

\def\physrep{Phys. Rep.}
\def\nat{Nature}

\def\apjs{ApJS}

\definecolor{mycolor1}{HTML}{D1BBD7}
\definecolor{mycolor2}{HTML}{AE76A3}
\definecolor{mycolor3}{HTML}{882E72}
\definecolor{mycolor4}{HTML}{1965B0}
\definecolor{mycolor5}{HTML}{5289C7}
\definecolor{mycolor6}{HTML}{7BAFDE}
\definecolor{mycolor7}{HTML}{4EB265}
\definecolor{mycolor8}{HTML}{90C987}
\definecolor{mycolor9}{HTML}{CAE0AB}
\definecolor{mycolor10}{HTML}{F7F056}
\definecolor{mycolor11}{HTML}{F6C141}
\definecolor{mycolor12}{HTML}{F1932D}
\definecolor{mycolor13}{HTML}{E8601C}
\definecolor{mycolor14}{HTML}{DC050C}

\title[Star-formation efficiencies in high-redshift galaxies]{The \thesanzoom project: Star-formation efficiencies in high-redshift galaxies}

\author[X. Shen et al.]{\parbox{17.5cm}{
Xuejian Shen,$^{1}$\thanks{E-mail: \href{mailto:xuejian@mit.edu}{xuejian@mit.edu}}
Rahul Kannan,$^{2}$
Ewald Puchwein,$^{3}$
Aaron Smith,$^{4}$
Mark Vogelsberger,$^{1}$
Josh Borrow,$^{5}$
Enrico Garaldi,$^{6,7}$
Laura Keating,$^{8}$
Oliver Zier,$^{9}$
William McClymont,$^{10,11}$
Sandro Tacchella,$^{10,11}$
Zihao Wang,$^{1}$
and
Lars Hernquist$^{9}$
}
\\ \vspace{0.2cm} \\
$^1$ Department of Physics, Kavli Institute for Astrophysics and Space Research, Massachusetts Institute of Technology, Cambridge, MA 02139, USA \\
$^2$ Department of Physics and Astronomy, York University, 4700 Keele Street, Toronto, ON M3J 1P3, Canada \\
$^3$ Leibniz-Institut f\"ur Astrophysik Potsdam, An der Sternwarte 16, 14482 Potsdam, Germany \\
$^4$ Department of Physics, The University of Texas at Dallas, Richardson, TX 75080, USA \\
$^5$ Department of Physics and Astronomy, University of Pennsylvania, 209 South 33rd Street, Philadelphia, PA 19104, USA \\
$^6$ Kavli Institute for the Physics and Mathematics of the Universe, The University of Tokyo, 5-1-5 Kashiwanoha, Kashiwa, 277-8583, Chiba, Japan \\
$^7$ Institute for Fundamental Physics of the Universe, via Beirut 2, 34151 Trieste, Italy \\
$^8$ Institute for Astronomy, University of Edinburgh, Blackford Hill, Edinburgh, EH9 3HJ, UK \\
$^9$ Center for Astrophysics | Harvard \& Smithsonian, 60 Garden St, Cambridge, MA 02138, USA\\
$^{10}$ Kavli Institute for Cosmology, University of Cambridge, Madingley Road, Cambridge CB3 0HA, UK \\
$^{11}$ Cavendish Laboratory, University of Cambridge, 19 JJ Thomson Avenue, Cambridge CB3 0HE, UK
}

\date{Accepted XXX. Received YYY; in original form ZZZ}

\pubyear{2025}

\begin{document}
\label{firstpage}
\pagerange{\pageref{firstpage}--\pageref{lastpage}}
\maketitle

\begin{abstract}
Recent James Webb Space Telescope (\textit{JWST}) observations hint at unexpectedly intense cosmic star-formation in the early Universe ($z\gtrsim 10$), often attributed to enhanced star-formation efficiencies (SFEs). Here, we analyze the SFE in \thesanzoom, a novel zoom-in radiation-hydrodynamic simulation campaign of high-redshift ($z \gtrsim 3$) galaxies employing a state-of-the-art galaxy formation model resolving the multiphase interstellar medium (ISM). The halo-scale SFE ($\epsilon^{\ast}_{\rm halo}$) -- the fraction of baryons accreted by a halo that are converted to stars -- follows a double power-law dependence on halo mass, with a mild redshift evolution above $M_{\rm halo} \gtrsim 10^{9.5}\msun$. The power-law slope is roughly $1/3$ at large halo masses, consistent with expectations when gas outflows are momentum-driven. At lower masses, the slope is roughly $2/3$ and is more aligned with the energy-driven outflow scenario. $\epsilon^{\ast}_{\rm halo}$ is a factor of $2-3$ larger than commonly assumed in empirical galaxy-formation models at $M_{\rm halo} \lesssim 10^{11}\msun$. On galactic (kpc) scales, the Kennicutt–Schmidt (KS) relation of neutral gas is universal in \thesanzoom, following $\Sigma_{\rm SFR} \propto \Sigma_{\rm gas}^2$, indicative of a turbulent energy balance in the ISM maintained by stellar feedback. The rise of $\epsilon^{\ast}_{\rm halo}$ with halo mass can be traced primarily to increasing gas surface densities in massive galaxies, while the underlying KS relation and neutral, star-forming gas fraction remain unchanged. These results are robust against variations in numerical resolution and the specifics of star formation and feedback recipes in simulations, depending mainly on the total feedback momentum budget. Although the increase in $\epsilon^{\ast}_{\rm halo}$ with redshift is relatively modest, it is sufficient to explain the large observed number density of UV-bright galaxies at $z \gtrsim 12$. However, reproducing the brightest sources at $M_{\rm UV} \lesssim -21$ may require extrapolating the SFE beyond the halo mass range directly covered by \thesanzoom.
\end{abstract}




\section{Introduction}

In the standard cosmological framework, galaxies form within gravitationally collapsed haloes of dark matter~\citep[DM; e.g.][]{Blumenthal1984,Davis1985}, where gas accreted by the halo cools down, collapses, fragments, and ultimately forms stars. Throughout this process, stellar feedback is a crucial component regulating star-formation, which would proceed rapidly and convert most of the baryons to stars in the absence of feedback and inevitably result in galaxies far more massive than observed~\citep[e.g.][]{White1991,Katz1996, Somerville1999, Cole2000, Springel2003b, Keres2009b}. Indeed, stellar feedback has been shown critical in reproducing the suppression of stellar-to-halo-mass ratios of low-mass galaxies~\citep[e.g.][]{Conroy2006, Moster2010, Behroozi2013, Behroozi2019} and the Kennicutt–Schmidt (KS) relation observed in local star-forming galaxies~\citep{Schmidt1959,Kennicutt1998}. On smaller spatial scales, stellar feedback is crucial in explaining the percent-level efficiency of baryon-to-star conversion during the lifecycle of giant molecular clouds~\citep[GMCs; e.g.][]{Zuckerman1974, Williams1997, Evans1999, Evans2009, Kennicutt2012,SunJY2023}. Meanwhile, stellar feedback is important in driving galactic-scale gas outflows found in observations~\citep[e.g.][]{Martin1999, Heckman2000, Steidel2010, Coil2011, Newman2012}, which bring metal-enriched gas into the circumgalactic and intergalactic medium~\citep[e.g.][]{Aguirre2001, Oppenheimer2006, Martin2010, Tumlinson2017}.

Despite the clear role of stellar feedback in regulating star-formation, the physical mechanism that dominates and how feedback energy/momentum couples to the interstellar medium (ISM) remain critical questions in galaxy formation theory. Multiple stellar feedback processes, such as supernovae (SNe), winds from OB and asymptotic giant branch (AGB) stars, protostellar jets, cosmic-rays, photoheating, and radiation pressure, all interact efficiently with the surrounding ISM (see e.g. \citealt{Draine2011,Hopkins2018}). Pre-processing of early stellar feedback (ESF) before SNe explosions results in rarefied environments where SNe act more efficiently~\citep[e.g.][]{Krumholz2007, Offner2009, Bate2012, Krumholz2019,Kannan2020feedback}. Therefore, these feedback processes can add non-linearly in dispersing star-forming GMCs, driving gas fountains and superwinds, and regulating star-formation on the galactic scale. Only in the past decade has the new generation of cosmological hydrodynamic simulations started to explicitly model the aforementioned feedback processes and resolve the multiphase structure of the ISM with adequate resolution~\citep[e.g.][]{Hopkins2014,Hopkins2018,Agertz2013,Agertz2021, Kim2017,Marinacci2019}. This class of models has successfully reproduced a wide range of observational properties of galaxies~\citep[e.g.][]{Hopkins2014,Angles2017,Hopkins2018,GK2019}, including the aforementioned constraints on cosmic inefficient star-formation~\citep[e.g.][]{Hopkins2014,Orr2018,Marinacci2019}. However, significant uncertainties remain in e.g. the numeric implementation of SNe feedback~\citep[e.g.][]{Hopkins2018feedback,Hopkins2024,Zhang2024} and radiative feedback from massive bright stars~\citep[e.g.][]{Hopkins2020-rad,Kannan2020smuggle,Deng2024} as well as star-formation recipes, which depend on the assumptions of the unresolved turbulence and density structure~\citep[e.g.][]{Padoan2011,Federrath2012,Semenov2016,Semenov2024b}. 

While most theoretical studies focused on star-formation laws in low-redshift galaxies ($z\lesssim 2$), the high-redshift regime could exhibit several qualitative differences and provide new insights into star-formation and feedback. For example, unlike the low-redshift mature galaxies with well-defined geometrically thin disks, high-redshift galaxies have more complicated morphology, with clumpy, irregular structures identified in rest-frame UV~\citep[e.g.][]{Bournaud2007, Elmegreen2009, Forster2011, Treu2023}, disk-like elongated structures in optical and near-infrared~\citep[e.g.][]{Ferreira2022,Ferreira2023,Robertson2023}, and dynamically cold gas disks found in ALMA observations~\citep[e.g.][]{Rizzo2020,Tsukui2021,RO2023,Rowland2024}. These geometrical factors could affect how gravitational instability develops and feedback energy couples~\citep[e.g.][]{Dekel2009,Dekel2023}. In addition, time variability (``burstiness'') of star-formation likely becomes important in the interpretation of observational results~\citep[e.g.][]{Tacchella2020,Mirocha2023,Shen2023,Sun2023a,Semenov2024a}. In dense star-forming regions at high redshifts, star-formation could proceed rapidly due to short free-fall time scales before the first SNe event, making ESF processes like stellar winds and radiative feedback increasingly important in such conditions~\citep{Dekel2023}. In extreme environments (total matter surface density $\Sigma_{\rm tot}\gtrsim 10^{4} \msun\pc^{-2}$), all forms of feedback may fail to regulate star-formation~\citep[e.g.][]{Grudic2018,Hopkins2022,Menon2024}, leading to extremely efficient star-formation. Moreover, external background radiation that is dynamically developed during the epoch of reionization (EoR) can substantially suppress star-formation in low-mass haloes~\citep[e.g.][]{Rees1986, Bullock2000, Shapiro2004, Iliev2005,Okamoto2008,Gnedin2014,Fitts2017,Katz2020}.

Recent observations from the James Webb Space Telescope (\textit{JWST}) offer a new channel to constrain the theory of star-formation at high redshifts. Several findings of the early \textit{JWST} observations have challenged the canonical picture of galaxy formation and appear more aligned with some of the theoretical ideas above. For example, \textit{JWST} has uncovered potentially overly-massive galaxies at $z\sim 5-12$~\citep[e.g.][]{Labbe2023,Xiao2024,Casey2024}, which imply an averaged star-formation efficiency (SFE) exceeding the largest values inferred from previous lower-redshift observations. Although many systematic uncertainties remain in interpreting these observations, it may suggest a distinct and more violent ``mode'' of star-formation in the earliest phase of galaxy evolution. In addition, \textit{JWST} has also revealed a large abundance of ultra-violet (UV)-bright galaxies at $z\gtrsim 10$~\citep[e.g.][]{Finkelstein2023,Finkelstein2024,Harikane2023} that exceeds most of the theoretical predictions from pre-\textit{JWST} models~\citep[e.g.][]{Mason2015,Tacchella2018,Behroozi2020,Kannan2023,Yung2024b}. These early photometric constraints have been verified by spectroscopic follow-ups up to $z\simeq 14$~\citep{Harikane2024a,Harikane2024b}. Among many theoretical explanations for this overabundance of bright galaxies, the enhanced SFE serves as an intuitive solution~\citep[e.g.][]{Mason2023,Dekel2023}. Despite plausible physical arguments for more efficient star-formation in the early Universe, numerical simulations have not yet reached a consensus~\citep[e.g.][]{Pallottini2023,Sun2023b,Ceverino2024,Feldmann2024,Dome2024}, partly due to significant uncertainties in star-formation and feedback models, as discussed earlier. 

In this paper, we analyze the SFE of high-redshift galaxies in the newly-developed \thesanzoom simulation suite (introduced in \citealt{Kannan2025}). This campaign is designed to provide realistic simulation counterparts to the plethora of high-redshift galaxies observed by \textit{JWST}. Building upon previous successful experiments of the \textsc{Smuggle} galaxy formation model~\citep{Marinacci2019}, the simulation suite includes explicit modeling of star-formation and multi-channel stellar feedback processes in the multiphase ISM while testing several variants of the ISM physics models. The simulations include on-the-fly sourcing and transfer of radiation~\citep{Kannan2019} in seven broad bands, coupling to hydrodynamics, and the thermochemistry of gas and dust grains~\citep{Kannan2020}. The zoom-in galaxies are selected from the large-volume radiation-hydrodynamical simulation \thesan~\citep{Kannan2022thesan,Garaldi2022,Smith2022,Garaldi2024}. We therefore capture realistic external reionization, self-shielding, as well as local radiative feedback processes

As part of the initial series of papers on \thesanzoom, this study focuses on the SFE of high-redshift galaxies, characterized by the baryon-to-star conversion efficiency on the halo scale and the gas depletion time in resolved patches of ISM. This paper is organized as follows. In Section~\ref{sec:sim}, we briefly summarize the numerical methods and the simulation suite. In Section~\ref{sec:result-epshalo}, we quantify the halo-scale SFE and discuss its dependence on halo mass and redshift. In Section~\ref{sec:result-epsgal}, we show the analyses on the KS relation of neutral gas and explore the impact of different physical/numerical variants of the galaxy formation model. We propose a simple analytical picture to understand the evolution of the KS relation and its connection to the halo-scale SFE. Discussions and conclusions are presented in Section~\ref{sec:discussion} and Section~\ref{sec:conclusion}. Throughout the paper, we assume the cosmological parameters from \citet{Planck2016} (obtained from their TT,TE,EE+lowP+lensing+BAO+JLA+H0 dataset), with $H_0=67.74\,\kms/\Mpc$, $\Omega_{\rm m}=0.3089$, $\Omega_{\Lambda}=0.6911$, $\Omega_{\rm b}=0.0486$, $\sigma_8=0.8159$, and $n_{\rm s}=0.9667$. 

\begin{figure*}
    \raggedright
    \hspace{0.06\linewidth} \textbf{UV-optical image} \hspace{0.15\linewidth} \textbf{Neutral gas} \hspace{0.16\linewidth} \textbf{LyC radiation} \hspace{0.13\linewidth} \textbf{Gas temperature}\\
    \includegraphics[width=0.2485\linewidth, trim={0.02cm 0.02cm 0.02cm 0.02cm}, clip]{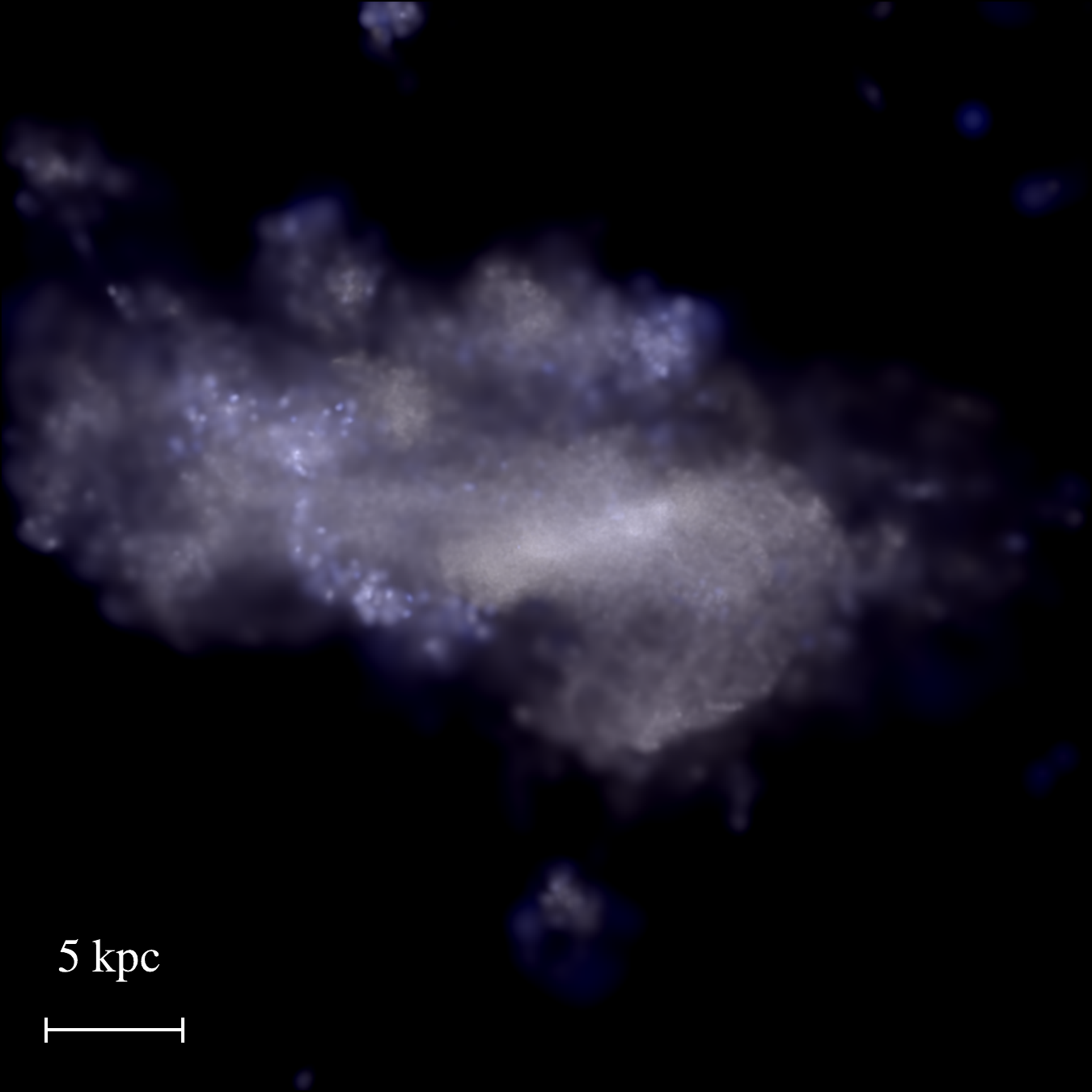}
    \hspace{-0.13cm}
    \includegraphics[width=0.2485\linewidth, trim={0.02cm 0.02cm 0.02cm 0.02cm}, clip]{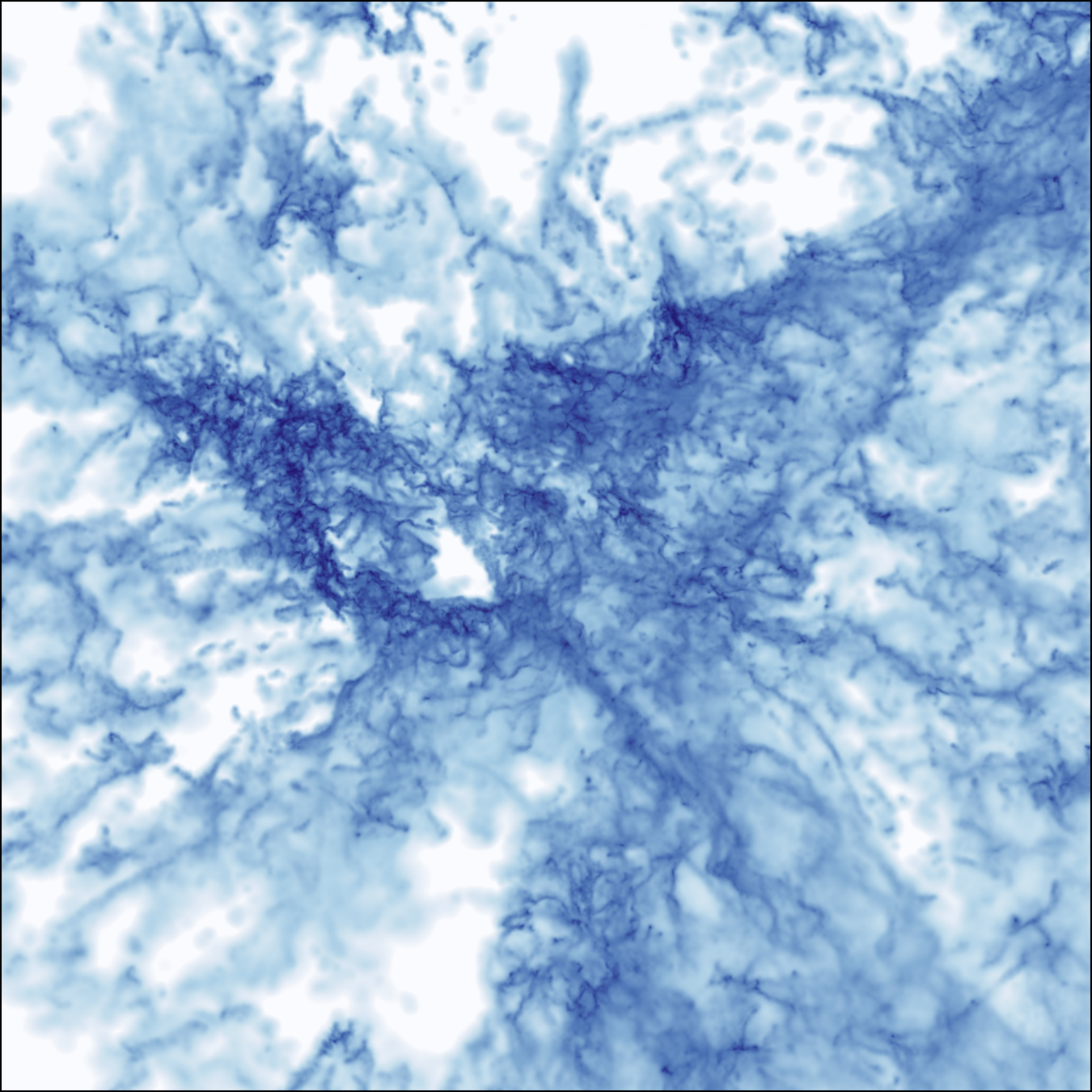}
    \hspace{-0.13cm}
    \includegraphics[width=0.2485\linewidth, trim={0.02cm 0.02cm 0.02cm 0.02cm}, clip]{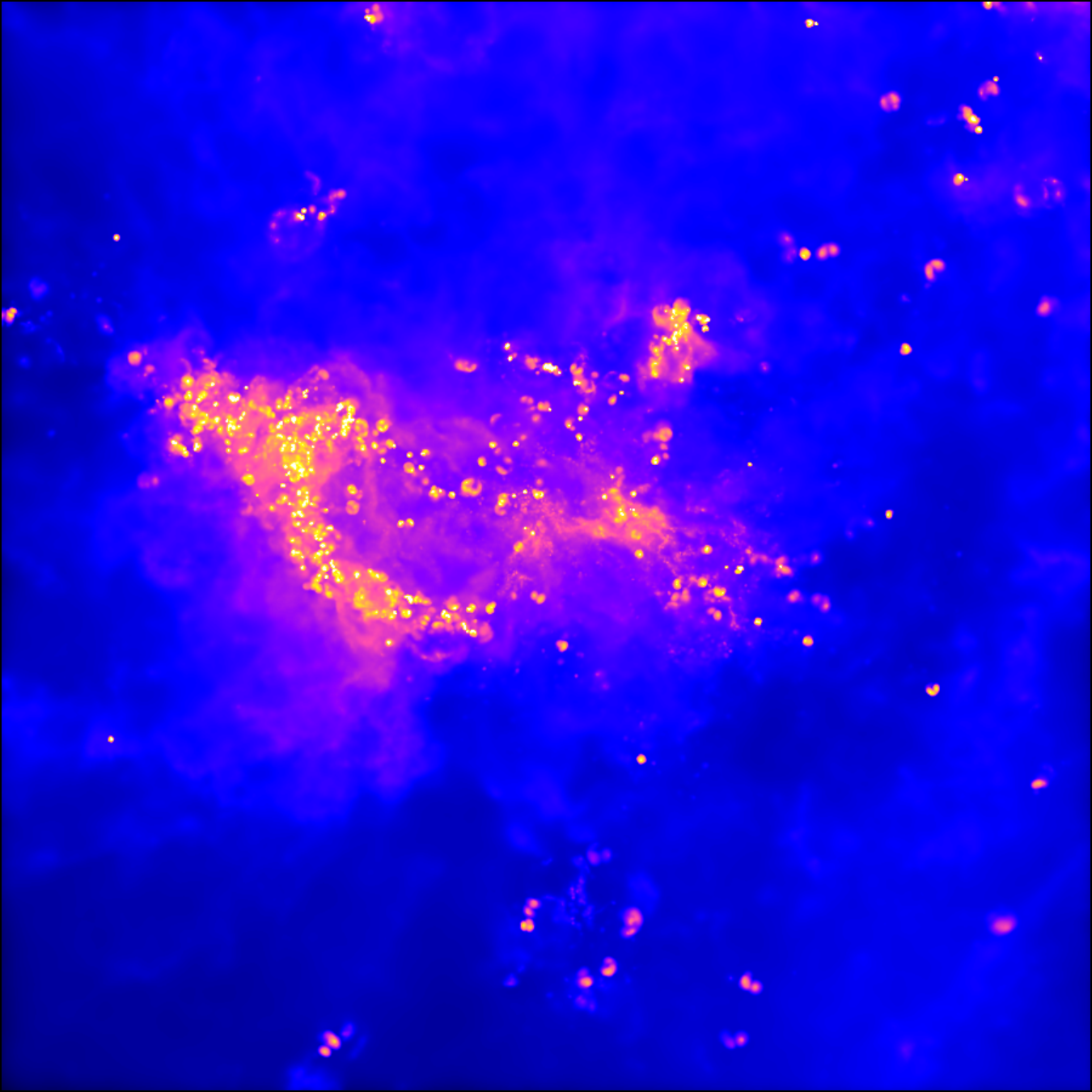}
    \hspace{-0.13cm}
    \includegraphics[width=0.2485\linewidth, trim={0.02cm 0.02cm 0.02cm 0.02cm}, clip]{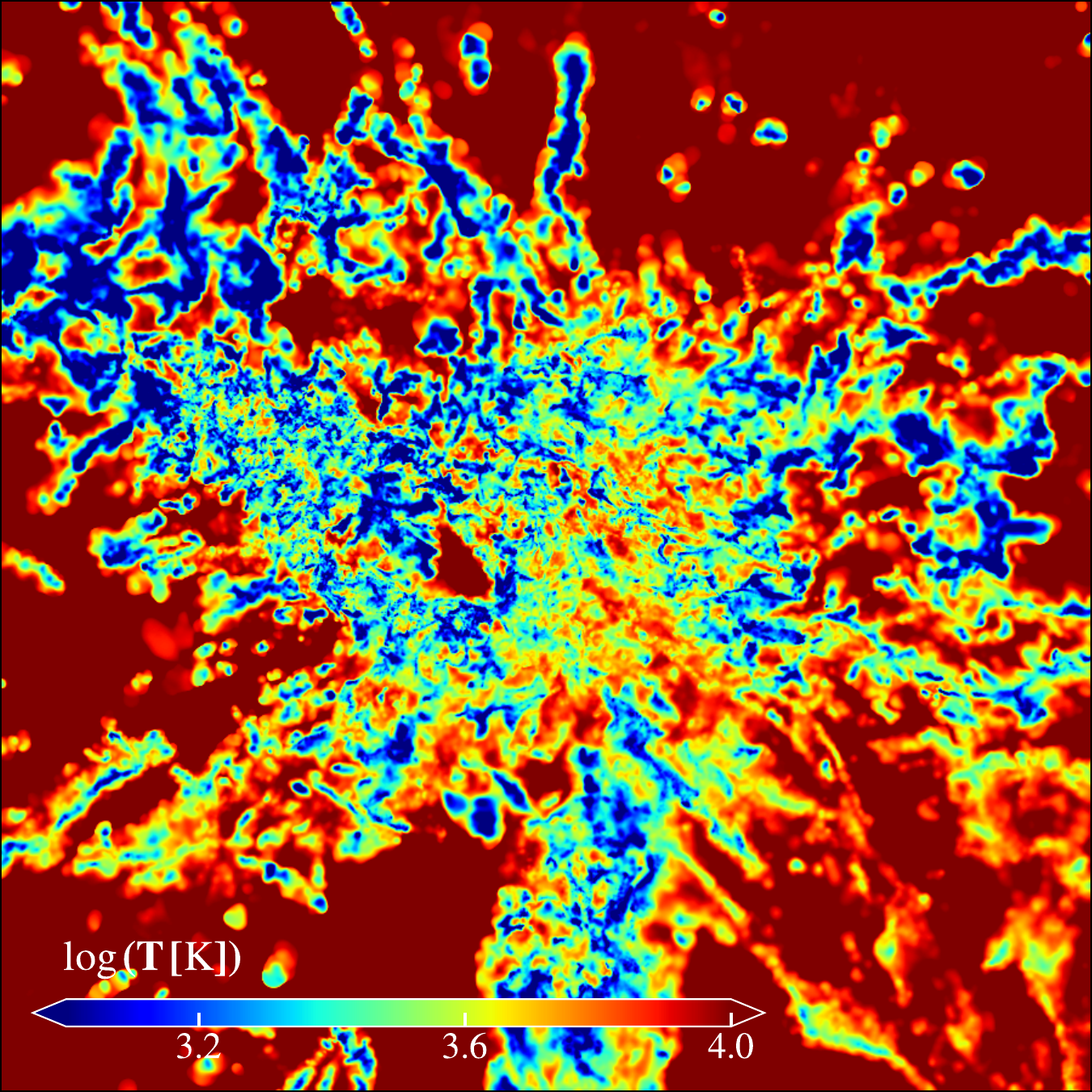} \\
    \vspace{-0.06cm}
    \includegraphics[width=0.498\linewidth, trim={0.04cm 0.04cm 0.04cm 0.04cm}, clip]{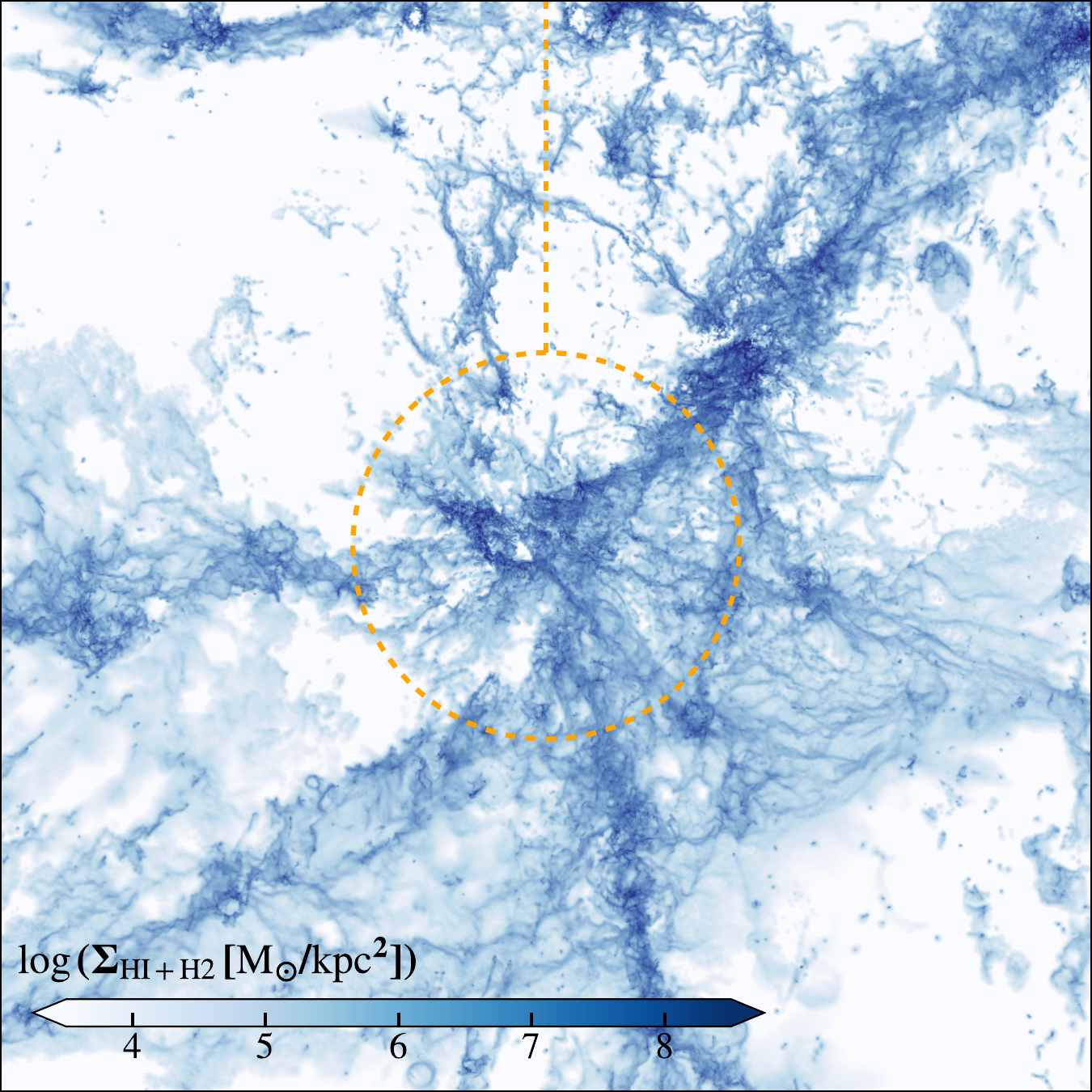}
    \hspace{-0.13cm}
    \includegraphics[width=0.498\linewidth, trim={0.04cm 0.04cm 0.04cm 0.04cm}, clip]{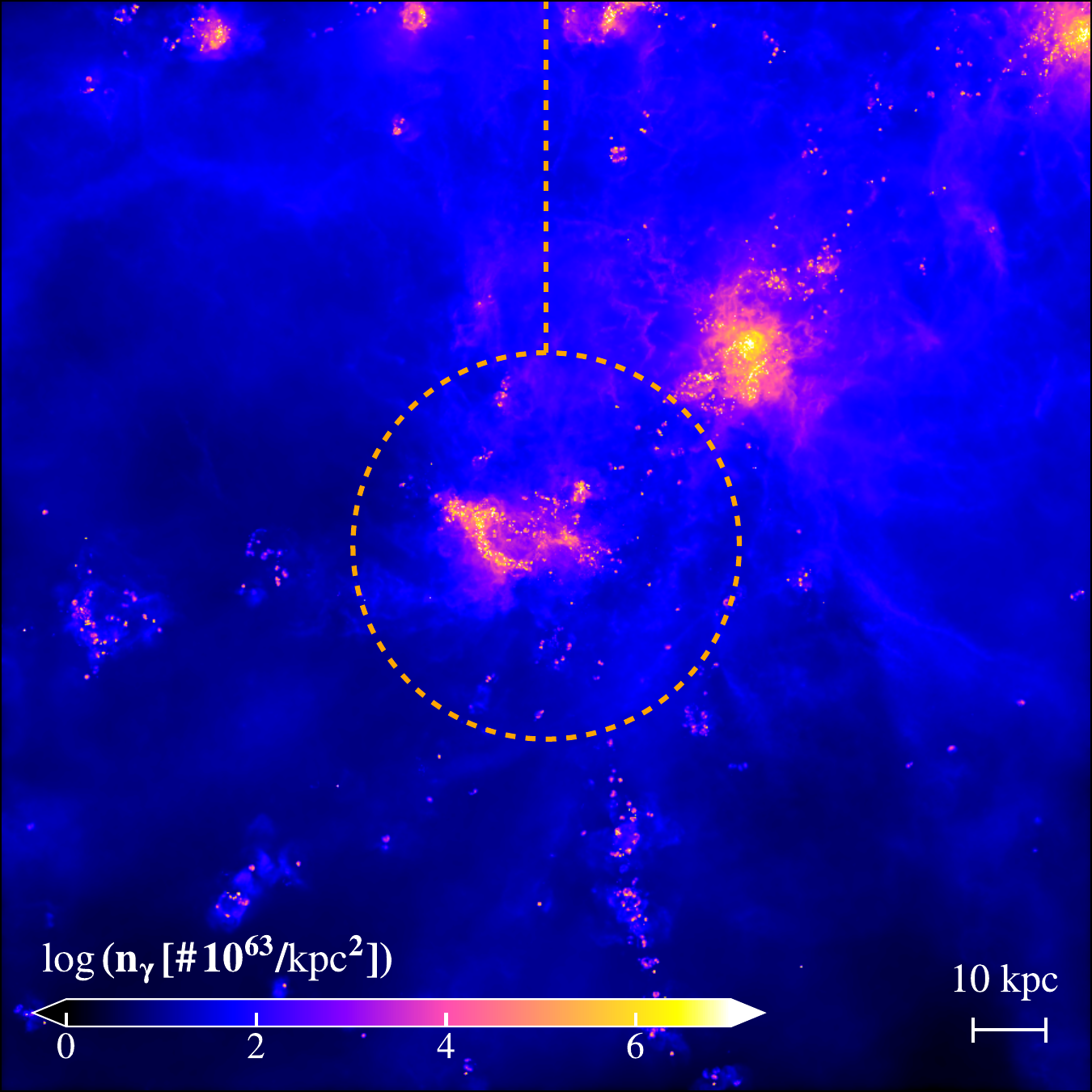}
    \label{fig:intro}
    \caption{A visual inspection of the surface density of neutral and molecular gas (bottom left) and Lyman-continuum (LyC) radiation field (bottom right) around the \thesanzoom galaxy ``m12.6'' at $z\simeq 6$. The orange circle shows the halo virial radius. In the top row from left to right, we show zoom-in views of galaxy stellar light in UV-optical bands, neutral gas distribution, LyC radiation field, and gas temperature. The galaxy has an irregular morphology with UV luminous clumps. The ISM of the galaxy is highly turbulent and displays a clear multiphase structure in the temperature map. The LyC radiation field strongly correlates with the distribution of young blue stars in the UV-optical image. Cold neutral gas fueling the galaxy comes from filamentary structures that are dense and self-shielded against the ionizing radiation background.}
\end{figure*}

\section{Simulations}
\label{sec:sim}

\begin{table}
    \centering
    \caption{Numerical parameters of the \thesanzoom simulation suite. From left to right, each column indicates: \\ 
    (1) Name of the resolution level. We note that all the target haloes have been simulated at the resolution level ``L4'' while only the low-mass ones were simulated at ``L8'' and ``L16''.  \\
    (2) Effective (total volume-equivalent) number of particles. \\
    (3,4) Mass of the high-resolution DM particles and gas cells. The later will also be referred to as the baryonic mass resolution $m_{\rm b}$. \\
    (5) The comoving softening length of star and DM particles. \\
    (6) The minimum comoving softening length of gas cells.}
    \label{table:res}
    \addtolength{\tabcolsep}{-0.2pt}
    \def\arraystretch{1.2}
    \begin{tabular}{lccccc} 
	\hline
	Name & $N_{\rm part}^{\rm eff}$ & $m_{\rm DM}$ & $m_{\rm gas}$ & $\epsilon_{\rm DM, stars}$ & $\epsilon_{\rm gas}^{\rm min}$\\  
		& & [$\mathrm{M}_\odot$] & [$\mathrm{M}_\odot$] & [cpc] & [cpc]\\
		\hline
            L4 & $2 \times 8400^3$ & $4.86 \times 10^4$ & $9.09 \times 10^3$ & $553.59$ & $69.20$\\
            L8 & $2 \times 16800^3$ & $6.09 \times 10^3$ & $1.14 \times 10^3$ & $276.79$ & $34.60$\\
            L16 & $2 \times 33600^3$  & $7.62 \times 10^2$ & $ 1.42 \times 10^2$ & $138.30$ & $17.30$\\
		\hline
	\end{tabular}
\end{table}

\begin{table}
    \centering
    \caption{The set of \thesanzoom simulations used for analyses in this paper. From left to right, each column indicates:\\ 
    (1) Names of the simulated galaxies, with the numbers representing the logarithm of halo mass at $z=3$ (in the corresponding DM-only run) and colors consistent with the ones adopted later in plots. \\
    (2) The halo mass, $M_{\rm crit,200}$, of the main target halo at $z=3$ (in the corresponding DM-only run).  \\
    (3) The maximum resolution level that the galaxy has been simulated with. \\
    (4,5,6) The model variants that have been experimented with (see Section~\ref{sec:sim-variants} for details).}
    \label{table:simtab}
    \addtolength{\tabcolsep}{0pt}
    \def\arraystretch{1.2}
    \begin{tabular}{lllccc} 
	\hline
    Name & $M_{\rm halo}(z=3)$ & max res. & no add.  & no $R_{\rm lim}$ & vary cell-\\
    & [$\msun$] & level & ESF & for SNe & level $\epsilon_{\rm ff}$\\
  \hline
  \textcolor{mycolor1}{m$13.0$} & $8.93 \times 10^{12}$ & L4 & \text{\sffamily x} & \text{\sffamily x} & \text{\sffamily x}\\
  \textcolor{mycolor2}{m$12.6$} & $4.07 \times 10^{12}$ & L4 & \text{\sffamily x} & \text{\sffamily x} & \text{\sffamily x}\\
  \textcolor{mycolor3}{m$12.2$} & $1.58 \times 10^{12}$ & L4 & \text{\sffamily x} & \text{\sffamily x} & \text{\sffamily x}\\
  \textcolor{mycolor4}{m$11.9$} & $7.70 \times 10^{11}$ & L4 & \checkmark & \text{\sffamily x} & \text{\sffamily x}\\
  \textcolor{mycolor5}{m$11.5$} & $3.28 \times 10^{11}$ & L4 & \checkmark & \text{\sffamily x} & \text{\sffamily x}\\
  \textcolor{mycolor6}{m$11.1$} & $1.40 \times 10^{11}$ & L8 & \checkmark & \checkmark & \checkmark \\
  \textcolor{mycolor7}{m$10.8$} & $5.93 \times 10^{10}$ & L8 & \checkmark & \text{\sffamily x} & \text{\sffamily x}\\ 
  \textcolor{mycolor8}{m$10.4$} & $2.53 \times 10^{10}$ & L8 & \checkmark & \checkmark & \checkmark \\ 
  \textcolor{mycolor9}{m$10.0$} & $1.07 \times 10^{10}$ & L8 & \text{\sffamily x}& \text{\sffamily x} & \text{\sffamily x} \\ 
  \textcolor{mycolor10}{m$9.7$ } & $4.58 \times 10^{9}$  & L16& \checkmark & \checkmark & \checkmark \\ 
  \textcolor{mycolor11}{m$9.3$ } & $1.95 \times 10^{9}$  & L16& \checkmark & \checkmark & \checkmark \\ 
  \textcolor{mycolor12}{m$8.9$ } & $8.29 \times 10^{8}$  & L16& \checkmark & \checkmark & \checkmark \\ 
  \textcolor{mycolor13}{m$8.5$ } & $3.51 \times 10^{8}$  & L16& \checkmark & \checkmark & \checkmark \\
  \textcolor{mycolor14}{m$8.2$ } & $1.52 \times 10^{8}$  & L16& \checkmark & \checkmark & \checkmark \\
\hline
 \end{tabular}%
\end{table}

\subsection{Radiation-hydrodynamics}

The analysis in this paper is based on the \thesanzoom simulation suite. The overview of the suite and the introduction of simulation methods are presented in \citet{Kannan2025}. This section briefly summarizes the key numeric techniques and galaxy formation model ingredients in this simulation suite. Target DM haloes for zoom-in simulations are drawn from the parent DM-only counterpart of the fiducial simulation in the \thesan project~\citep{Kannan2022thesan,Garaldi2022,Smith2022,Garaldi2024}. In total, we select $14$ galaxies at $z=3$ covering the halo mass range $10^{8}$ to $10^{13}\msun$. The zoom-in initial conditions were created with a new code that allows arbitrarily shaped high-resolution regions (Puchwein et al. in prep.). The \thesanzoom simulations utilized the massively parallel, multi-physics simulation code \textsc{Arepo}~\citep{Springel2010,Pakmor2016,Weinberger2020arepo} with its radiation hydrodynamic extension~\citep{Kannan2019,Zier2024}. The gravitational forces are calculated using a hybrid approach, with the short-range forces computed using a hierarchical Tree algorithm~\citep{Barnes1986} and the long-range forces calculated using the Particle Mesh method~\citep{Aarseth2003}. Radiation-hydrodynamic (RHD) equations are solved using a quasi-Lagrangian Godunov scheme on a moving, unstructured Voronoi mesh following the motion of the gas. Radiation fields are simulated by casting the radiative transfer (RT) equation into a set of hyperbolic conservation laws of its zeroth and first moments, i.e. photon number density and flux~\citep{Kannan2019}, closed using the M1 scheme~\citep{Levermore1984,Dubroca1999}. We discretize the radiation field in seven radiation bins, including the infrared (IR), optical, far-UV, Lyman-Werner, H ionizing bands, and the two He ionizing bands. The numeric parameters and information of the main target halos for zoom-in simulations are listed in Table~\ref{table:res} and Table~\ref{table:simtab}, respectively.

One of the key innovations of \thesanzoom is the modeling of external radiation fields, developed during the EoR with an inhomogeneous geometry. The external radiation is an important source of feedback for low-mass haloes that can suppress gas inflows and star-formation~\citep[e.g.][]{Rees1986, Shapiro2004, Okamoto2008}. A strong external radiation field can reduce the number of low-mass galaxies and the production of ionizing photons from these sources, which weaken the local contribution to the radiation field. Self-consistently modeling this complicated cyclical feedback loop requires radiation hydrodynamics coupled with accurate galaxy formation models~\citep{Pawlik2017,Borrow2023}. However, most of the previous cosmological simulations typically adopt a redshift-dependent but spatially uniform UV background~\citep[e.g.][]{FG2009,Haardt2012} to simulate the impact of the large-scale radiation fields. These models can result in an artificial sharp transition between a fully neutral and a fully ionized Universe~\citep{Puchwein2019,Borrow2023} and miss the patchiness of reionization and the radiation field intensity~\citep[e.g.][]{Puchwein2023}. The self-shielding approximations~\citep{Rahmati2013} adopted in these models can also break down during the development of radiation fields during the process of reionization. We take a different approach since the \thesanzoom galaxies are selected from the parent \thesanone simulation, which simultaneously models the large-scale structure, the properties of the galaxies, and the evolution of the radiation field around the selected haloes. We save the radiation field maps from the parent simulation with a high cadence and use it as the boundary condition of the zoom-in runs, by interpolating them in space and time. Inflowing radiation is then propagated into the high-resolution region. This allows for a more realistic treatment of reionization for objects that are inefficient sources of ionizing radiation and reionize outside-in~\footnote{The parent \thesanone simulation adopted a sub-grid model for the unresolved ISM~\citep{Springel2003a}. We note that inconsistencies could exist in the external radiation fields as the zoom-in regions are simulated with the more advanced ISM model. The reionization history in Thesan is however in good agreement with observational constraints, suggesting that the radiation field in the IGM is reasonably realistic.}. At low redshift, after the final output of the parent simulation ($z=5.5$), we smoothly switch to setting the external radiation field based on a homogeneous UV background model~\citep{FG2009}.

In Figure~\ref{fig:intro}, we present a visual inspection of a \thesanzoom galaxy ``m12.6'' at $z\simeq 6$. We show the surface density of neutral and molecular gas and Lyman-continuum (LyC) radiation field\footnote{To be specific, we show the photon density in the H ionizing band, which has been corrected for the reduced speed of light in the simulation. The physical photon density should be $100$ times smaller than the values recorded here.} around this galaxy. The thickness of the layer for projection is set the same as the field-of-view. In the top row from left to right, we show zoom-in views of galaxy stellar light in UV-optical bands, neutral gas distribution, LyC radiation field, and gas temperature. 

\subsection{Cooling, star-formation, and feedback}

Gas is coupled to the radiation fields using a non-equilibrium thermochemical network~\citep{Kannan2020}, which calculates the non-equilibrium abundance of ${\rm H}_{2}, \HI, \HII, \HeI, \HeII,$ and $\HeIII$ as well as the resulting primordial cooling rates. In addition, we include tabulated cooling rates for metals, photoelectric heating, cooling from dust-gas-radiation interactions, and Compton cooling/heating from the cosmic microwave background. Star-formation happens in dense (limited to $n_{\rm H}>10\,\cm^{-3}$ but typically severl orders of magnitude denser), self-gravitating~\citep{Hopkins2013}, Jeans-unstable~\citep{Truelove1997} gas. The cell-level SFE per free-fall is assumed to be 100\%. According to this rate, collisionless particles representing stellar populations are spawned stochastically from gas cells with the probability drawn from a Poisson distribution. 

Following \citet{Marinacci2019}, we model stellar feedback from SNe and stellar winds from young massive OB stars and AGB stars. Assuming a \citet{Chabrier2003} stellar initial mass function, we compute the SNe rate for each stellar particle and model them as discrete events with additional time-stepping constraints such that the expected value for the number of SNe events per time-step is of the order of unity. Each SNe explosion injects the canonical $10^{51}\,\erg$ energy into the surrounding ISM within a coupling radius, which accounts for the fact that the energy/momentum from SNe is not expected to have a strong impact on ISM properties beyond the superbubble radius~\citep{Hopkins2018} and is taken to be $2$ physical kpc in the fiducial runs. Additional corrections for the unresolved Sedov-Taylor phase of the SNe blast wave are included with a terminal momentum~\citep{Hopkins2018feedback,Marinacci2019}. On the other hand, the mass, momentum, and energy injection rates of stellar winds are calculated using the analytical prescriptions in \citet{Hopkins2018,Hopkins2023fire3} which are based on the \textsc{Starburst99}~\citep{Leitherer1999} stellar evolution model. To model the radiative feedback from young massive stars, the luminosity and spectral energy density of stars as a function of age and metallicity are taken from the Binary Population and Spectral Synthesis models (BPASS; \citealt{Eldridge2017}), which are then tracked by the RT method described above. Photoionization, radiation pressure, and photoelectric heating are handled by the non-equilibrium thermochemical network~\citep{Kannan2020,Kannan2021}. An additional empirical ESF is included in the first $5\Myr$ after the formation time of a stellar particle with total momentum injection rates comparable to the SNe feedback. As discussed in detail in \citet{Kannan2025}, this gives better agreement with the stellar-mass-halo-mass relations at high redshifts~\citep[e.g.][]{Tacchella2018,Behroozi2019} and may represent missing physics in the simulations. Dust is modelled as a scalar property of the gas cells as outlined in \citet{McKinnon2016, McKinnon2017}, which includes the production of dust from SNe and AGB stars, the growth in dense ISM, and the destruction by SNe shocks and sputtering. One caveat of the model is the physics of supermassive black holes (SMBHs) and feedback from active galactic nuclei (AGN) are not included, of which the impact on high-redshift galaxies is debated~\citep[e.g.][]{Kimmig2025-agn,Kokorev2024-agn,Maiolino2024-agn,Silk2024-agn} and may affect the most massive galaxies in \thesanzoom when they reach $M_{\rm halo}\gtrsim 10^{12}\msun$. This will be explored in follow-up simulations.

\subsection{Simulations with model variations}
\label{sec:sim-variants}

In addition to the fiducial setup introduced above, \thesanzoom includes several sets of runs with variants of the physics models. Here, we will discuss the ones relevant to this paper and refer readers to \citet{Kannan2025} for details of the full suite.

\vspace{0.1cm}

\noindent (1) \textbf{Varying cell-level SFE:} In the fiducial runs, we convert gas cells that satisfy the star-forming criteria with a cell-level SFE per free-fall time $\epsilon_{\rm ff}=100\%$. The star-formation rate (SFR) of a gas cell is therefore computed as ${\rm SFR}=\epsilon_{\rm ff}\,m_{\rm cell}/t_{\rm ff}(\rho_{\rm cell})$, where $t_{\rm ff}(\rho_{\rm cell})\equiv \sqrt{3\pi/32\,{\rm G}\,\rho_{\rm cell}}$ is the free-fall time at the density of the cell. The underlying assumption of taking $\epsilon_{\rm ff}$ to unity is the rapid dissipation of turbulence to the viscous scale and therefore negligible time for further fragmentation and star-formation below the resolution scale~\citep[e.g.][]{Hopkins2014}. This is connected to the results of higher-resolution simulations of turbulent clouds~\citep[e.g.][]{Padoan2011}. However, smaller values of $\epsilon_{\rm ff}$ are suggested in studies accounting for the unresolved density and turbulence fields~\citep[e.g.][]{Krumholz2005,Hennebelle2011,Semenov2016} and appear more consistent with observations of star-formation in dense gas~\citep[e.g.][]{Krumholz2012,SunJY2023}. It has been shown in many works that the choice of $\epsilon_{\rm ff}$ can significantly affect the distribution functions of dense gas in the ISM, the physical properties of GMCs captured in simulations, as well as potential observables that are sensitive to dense gas. Motivated by these, in one set of runs, $\epsilon_{\rm ff}$ is changed from 100\% to a variable value. It starts off small with 1\% at the threshold density of $n_{\rm H} = 10 \cm^{-3}$ and scales linearly with the density of the gas to reach a maximum value of 100\% at $n_{\rm H} \geq 10^{3} \cm^{-3}$.

\vspace{0.1cm}

\noindent (2) \textbf{Remove the additional ESF:} In the fiducial runs, we impose an additional ESF acting in the first 5 Myr after the birth of a single stellar population. The momentum injection rate is set to $1000\kms \Myr^{-1}$ per stellar mass formed. This is an empirical but necessary component to match observational constraints when ESF from stellar winds is suppressed in low-metallicity environments at high redshifts. This may represent missing additional physics in our simulations, like cosmic-rays~\citep[e.g.][]{Pakmor2016b,Buck2020,Hopkins2020-cr}, magnetic fields~\citep[e.g.][]{Marinacci2016,Hopkins2020-cr}, Lyman-$\alpha$ radiation pressure~\citep[e.g.][]{Smith2017, Kimm2018, Nebrin2025} or other numerical uncertainties. Nevertheless, in one set of runs, we experiment with removing this additional component and investigate its impact on galaxy properties.

\vspace{0.1cm}

\noindent (3) \textbf{Remove the limiting radius of SNe feedback coupling:} In the fiducial model, we assume that the energy/momentum from SNe does not have a strong impact on ISM beyond the superbubble radius~\citep{Hopkins2018}. This radius depends on the energy of SNe and other ISM properties. For simplicity, following \citet{Marinacci2019}, we use a constant value of the limiting radius 2 kpc for our fiducial model. In one set of runs, we experiment with removing this limiting radius.

\subsection{Halo catalog and merger trees}

In \thesanzoom, the DM haloes are identified using the friends-of-friends~\citep[FOF;][]{Davis1985} algorithm with a linking length of $0.2$ times the initial mean inter-particle distance. Stellar particles and gas cells are attached to these FOF primaries in a secondary linking stage. The SUBFIND-HBT algorithm~\citep{Springel2021} is then used to identify gravitationally bound subhaloes. We trace the progenitors of subhaloes over time using the SUBFIND-HBT algorithm and document the growth histories of the most massive subhalo progenitor (and its host halo). In this work, we focus on the central galaxies, defined as the most massive galaxy (i.e. having the largest total mass of stellar particles associated with the subhalo) in a DM halo. Halo virial mass and radius are defined using spherical overdensity criterion as $M_{\rm halo} = M_{\rm crit,200}$ and $R_{\rm vir} = R_{\rm crit,200}$. Galaxy stellar mass ($M_{\ast}$) is defined as the sum of stellar particle masses within twice the stellar-half-mass radius. Additionally, all DM haloes that do not contain any low-resolution DM, stellar particles, or gas cells within $R_{\rm vir}$ are designated as uncontaminated and included in our analysis. 

\begin{figure}
    \includegraphics[width=\linewidth]{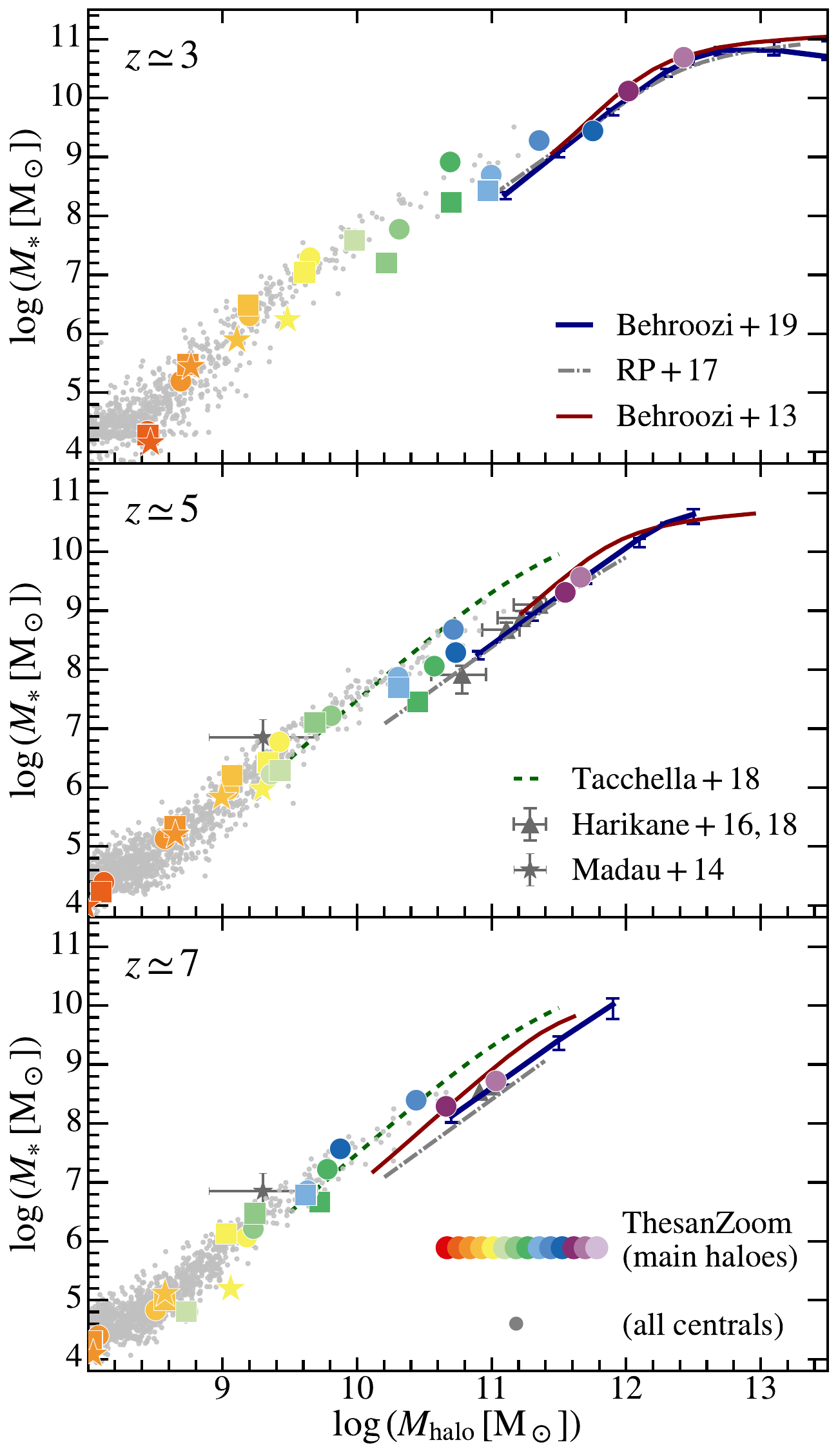}
    \caption{Stellar mass versus halo mass of simulated galaxies in \thesanzoom at $z=3, 5, 7$. We measure the median halo mass and stellar mass of a galaxy during its evolution within $z\pm 0.5$. The main target galaxies are shown with colors consistent with the choices in Table~\ref{table:simtab} and results from L4, L8, and L16 runs are shown in circles, squares, and stars, respectively. All central galaxies in the zoom-in region are shown as gray points in the background. We compare the results with empirical/abundance matching constraints from \citet{Behroozi2013, Behroozi2019, RP2017, Harikane2016, Harikane2018, Tacchella2018}. The gray stars with error bars show indirect constraints from the star-formation histories of local dwarf galaxies in \citet{Madau2014} at $z\sim 6$. The \thesanzoom results are in excellent agreement with these (indirect) observational constraints in overlapped mass ranges, (at least partially) as a consequence of the calibration of the additional ESF in the model.}
    \label{fig:mstar-mhalo}
\end{figure}

\begin{figure}
    \includegraphics[width=\linewidth]{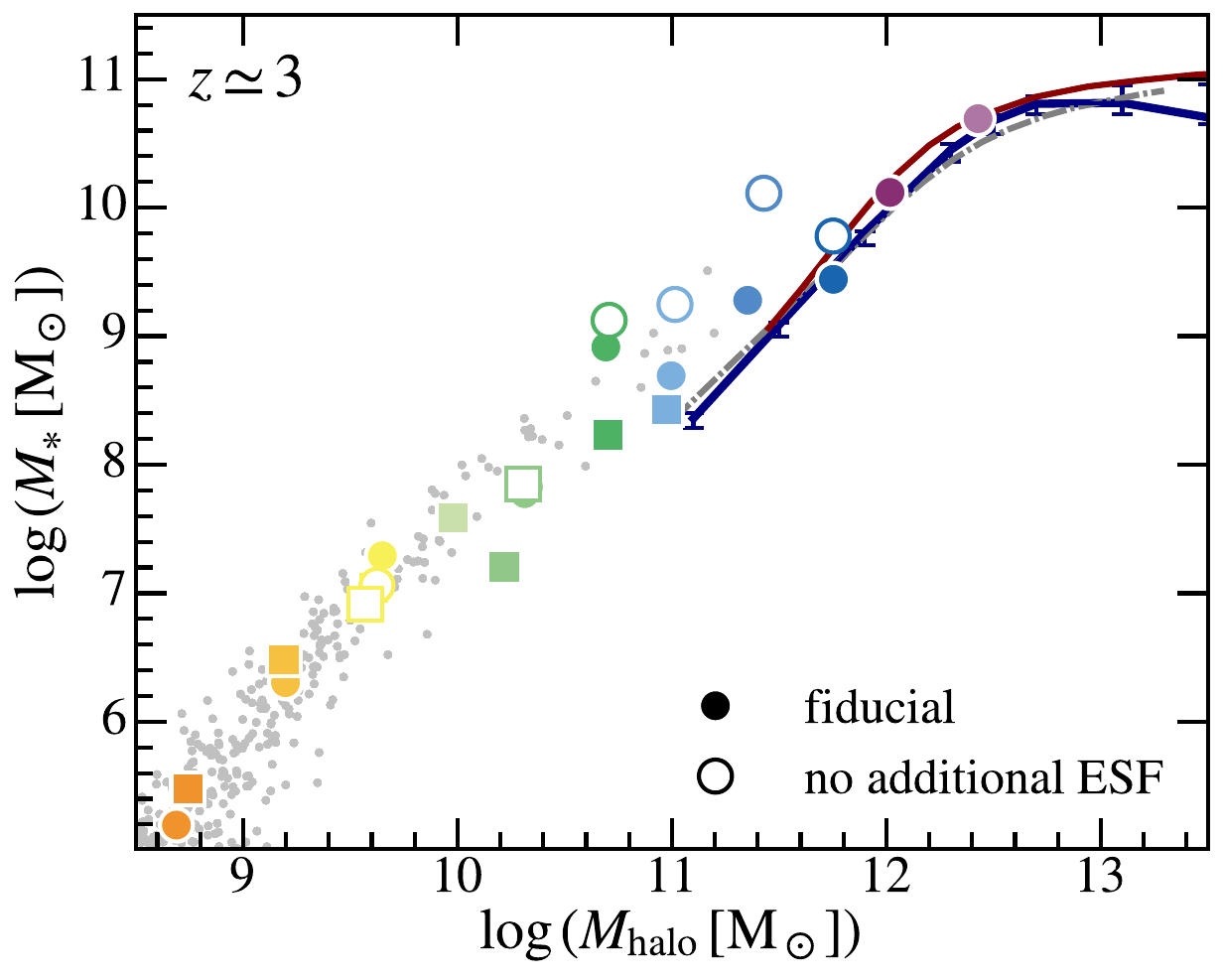}
    \caption{Stellar mass versus halo mass of simulated galaxies in runs with physics variants at $z\simeq 3$. We compare results in the fiducial runs and the runs with no additional ESF, since this is the only physics variant that gives rise to substantial differences in galaxy stellar-to-halo mass ratios. Results from L4 and L8 runs are shown in circles and squares, respectively. Without the additional ESF, galaxy stellar masses can be overpredicted by up to an order of magnitudes at $M_{\rm halo}\gtrsim 10^{11}\msun$. The additional ESF component does not affect the results in lower-mass galaxies or at higher redshifts.}
    \label{fig:mstar-mhalo-compare-noesf}
\end{figure}

\begin{figure*}
    \includegraphics[width=\linewidth]{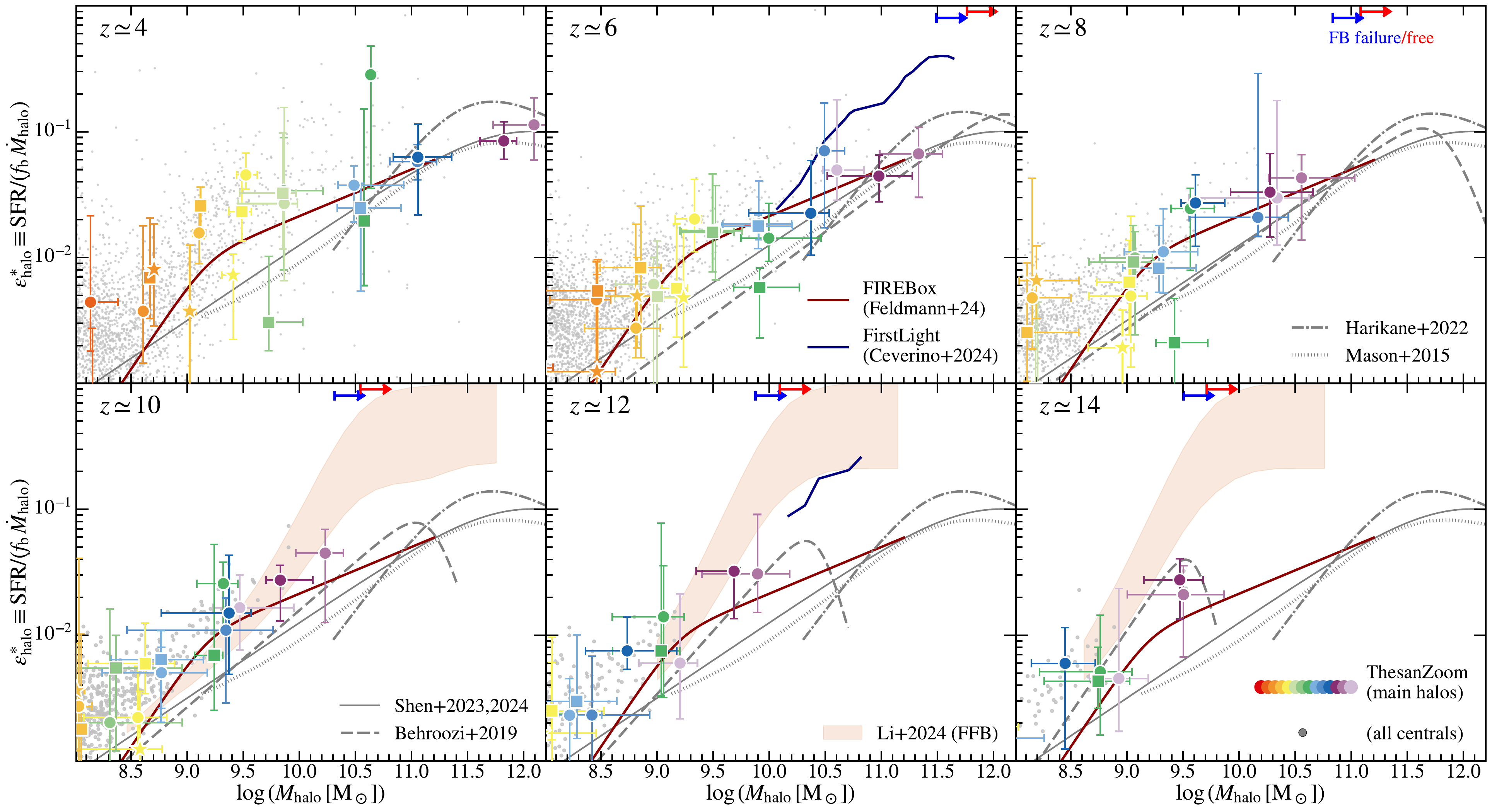}
    \caption{Halo-scale SFE of \thesanzoom galaxies from $z\simeq 4$ to $z\simeq 14$. As discussed in the main text, we take the growth track of galaxies within $2\times t_{\rm dyn}$ around the target redshift and measure SFE for each segment of a duration $t_{\rm dyn}$. The circles and vertical error bars show the median and 1$\sigma$ scatters of the SFE while the horizontal error bars show the 1$\sigma$ scatters of the halo mass during the subtracted growth track. Results from L4, L8, and L16 runs are shown in circles, squares, and stars, respectively. The gray points in the background are the median SFE of all central galaxies in the zoom-in regions of \thesanzoom. We compare our results with constraints through empirical models/abundance matching from \citet{Mason2015,Behroozi2019,Harikane2022,Shen2023,Shen2024b-ede} shown in gray lines. We also compare them with results from other cosmological hydrodynamic simulations, the \textsc{FIREBox}~\citep{Feldmann2024} in red and the \textsc{FirstLight}~\citep{Ceverino2017,Ceverino2024} in blue. The red and blue arrows on the top show the halo mass thresholds of the feedback-free~\citep{Dekel2023} and feedback-failure~\citep{BK2024} scenarios, which could lead to order-unity halo-scale SFE. The orange-shaded region shows a relaxed form of the feedback-free scenario considering the duty cycles of starbursts~\citep{Li2024}.}
    \label{fig:halo-sfe}
\end{figure*}

\section{Halo-scale star-formation efficiency}
\label{sec:result-epshalo}

In this section, we study the efficiency of converting baryons to stars at the scale of a DM halo. We define the instantaneous halo-scale SFE as
\begin{equation}
    \label{eq:epshalo}
    \epsilon^{\ast}_{\rm halo} \equiv {\rm SFR}/\dot{M}_{\rm gas} \simeq {\rm SFR}/(f_{\rm b}\,\dot{M}_{\rm halo}),
\end{equation}
where $f_{\rm b} \equiv \Omega_{\rm b}/\Omega_{\rm m} \simeq 0.16$ is the baryon mass fraction of the Universe, and $\dot{M}_{\rm halo}$ ($\dot{M}_{\rm gas}$) is the total (gas) mass accretion rate of the halo. To measure this in practice, we take a segment of a target galaxy's main progenitor evolution history with a duration of $t_{\rm dyn} \equiv R_{\rm vir}/V_{\rm vir}$, compute the change in galaxy stellar mass and host halo mass, and obtain the averaged $\epsilon^{\ast}_{\rm halo}$ as $\Delta M_{\ast}/(f_{\rm b}\,\Delta M_{\rm halo})$. We repeat this for all segments found within $2\times t_{\rm dyn}$ around a target redshift and report the median and 1$\sigma$ scatters of $\epsilon^{\ast}_{\rm halo}$. In principle, $\Delta M_{\ast}$ and $\Delta M_{\rm halo}$ here can include mass growth through mergers, which can contaminate the measured $\epsilon^{\ast}_{\rm halo}$ if strong halo mass or redshift dependence exist. However, as will be shown later in Section~\ref{sec:finsitu}, the majority of the stellar mass is built in-situ in \thesanzoom galaxies.

\subsection{Stellar-to-halo mass relations -- integrated halo-scale SFE}

If we integrate Equation~(\ref{eq:epshalo}), we can obtain the galaxy stellar mass
\begin{equation}
    M_{\ast} = \int \epsilon^{\ast}_{\rm halo}(M_{\rm halo}, z)\,f_{\rm b}\,{\rm d}M_{\rm halo}.
    \label{eq:epshalo-integral}
\end{equation} 
This integral gives the stellar versus halo mass relation of galaxies, representing an integrated version of the halo-scale SFE. Since this relation is more often used in literature for model calibration, we will first investigate it before moving back to the instantaneous halo-scale SFE.

In Figure~\ref{fig:mstar-mhalo}, we present the stellar versus halo mass relations of \thesanzoom galaxies at three representative redshifts. We show results for both the main target galaxies and all central galaxies in the zoom-in region (with $M_{\rm halo}\geq 10^{8}\msun$ and uncontaminated from low-resolution particles). The galaxy stellar mass exhibits a tight correlation with the host halo mass, which does not display apparent evolution through redshifts. The stellar-to-halo-mass ratios are always significantly lower than the universal baryon fraction and are decreasing towards lower halo masses, which is a result of both the internal stellar feedback and the external feedback from radiation backgrounds. The stellar-to-halo-mass ratios we obtain agree with the observational constraints using empirical model fitting or abundance matching~\citep[e.g.][]{Behroozi2013, Behroozi2019, RP2017,Harikane2016, Harikane2018} in relatively massive haloes probed by observations. At $M_{\rm halo}\sim 10^{9}\msun$, we compare our results with the indirect constraints in \citet{Madau2014} based on the star-formation histories of local dwarf galaxies and also find decent agreement. In Figure~\ref{fig:mstar-mhalo-compare-noesf}, we show the same relation at $z\simeq 3$ in the fiducial runs versus runs without the additional ESF. Out of all the physics variants we have tested, this is the only one that substantially affects the stellar-to-halo-mass ratios. At $z\simeq 3$, without the additional ESF, stellar masses can be overpredicted by up to an order of magnitude at $M_{\rm halo} \gtrsim 10^{11}\msun$. The differences are negligible at higher redshifts but potentially due to the limited halo mass range covered by \thesanzoom. Since the discrepancy only shows up in the massive end, it hints that the dominant ESF in low-mass galaxies ($M_{\rm halo} \gtrsim 10^{11}\msun$) at high redshifts is not the additional ESF we include. Given the weak contribution from stellar winds in low-metallicity environments at these redshifts~\citep[e.g.][]{Hirschi2007,Dekel2023}, the dominant ESF mechanism is likely the radiative heating from young massive stars and external radiation fields. The agreement with observational constraints in the massive end should be understood as a consequence of the calibration we perform by varying the strength of the additional ESF component in the model.

\begin{figure}
    \centering
    \includegraphics[width=\linewidth]{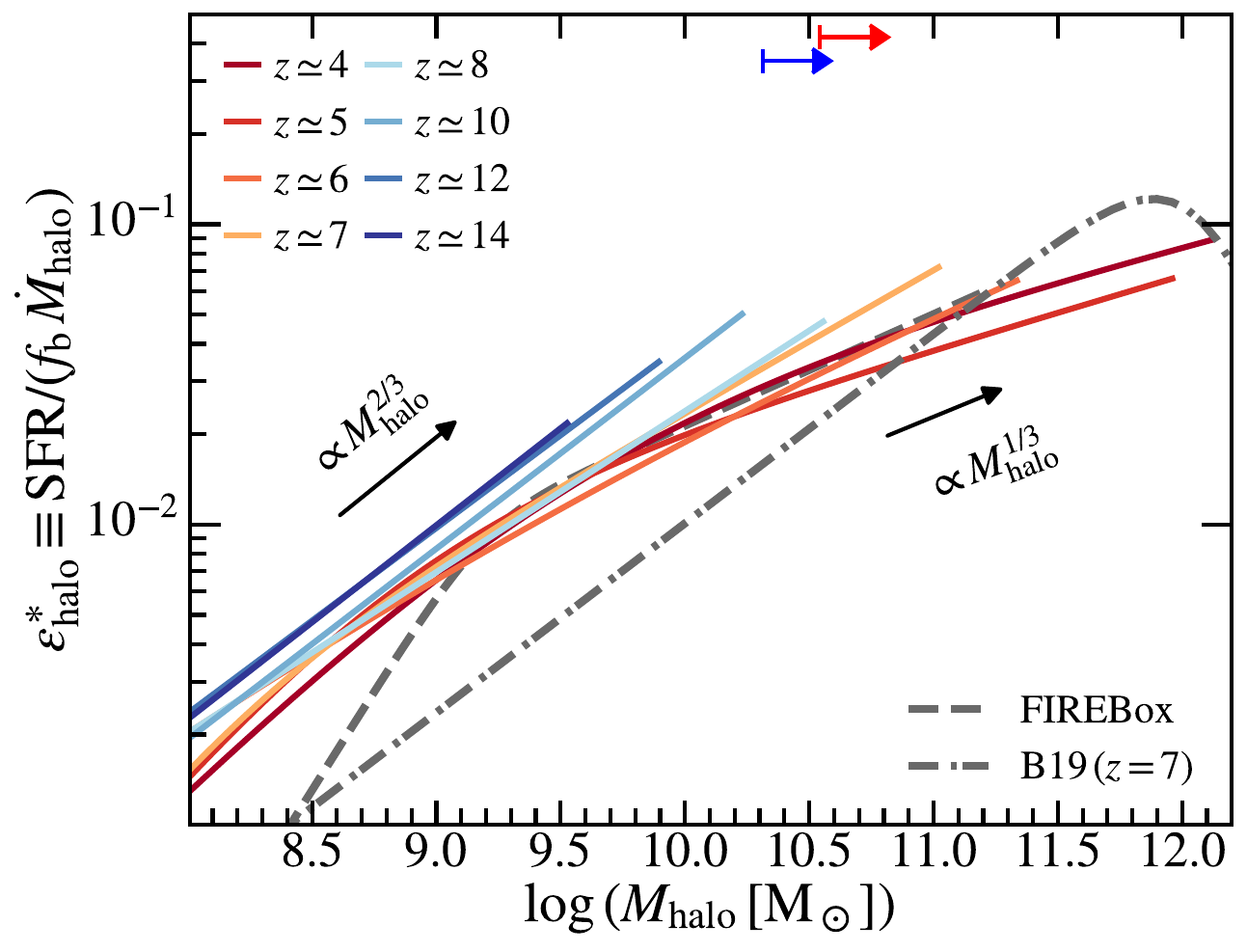}
    \hspace*{0.008\linewidth}\includegraphics[width=0.995\linewidth]{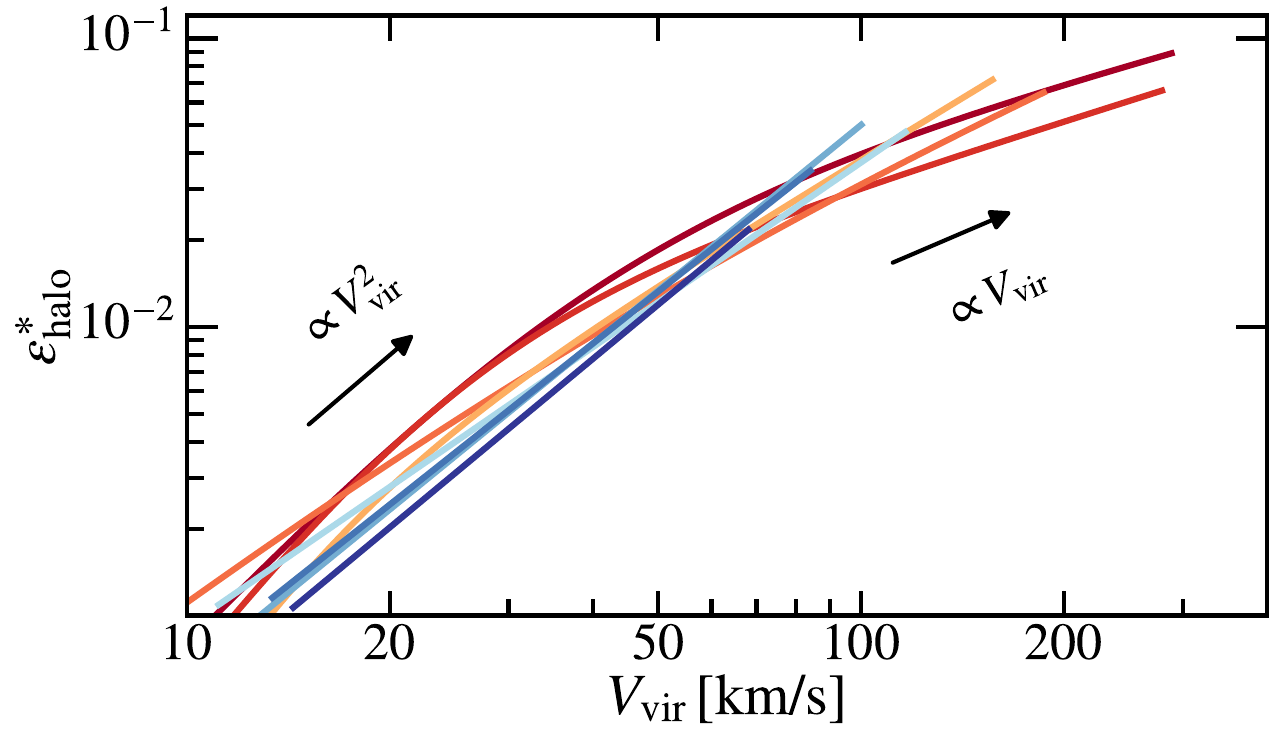}
    \caption{Halo-scale SFE versus halo mass (top) and virial velocity (bottom) of \thesanzoom galaxies at different redshifts. We show the best-fit double power-law relations from $z\simeq 4$ to $z\simeq 14$. In the top panel, red and blue arrows show the same feedback-free/failure thresholds (at $z=10$) as in Figure~\ref{fig:halo-sfe}. For clarity, we only keep the results from FIREBox~\citep[dashed line;][]{Feldmann2024} and UniverseMachine~\citep[dot-dashed line;][]{Behroozi2019} for comparison. In the low-mass end ($M_{\rm halo}\lesssim 10^{9}\msun$), there is almost no redshift evolution but the SFEs are higher than the redshift-independent values found in FIREBox. Compared to UniverseMachine, we predict up to half a dex higher SFE at $M_{\rm halo}\lesssim 10^{11}\msun$. At low redshifts, the relation has a double power-law shape with a characteristic slope of roughly $1/3$ ($2/3$) in the high-mass (low-mass) end. This transition to a single power-law of slope $\sim 2/3$ at higher redshifts with a mild increase of SFE at $M_{\rm halo}\gtrsim 10^{9.5}\msun$. However, due to the limited halo mass range covered by \thesanzoom, we cannot conclude (with statistical significance) if the same scaling extends to more massive haloes. The corresponding dependence on virial velocity is $\propto V_{\rm vir}$ and $\propto V^2_{\rm vir}$ at high and low masses, respectively, which likely reflects the nature of gas outflows (see discussions in the main text).}
    \label{fig:halo-sfe-compare-z}
\end{figure}

\subsection{Instantaneous halo-scale SFE}

In Figure~\ref{fig:halo-sfe}, we present the instantaneous halo-scale SFE of \thesanzoom galaxies versus host halo mass from $z\simeq 4$ to $z\simeq 14$. The median and 1$\sigma$ scatter of the SFE (along the segment of halo growth histories discussed at the beginning of Section~\ref{sec:result-epshalo}) is represented by points and their vertical error bars. The horizontal error bars indicate the halo mass range along the segment. The halo-scale SFE shows a clear dependence on halo mass, with a mild increase from $\lesssim 1\%$ in low-mass dwarf haloes ($\lesssim 10^{9}\msun$) to $\sim 10\%$ in Milky Way-mass haloes ($\sim 10^{12}\msun$). The redshift dependence of halo-scale SFE is limited in the halo mass range probed by \thesanzoom. We compare the results to constraints from abundance matching or empirical model calibrations~\citep{Mason2015,Behroozi2019,Harikane2022,Shen2023,Shen2024b-ede}. At $M_{\rm halo}\lesssim 10^{10}\msun$, the halo-scale SFE in \thesanzoom is about a factor of $2-3$ higher than most of the empirical model constraints. This could be due to either limited observational data used for calibration that biases SFE towards low-redshift results or ansatz about halo mass functions and accretion histories in these works that are different from our simulation predictions. The mild increase of $\epsilon^{\ast}_{\rm halo}$ at $z\gtrsim 12$ agrees better with the constraints from the UniverseMachine~\citep{Behroozi2019}. Since these empirical constraints of halo-scale SFE are often the backbones of canonical empirical/semi-analytical models of galaxy formation, the enhanced SFE we find in \thesanzoom may have implications for the ``overabundance'' of bright galaxies at $z\gtrsim 10$ revealed by \textit{JWST}, which will be discussed in Section~\ref{sec:discussion}. 

We also compare our results to other cosmological hydrodynamic simulations, FIREBox(-HR)~\citep[``100 Myr-average'' results;][]{Feldmann2024} and FirstLight~\citep{Ceverino2024}. In general, our results are in good agreement with the ``redshift-independent'' relation found in FIREBox despite higher SFE in low-mass haloes at $\lesssim 10^{9}\msun$. Compared to the FirstLight simulations, we find at $\gtrsim 10^{11}\msun$ halo-scale SFE smaller by roughly half an order of magnitude at $z\simeq 6$. At $z\simeq 12$, their results are outside the halo mass range covered by \thesanzoom but would more or less lie on the extrapolation of our $\epsilon^{\ast}_{\rm halo}$ if a single power-law dependence on $M_{\rm halo}$ is assumed (see also the fitting later in this section). 

Motivated by recent \textit{JWST} observations of bright and potentially massive galaxies at high redshifts, many mechanisms have been proposed that can drive efficient star-formation in extreme environments at these redshifts. For example, \citet{Dekel2023} and \citet{Li2024} discussed a feedback-free starburst scenario, where the free-fall times of star-forming regions become short enough to evade SNe feedback while ``early'' feedback from stellar winds is suppressed in low-metallicity environments at high-redshift. \citet{BK2024} considered a feedback-failure scenario: enhanced DM density in massive haloes at high redshifts provides deep gravitational potentials such that any form of feedback would fail to dissociate dense giant molecular clouds~\citep[e.g.][]{Grudic2018,Hopkins2022,Menon2024} and regulate star-formation. We over-plot the mass thresholds for these mechanisms to operate in Figure~\ref{fig:halo-sfe}. In the halo mass range probed by \thesanzoom, we find no evidence for a sharp transition of $\epsilon^{\ast}_{\rm halo}$ to order unity at these masses. This could be due to the limited duty cycle of the phase of efficient star-formation in these scenarios, which stays as an unconstrained free parameter. We also show the feedback-free starburst predictions from \citet{Li2024}, which considered the duty cycle of feedback-free starburst and constrained this free parameter using observations. However, with the limited halo mass range covered by \thesanzoom, we cannot draw conclusions about haloes deep enough in the feedback-free/failure regime to support or challenge these scenarios.

\begin{table}
\centering
\addtolength{\tabcolsep}{0.3pt}
\def\arraystretch{1.3}
\caption{Best-fit double power-law parameters for $\epsilon^{\ast}_{\rm halo}(M_{\rm halo})$ at each redshift (see Equation~\ref{eq:dpl}). In practice, the fits reduce to a single power-law at $z\geq 8$ (but could be due to an insufficient number of massive haloes). Therefore, we fix $M_0$ and the low-mass end slope $\beta$ becomes a redundant parameter colored in gray.}
\begin{tabular}{lclllll}
\hline
z  &  $\log{(M^{\rm max}_{\rm halo}/\msun)}$ &  $\log{(M_{0}/\msun)}$  &  $\log{(\epsilon_{0})}$ &  $\alpha$  &  $\beta$  \\
\hline
$4 $ &  $12.12$ &  $9.50\,({\rm fixed})$ &  $-1.59$ &  $0.21$ &  $0.83$ \\
$5 $ &  $11.96$ &  $8.77$ &  $-1.96$ &   $0.24$ &  $1.03$ \\
$6 $ &  $11.33$ &  $9.32$ &  $-1.73$ &   $0.31$ &  $0.62$ \\
$7 $ &  $11.02$ &  $7.89$ &  $-2.62$ &   $0.47$ &  $1.33$ \\
$8 $ &  $10.56$ &  $9.00\,({\rm fixed})$ &  $-1.86$ &   $0.54$ &  \textcolor{gray}{$0.54$} \\
$10$ &  $10.23$ &  $9.00\,({\rm fixed})$ &  $-1.78$ &   $0.64$ &  \textcolor{gray}{$0.64$} \\
$12$ &  $ 9.90$ &  $9.00\,({\rm fixed})$ &  $-1.72$ &   $0.62$ &  \textcolor{gray}{$0.62$} \\
$14$ &  $ 9.53$ &  $9.00\,({\rm fixed})$ &  $-1.73$ &   $0.65$ &  \textcolor{gray}{$0.65$} \\
\hline
\end{tabular}
\label{apptab:bestfit}
\end{table}

At each redshift, we fit the $\epsilon^{\ast}_{\rm halo}-M_{\rm halo}$ relation with a double power-law function
\begin{equation}
    \epsilon^{\ast}_{\rm halo} = \dfrac{\epsilon_0}{(M_{\rm halo}/M_0)^{-\alpha} + (M_{\rm halo}/M_0)^{-\beta}},
    \label{eq:dpl}
\end{equation}
where $\alpha$ and $\beta$ are the high- and low-mass end slopes, if we force $\alpha \leq \beta$. For the fitting, we combine the SFEs of the main target haloes and all central galaxies in the zoom-in region (with equal weights between the two groups of data). The best-fit parameters are summarized in Table~\ref{apptab:bestfit}. In the top panel of Figure~\ref{fig:halo-sfe-compare-z}, we show the best-fit $\epsilon^{\ast}_{\rm halo}-M_{\rm halo}$ relations from $z\simeq 4$ to $z\simeq 14$. We find that $\epsilon^{\ast}_{\rm halo}\sim M^{1/3}_{\rm halo}$ at $M_{\rm halo}\gtrsim 10^{9}\msun$ and the slope becomes steeper to about $\epsilon^{\ast}_{\rm halo}\sim M^{2/3}_{\rm halo}$ at lower masses. For reference, we show the FIREBox result as in Figure~\ref{fig:halo-sfe} again, which almost overlaps with our $z\lesssim 6$ relations at $M_{\rm halo}\gtrsim 10^{9}\msun$. The agreement is probably not surprising given the similar numerical hydrodynamic methods and architecture of the galaxy formation model adopted. However, at lower masses, \thesanzoom predicts a steeper decline of $\epsilon^{\ast}_{\rm halo}$, which likely stems from the more appropriate RT and treatments of UV background and thermochemistry. Unlike FIREBox, which reported a redshift-independent relation, we find that the $\epsilon^{\ast}_{\rm halo}-M_{\rm halo}$ relation becomes steeper approaching higher redshifts and approximates to a single power-law at least in the mass range covered by \thesanzoom. We note that the fits at these redshifts are highly sensitive to the limited number of massive haloes in \thesanzoom, and should therefore be interpreted with caution. Meanwhile, in the low-mass end, no significant redshift dependence is found. 

What might determine the slope of the $\epsilon^{\ast}_{\rm halo}-M_{\rm halo}$ relation? A simple analytical relation can be derived from the nature of feedback-driven winds~\citep[e.g.][]{Feldmann2024}. From the conservation of baryon mass in the halo~\citep[e.g.][]{Bouche2010,Dave2012}, we first have ${\rm SFR}\,(1+\eta) \simeq f_{\rm b}\,\dot{M}_{\rm halo}$, where $\eta$ is the mass-loading factor of the feedback-driven wind. Reorganizing the terms, we obtain $\epsilon^{\ast}_{\rm halo}\propto (1+\eta)^{-1} \sim \eta^{-1}$, where we assume $\eta \gg 1$. If the wind is momentum-driven, we expect~\citep[e.g.][]{Murray2005} 
${\rm SFR}\,(P_{\ast}/m_{\ast}) = \dot{M}_{\rm w}\,V_{\rm w} = \eta\,{\rm SFR}\,V_{\rm w}$, where $(P_{\ast}/m_{\ast})$ is a roughly constant momentum injection from stellar feedback per unit stellar mass formed, $V_{\rm w}$ is the terminal velocity of the wind. This implies $\eta \sim V^{-1}_{\rm w}$. If we further assume that $V_{\rm w}$ scales with the escape velocity of the halo ($V_{\rm esc} \propto V_{\rm vir}\propto M^{1/3}_{\rm halo}$), we obtain $\epsilon^{\ast}_{\rm halo}\propto M^{1/3}_{\rm halo}$. Such a scaling applies to massive haloes\footnote{Further flattening or a reversed scaling at even higher halo masses could exist due to AGN feedback, which is not included in the \thesanzoom simulations.} where the superbubbles generated by clustered SNe feedback do not break out the ISM while radiative cooling is efficient at the shell of ejecta~\citep[e.g.][]{Kim2017-wind,Fielding2018-wind}. Feedback momentum is conserved in driving galactic-scale wind. The halo mass dependence is consistent with the findings from relatively massive haloes in \thesanzoom. In Section~\ref{sec:result-epsgal}, we will derive the same scaling but instead from the ``microscopic'' features of star-forming complexes assuming an equilibrium of turbulent energy dissipation and injection from feedback.

On the other hand, if the wind is energy-driven, we expect~\citep[e.g.][]{Chevalier1985-wind,Murray2005} 
${\rm SFR}\,(E_{\ast}/m_{\ast}) = \eta\,{\rm SFR}\,V^{2}_{\rm w}$, where $(E_{\ast}/m_{\ast})$ is a constant energy output from stellar feedback per unit stellar mass formed. Similarly, we obtain $\epsilon^{\ast}_{\rm halo} \sim V^{2}_{\rm vir} \propto M^{2/3}_{\rm halo}$, which better applies to low-mass haloes in \thesanzoom. In this regime, superbubbles driven by feedback can break out, and most of the SNe ejecta can freely vent out the ISM in an energy-conserving fashion~\citep[e.g.][]{Fielding2018-wind}. Another possibility for energy-driven wind is that radiative heating becomes most important and evaporates gas out of the halo~\citep{Rees1986,Shapiro2004,Okamoto2008}. This happens when the virial temperature drops below roughly $10^{4}\,{\rm K}$, corresponding to $M_{\rm halo}\sim 10^{9}\msun$ at these redshifts, which roughly agrees with the break halo mass we find. Therefore, the $\epsilon^{\ast}_{\rm halo}-M_{\rm halo}$ relation in \thesanzoom can be interpreted as an energy-driven mode in low-mass haloes and a momentum-driven mode in massive haloes. In the bottom panel of Figure~\ref{fig:halo-sfe-compare-z}, we illustrate this idea more directly by showing $\epsilon^{\ast}_{\rm halo}$ versus $V_{\rm vir}$. The transition between the two modes has previously been found in simulations of low-redshift galaxies using similar galaxy formation models~\citep[e.g.][]{Hopkins2012,Muratov2015,Angles2017}. For example, in \citet{Muratov2015}, the transition is at $V_{\rm vir}\sim 60\kms$, which would correspond to $M_{\rm halo} \sim 10^{10}\msun$ at $z\simeq 4$. It is close but slightly larger than where we find the break of the $\epsilon^{\ast}_{\rm halo}-M_{\rm halo}$ relation. 

\begin{figure}
    \centering
    \includegraphics[width=\linewidth]{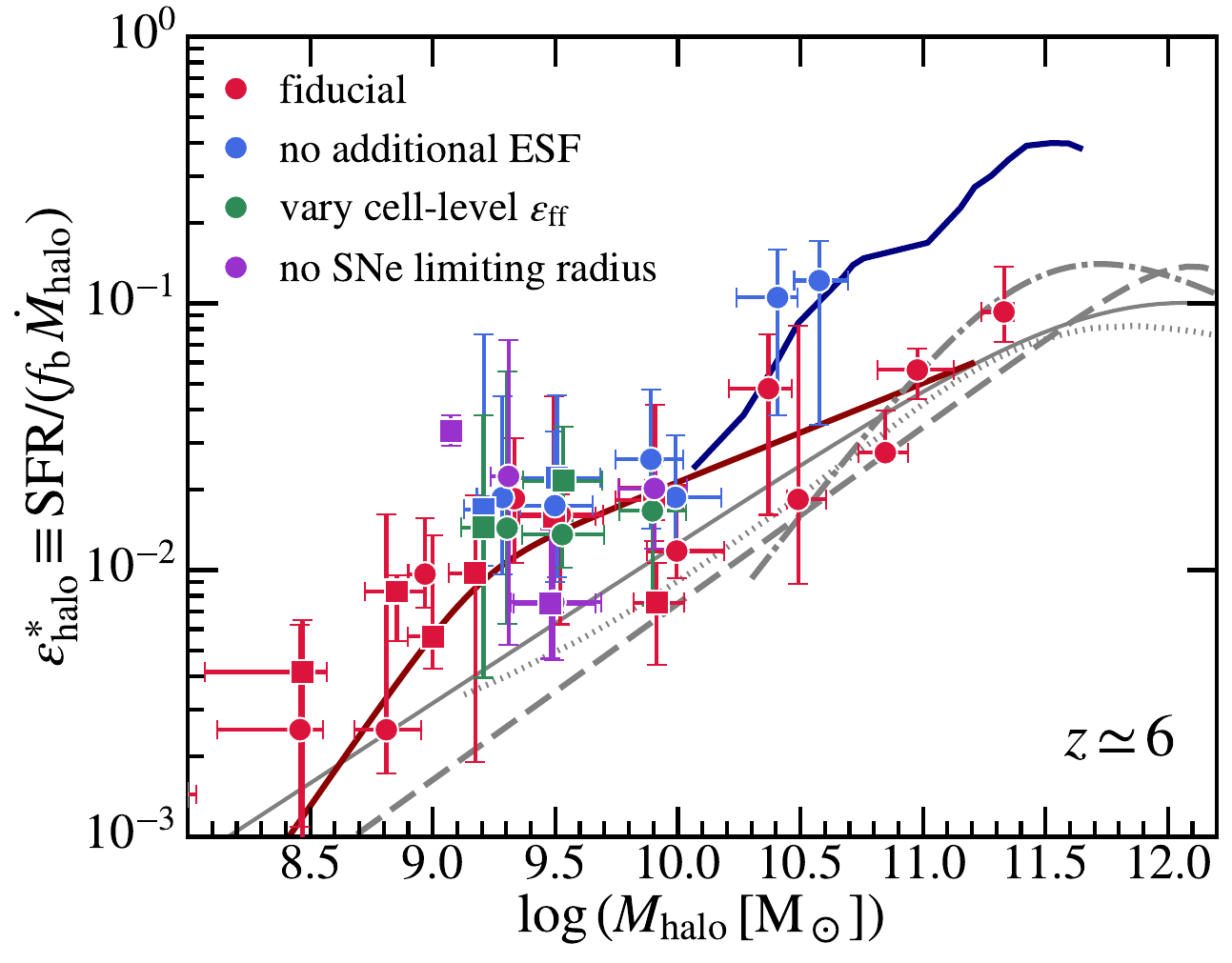}
    \caption{Halo-scale SFE of \thesanzoom galaxies in runs with different physics at $z\simeq 6$. We compare the results of the fiducial runs to the runs with no additional ESF, with varying cell-level SFE, and with no limiting radius for SNe feedback (see Section~\ref{sec:sim} for details). The gray lines in the background are the same constraints from empirical models shown in Figure~\ref{fig:halo-sfe} above while the red and blue lines show predictions from FIREBox and Firstlight as in Figure~\ref{fig:halo-sfe}. At $M_{\rm halo}\lesssim 10^{10}\msun$, no substantial differences in halo-scale SFE are found between runs with physics variants. At $M_{\rm halo}\gtrsim 10^{10}\msun$, the suite only has variants with no additional ESF. The halo-scale SFEs are about half a dex higher than the values in the fiducial run, which are more aligned with the predictions from the \textsc{FirstLight} simulation. Such a difference is consistent with the larger stellar-to-halo-mass ratios in runs with no additional ESF at lower redshifts, as shown in Figure~\ref{fig:mstar-mhalo-compare-noesf}.}
    \label{fig:halo-sfe-compare-phy}
\end{figure}

\subsection{Halo-scale SFE in simulations with model variations}

In Figure~\ref{fig:halo-sfe-compare-phy}, we show the instantaneous halo-scale SFE of main target galaxies at $z\simeq 6$ in runs with model variations. We note that the physics variants have only been run for a subset of haloes as listed in Table~\ref{table:simtab} and therefore do not cover the same halo mass range as the fiducial runs. In low-mass haloes ($M_{\rm halo}\lesssim 10^{10}\msun$), none of the physics variants affect the halo-scale SFE. The independence of cell-level SFE and numerical resolution indicates that star-formation in these galaxies has reached self-regulation. Detailed star-formation recipes chosen at the resolution scale no longer affect halo-scale star-formation. $\epsilon^{\ast}_{\rm halo}$ is also independent of details of SNe feedback coupling and the additional ESF at these halo masses. In more massive haloes ($M_{\rm halo}\gtrsim 10^{10}\msun$), we only have simulations without additional ESF tested. In these runs, the halo-scale SFE starts to ramp up and appears to be more consistent with the results from the FirstLight simulations. These are consistent with the impacts we found for the stellar-to-halo-mass relations above. It highlights the importance of the treatment of ESF in regulating star-formation in massive dwarf to Milky Way-mass haloes.

\begin{figure}
    \centering
    \includegraphics[width=\linewidth]{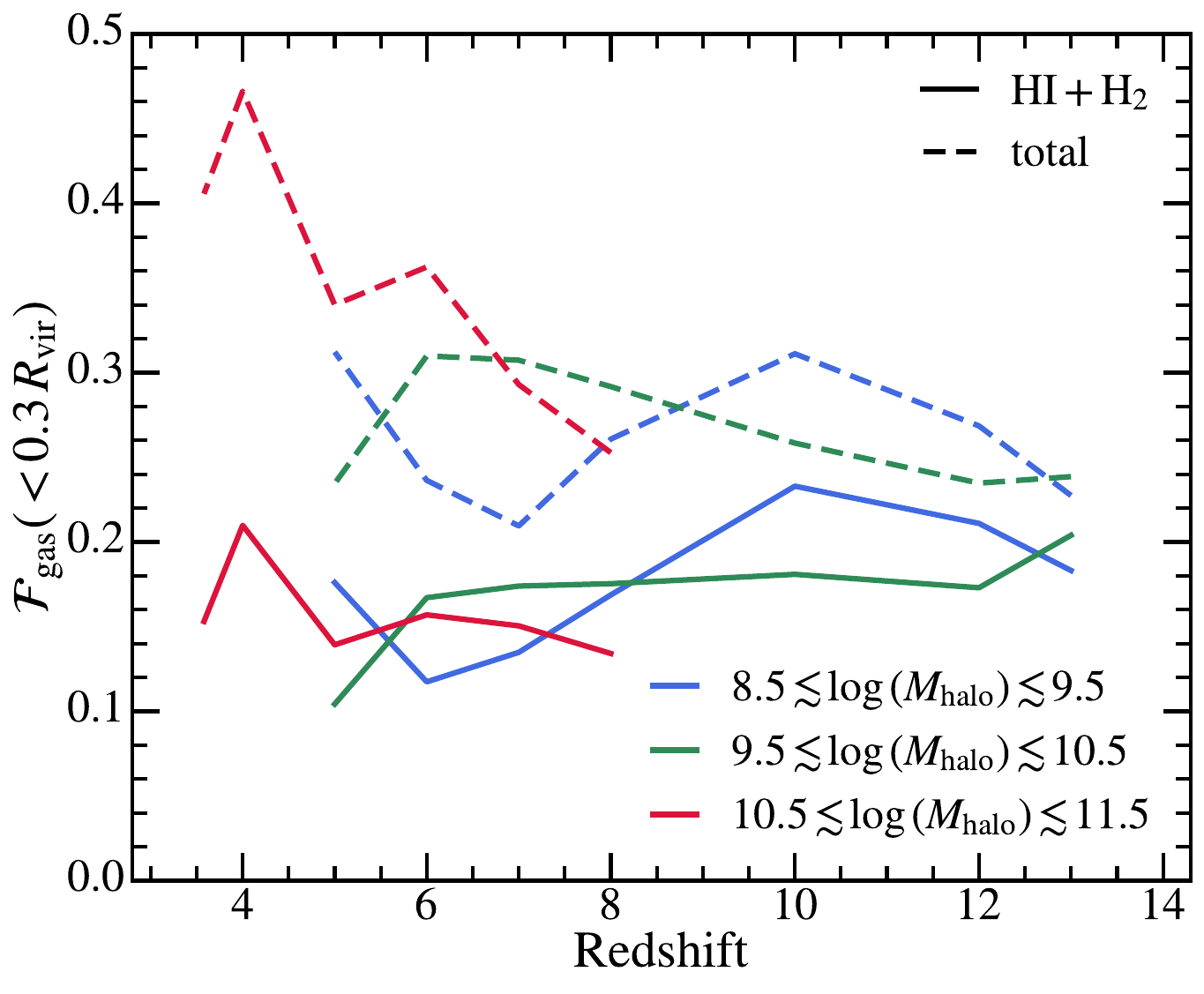}
    \caption{Mass fraction of the neutral-plus-molecular and total gas reservoir within  $0.3\,R_{\rm vir}$ versus redshift. We compute the neutral-plus-molecular and total gas mass within the same aperture as we measure the KS laws ($0.3\,R_{\rm vir}$) and divide it by the total gas abundance expected in the halo ($f_{\rm b}\,M_{\rm halo}$). The measurements are done in three halo mass bins as labelled. While the total gas abundance in the central star-forming region is increasing at lower redshifts, the abundance of neutral and molecular gas stays stable across a wide redshift range and is independent of halo mass.}
    \label{fig:fgas}
\end{figure}

\begin{figure*}
    \centering
    \includegraphics[width=\linewidth, trim={0 1cm 0 0}]{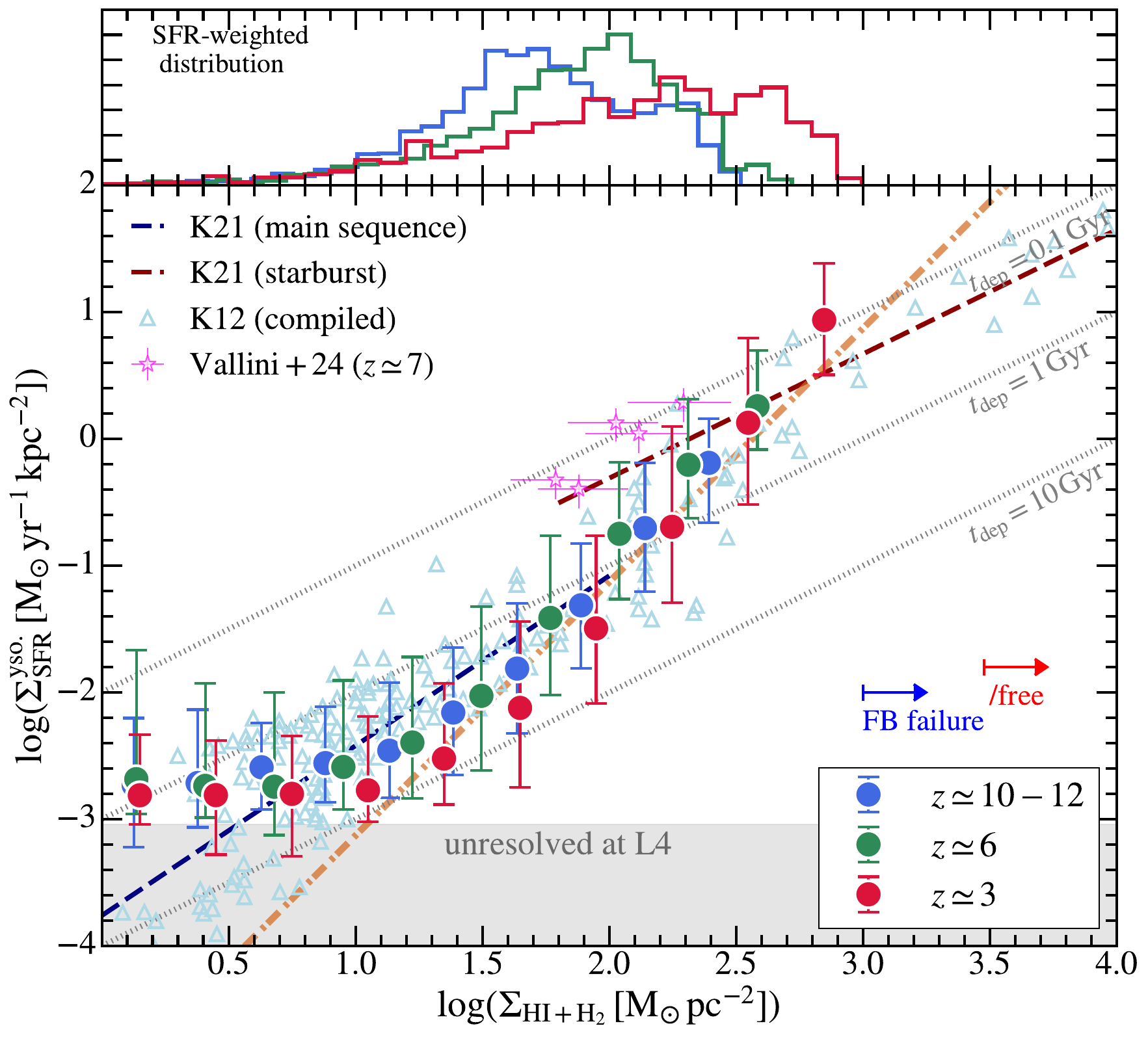}
    \caption{KS relation of neutral gas at $z\simeq 3,\,6,$ and $10-12$ of \thesanzoom galaxies resolved at $1\kpc$ scale. SFR is calculated using the young stellar objects (YSO; the sum of masses of young stellar particles with age $<10\Myr$ in simulations) as the tracer. The top subpanel shows the SFR-weighted distribution of pixels that we used in this analysis (we note that this does not reflect the surface density's dependence on redshifts as the target halo masses vary with redshifts).
    We compare the KS relation measured in simulations to the local observed ones of main-sequence and starburst galaxies from \citet{Kennicutt2021}, individual data points compiled in \citet{Kennicutt2012}, and a recent ALMA study of $z\simeq 7$ galaxies~\citep{Vallini2024}. The orange dot-dashed line shows the $\Sigma_{\rm SFR}\sim \Sigma^{2}_{\rm HI+H_2}$ scaling we derived from simple analytical calculations in Section~\ref{sec:result-epsgal-analytical}. The shaded region shows the regime where SFR surface density is not properly resolved, i.e. only one young stellar particle in a pixel. Before reaching the resolution limit, the simulation results agree with the simple equilibrium solution. Roughly, the KS law agrees with local observations, although the slope in the starburst regime is steeper. We find no prominent redshift dependence of the KS relation. The maximum surface densities in \thesanzoom galaxies have not yet reached the feedback-free/failure regime discussed in \citet{Dekel2009,BK2024}.}
    \label{fig:kslaw-yso}
\end{figure*}

\section{Galaxy-scale star-formation efficiency}
\label{sec:result-epsgal}

In observations, the efficiency of star-formation is often expressed as the depletion time ($t_{\rm dep}$) of certain phases of gas resolved over roughly kpc scales. The phase of gas traced can be neutral (${\rm HI}+{\rm H}_{2}$), molecular (${\rm H}_{2}$), or other dense gas tracers. We focus on neutral gas in this paper. One can rewrite the halo-scale SFE as
\begin{equation}
    \epsilon^{\ast}_{\rm halo} \simeq \dfrac{1}{t_{\rm dep}}\, \dfrac{M_{\rm gas}}{f_{\rm b}\,M_{\rm halo}}\, \dfrac{M_{\rm halo}}{\dot{M}_{\rm halo}} 
\end{equation}
where $M_{\rm gas}$ is the neutral gas mass in the central star-forming region. In cosmological N-body simulations, the specific accretion rates of DM haloes have the following scaling
\begin{equation}
    \dot{M}_{\rm halo} \simeq \beta(z)\,\left(M_{\rm halo}\, H(z)\right)^{\alpha(z)},
    \label{eq:acc}
\end{equation}
where we use the formula from the most recent calibration in \citet{Yung2024}, but many previous works~\citep[e.g.][]{Fakhouri2010,Behroozi2015,RP2016} found a similar scaling. \citet{Yung2024} found that $\alpha(z)\simeq 1 \pm 0.05$ in the redshift range $6\lesssim z \lesssim 14$, so we approximate it as unity here. $\beta(z)$ in the redshift range we care about can be roughly approximated as a broken power-law, $\beta(z) \propto (1+z)^{\gamma}$, where $\gamma\simeq 0.5$ at $z\lesssim 8$ and $\gamma\simeq 0.2$ at $z\gtrsim 8$~\citep[but see the original fits in][]{RP2016,Yung2024}. Combining these, we obtain
\begin{equation}
    \epsilon^{\ast}_{\rm halo} \propto \left(\dfrac{M_{\rm gas}}{f_{\rm b}\,M_{\rm halo}}\right)\, \left( \dfrac{t_{\rm dep}}{t_{\rm H}}\right)^{-1}\,(1+z)^{-\gamma},
    \label{eq:epshalo-decompose}
\end{equation}
where $t_{\rm H}\equiv 1/H(z)$ is the Hubble time. Here, $\epsilon^{\ast}_{\rm halo}$ is decomposed into two parts, where the first represents the fraction of a certain phase of gas locked in the central star-forming region of the halo (``supply'') and the second represents the depletion rates of gas due to star-formation (``consumption''). This will allow us to understand the underlying drivers for the halo mass and redshift dependence of $\epsilon^{\ast}_{\rm halo}$ seen in the previous section. 

\subsection{Supply of star-formation fuel from large-scale environments}

We first focus on the ``supply'' of cold neutral gas as fuel for star-formation. We compute the total gas mass and the neutral gas within an aperture of $0.3\,R_{\rm vir}$ for a subset of well-resolved central galaxies in zoom-in regions of \thesanzoom simulations. The selection criteria will be introduced in detail in the following section since the same set of galaxies will be used to study resolved star-formation scaling relations.

In Figure~\ref{fig:fgas}, we show the redshift evolution of the mass fraction of neutral gas (HI + ${\rm H}_2$) and total gas within $0.3\,R_{\rm vir}$ with respect to the expected total baryon reservoir in the halo, $f_{\rm b}\,M_{\rm halo}$. We present results in three halo mass bins as labelled. The total gas fraction within $0.3\,R_{\rm vir}$ is increasing with decreasing redshift and is roughly independent of halo mass, indicating more compact gas distributions established at later times. However, the fraction of neutral and star-forming gas within $0.3\,R_{\rm vir}$ remains stable at $\sim 20\%$ across a wide redshift range. This equilibrium could be driven by more intense radiation from local sources when enhanced star-formation is stimulated by stronger gas inflow. Indeed the neutral fraction of gas is decreasing during the EoR but the trend continues even at $z\lesssim 5$ when the global reionization is completed in \thesan~\citep{Kannan2022thesan,Garaldi2024}. In Zier et al. (2025), we will show that the ionization status and self-shielding fraction of gas can vary significantly even after the end of the reionization of the intergalactic medium. Motivated by the findings here, in the later analysis, we will assume that the $M_{\rm gas}/f_{\rm b}\,M_{\rm halo}$ term in Equation~(\ref{eq:epshalo-decompose}) is independent of halo mass and redshift.

\begin{figure}
    \centering
    \includegraphics[width=\linewidth]{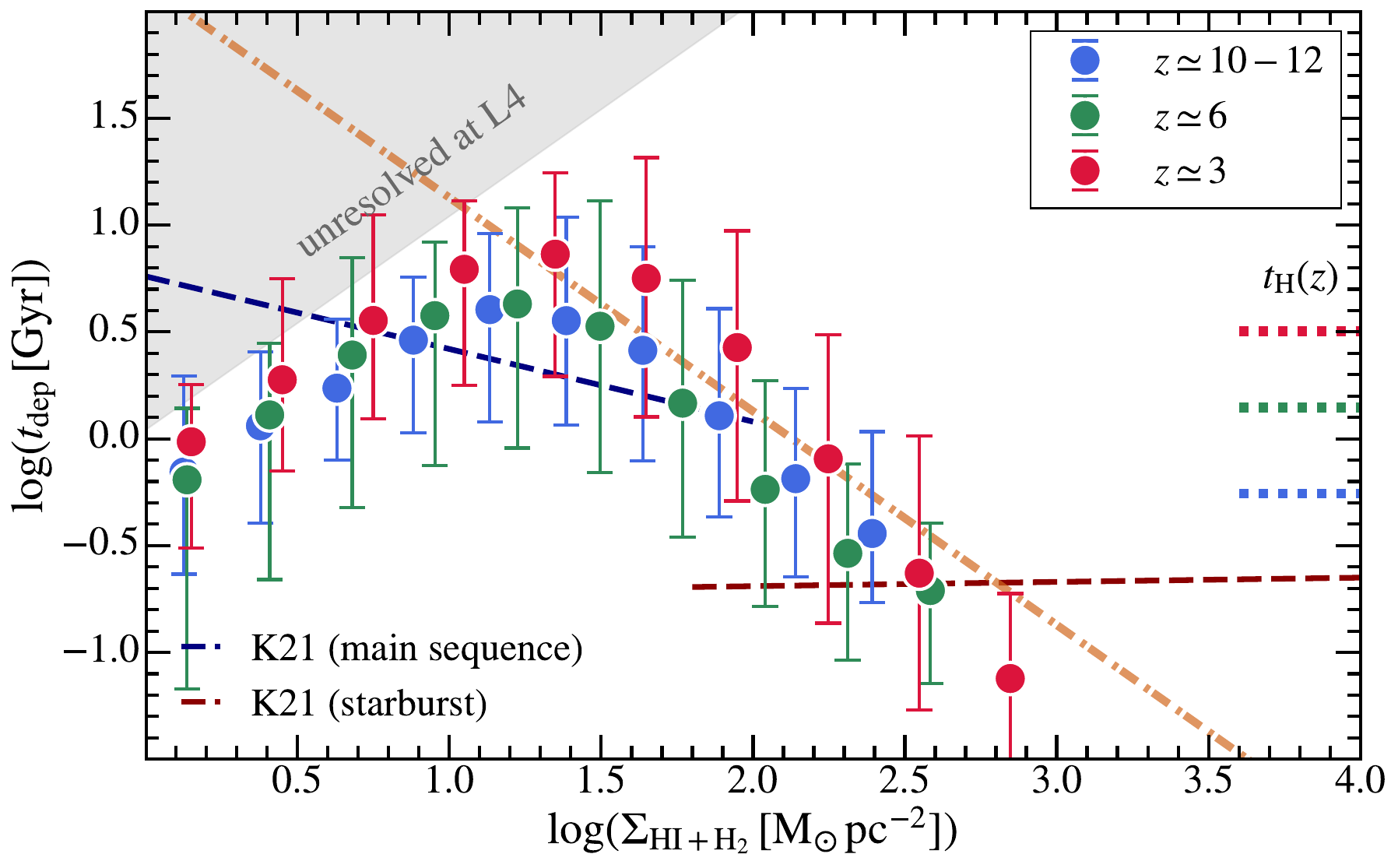}
    \caption{Depletion time of neutral gas ($t_{\rm dep}\equiv \Sigma_{\rm HI+H_2}/\Sigma_{\rm SFR}$) versus neutral gas surface density. The colored points show the results at $z\simeq 3$, $6$, and $10-12$ from \thesanzoom with the same notations as in Figure~\ref{fig:kslaw-yso}. The orange dashed line shows the equilibrium solution we obtain analytically. The red and blue lines show the same local fitting from \citet{Kennicutt2021}. Before reaching the resolution limit, $t_{\rm dep}$ shows a $\Sigma^{-1}_{\rm HI+H_2}$ dependence and is consistent with our analytical model. The short dotted lines on the right edge show the Hubble times at the corresponding redshifts. The change of $t_{\rm dep}$ over redshifts is much smaller than the change of the Hubble time.}
    \label{fig:kslaw-tdep}
\end{figure}

\subsection{The Kennicutt--Schmidt relation at high redshifts}

We next focus on the ``consumption'' side and measure the KS relations of the neutral and molecular gas. For better statistics, we include all central galaxies in the zoom-in region with $M_{\rm halo}$ larger than 50\% that of the main target halo (and without contamination from low-resolution particles) in this analysis. Most of \thesanzoom galaxies at high redshifts do not display a well-defined disk structure but have clumpy irregular distributions of gas. Therefore, we project the stellar and neutral gas distributions in three sightlines corresponding to the positive $x$, $y$, and $z$ directions of the simulation box and measure the projected surface densities of gas and SFR in pixels of size $\sim 1\,{\rm kpc}$ (physical units) in a field of view of $2 \times 0.3\,R_{\rm vir}$. This is our rough definition of a central star-forming region. We find no dependence of our results on sightline choices or pixel sizes in practice. Gas cells are smoothed~\footnote{We utilize the \textsc{swiftsimio} package~\citep{Borrow2020,Borrow2021} to accelerate the smoothing and projection process.} using a Wendland-C2 kernel (as described in e.g. \citealt{Dehnen2012}) with the smoothing length taken to be $2.5\,r_{\rm cell}$, where $r_{\rm cell}\equiv (m_{\rm cell}/\rho_{\rm cell}/(4\pi/3))^{1/3}$ is the effective size of a gas cell. We note that overestimation of $\Sigma_{\rm HI+H_2}$ and $\Sigma_{\rm SFR}$ could occur when multiple star-forming complexes are projected to the same pixel or when there is a disk structure involving an additional geometrical factor from projection. However, in either case, the depletion time of gas will not be affected. 

In Figure~\ref{fig:kslaw-yso}, we show the KS relation of the neutral gas of \thesanzoom galaxies at three redshifts. We compare them to the scaling relations found in local observations compiled in \citet{Kennicutt2012} and \citet{Kennicutt2021}. Here, SFR is computed using the young stellar objects (YSO) in simulations (i.e. the total mass of stellar particles younger than $10\Myr$ divided by the same time). The SFR surface density computed in this way is subject to numerical resolution limits, and the shaded region in the figure shows where only one young stellar particle in a pixel is present at the resolution level L4 (we explore the impact of numeric resolution in Section~\ref{sec:result-epsgal-var} below). Interestingly, before reaching the plateau caused by numerical effects, the \thesanzoom results agree well with local observations, in terms of the typical depletion time on the main sequence and the transition to the starburst regime at slightly above $100\msun\,{\rm pc}^{-2}$. However, the KS relation in \thesanzoom follows a simple power-law form with no break and the slope of the KS relation is in general larger than the values found in local observations, regardless of the details in dataset, fitting methods, and molecular gas conversion factors~\citep[e.g.][]{Daddi2010,Liu2015,Kennicutt2021}. We will revisit the origin of a steeper KS law in Section~\ref{sec:result-epsgal-analytical} below, and for reference here, the orange dashed line shows the prediction of this analytical model. Spatially resolved studies of the KS relation have recently been extended to $z\simeq 7$ with ALMA observations on sub-kpc scale~\citep{Vallini2024}. We show their measured $\Sigma_{\rm HI+H_2}$ and $\Sigma_{\rm SFR}$ in Figure~\ref{fig:kslaw-yso} for comparison. The \thesanzoom predictions are slightly below these observational constraints, but we note the great uncertainties in molecular gas conversion factors in observational studies. 

We find almost no redshift dependence of the KS relation, although the absolute $\Sigma_{\rm SFR}$ and $\Sigma_{\rm \rm HI+H_2}$ values do change in \thesanzoom galaxies. In the top subpanel of Figure~\ref{fig:kslaw-yso}, we show the SFR-weighted distribution of $\Sigma_{\rm \rm HI+H_2}$. The typical $\Sigma_{\rm \rm HI+H_2}$ is higher at lower redshifts but is mainly due to larger halo masses of target galaxies evolved to lower redshifts in \thesanzoom, which will be revisited in Section~\ref{sec:result-epsgal-analytical}. The majority of the star-formation takes places in pixels that are not subjected to numerical effects, and the global aggregated SFR of a galaxy is not affected by numerical resolution. In Figure~\ref{appfig:kslaw-inst} in Appendix, we also show the results using an alternative way to compute SFR, directly using the instantaneous SFR in gas cells in simulations. This yields consistent results but allows the measured $\Sigma_{\rm SFR}$ to extend to lower values compared to the fiducial method. The model dependence of the instantaneous SFR will also be investigated in Figure~\ref{fig:kslaw-compare} below.

In Figure~\ref{fig:kslaw-tdep}, we show the depletion time of neutral gas as a function of $\Sigma_{\rm \rm HI+H_2}$ and plot the Hubble time at the corresponding redshifts for reference. Before reaching the resolution limit, the depletion time follows a simple power-law dependence on neutral gas surface density. The gas depletion times are overall of the same order as the Hubble time, a proxy for the age of the Universe at the corresponding redshift. However, the change of depletion time over redshifts is apparently smaller than the change in Hubble time.

\begin{figure}
    \centering
    \includegraphics[width=\linewidth]{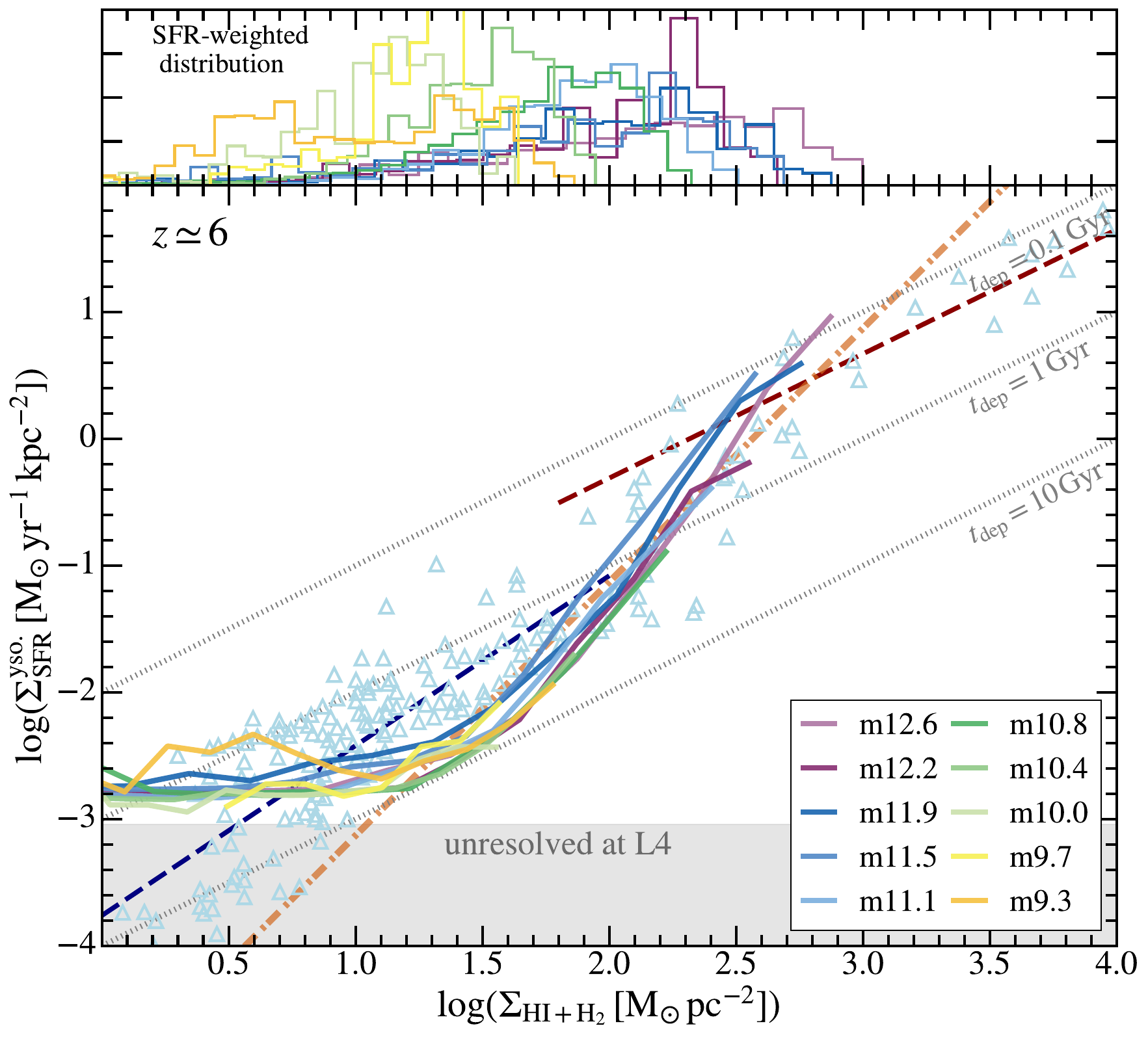}
    \caption{KS relation of neutral gas (same as Figure~\ref{fig:kslaw-yso}) at $z\simeq 6$ divided into galaxy groups found in the zoom-in regions of each of the \thesanzoom runs. Since haloes used for the analysis here have masses at least $50\%$ of the main target halo, different groups should roughly represent different halo mass ranges. The curves are labelled by the mass of the main target halo at $z\simeq 3$ as in Table~\ref{table:simtab}. We find no dependence of depletion time on halo mass at fixed $\Sigma_{\rm \rm HI+H_2}$. However, as shown in the top subpanel, the surface density distributions move to higher values in more massive haloes, which can effectively reduce the depletion time.}
    \label{fig:kslaw-byhalo}
\end{figure}

In Figure~\ref{fig:kslaw-byhalo}, we show the KS relations of neutral gas at $z\simeq 6$ divided into different groups of galaxies found around each of our main target galaxies. Since we force a mass cut of $50\%$ of the main target halo mass during the selection, each group of galaxies should be roughly in the same halo mass range as the main target. Here we find essentially no dependence of the KS law on host halo mass, and this also implies little dependence on stellar mass, metallicity, etc. given the strong correlations between these galaxy properties. Nevertheless, the surface density ranges spanned by pixels in the star-forming region depend on halo mass and this can indirectly lead to different halo-scale gas depletion times as will be discussed in the following section.

\subsection{Simple analytical picture of turbulent ISM in equilibrium}
\label{sec:result-epsgal-analytical}

What determines the slope of the KS relation? Following approaches in e.g. \citet{OstrikerShetty2011,FG2013,Orr2018}, we consider a star-forming complex of size $R$, mean gas density $\rho_{\rm gas}$ (here we assume that gas is all in neutral phase), and turbulent velocity $\sigma$ (at the scale of $R$). The turbulent Jeans mass is
\begin{equation}
    M_{\rm J} \simeq \dfrac{\pi\,(\sigma^{2}+c^{2}_{\rm s})^{3/2}}{6\,G^{3/2}\,\rho_{\rm gas}^{1/2}} \sim \dfrac{\pi\,\sigma^{3}}{6\,G^{3/2}\,\rho_{\rm gas}^{1/2}},
    \label{eq:jeansmass}
\end{equation}
where we assume a turbulence-dominated regime $\sigma \gg c_{\rm s}$. This mass is of the same order as the Bonnor-Ebert mass, the largest mass of an isothermal, non-magnetized gas sphere embedded in a pressurized medium that can maintain hydrostatic equilibrium. If we assume $M_{\rm J}$ is the typical mass of the star-forming complex, i.e. it is marginally stable, we have
\begin{equation}
    \sigma = 2\,G^{1/2}\,\rho^{1/2}_{\rm gas}\,R.
    \label{eq:sigma-rho-R}
\end{equation}
Further fragmentation of this sphere is expected. Stellar feedback as a consequence of star-formation in denser cores of those fragments can supply the turbulent energy on the scale of $R$ and maintain the marginal stable state of the sphere.

Once the equilibrium state is reached, the amount of turbulent energy injected from stellar feedback should balance its dissipation rate
\begin{equation}
    \dot{E}_{\rm -} = \dfrac{1}{2}\,\left(\dfrac{4\pi}{3}\,\rho_{\rm gas}\,R^{3}\right)\,\dfrac{\sigma^2}{t_{\rm diss}} = \dfrac{2\,\pi}{3}\,\rho_{\rm gas}\,R^{2}\,\sigma^3,
    \label{eq:Eminus}
\end{equation}
where the dissipation time scale of turbulent energy is related to the coherence time scale of turbulent eddies, $t_{\rm diss} \sim R/\sigma$~\citep[e.g.][]{Stone1998}. Meanwhile, the amount of turbulent energy injected through stellar feedback per unit time is
\begin{equation}
    \dot{E}_{\rm +} = \dfrac{1}{2}\,\dot{M}_{\ast}\,\left(\dfrac{P_{\ast}}{m_{\ast}}\right)\,v_{\rm fb},
    \label{eq:Eplus}
\end{equation}
where $(P_{\ast}/m_{\ast})$ is the amount of momentum injection into the ISM per young stellar mass formed and we assume the typical velocity of SNe-driven wind is $v_{\rm fb}\simeq \sigma$, that is when they become indistinguishable from the ISM. We neglect external turbulence driven by processes such as gas accretion through cold streams. This assumption appears justified by our finding above that the KS relation shows essentially no dependence on halo mass. The depletion time is therefore governed by local ISM properties rather than large-scale halo properties or gas inflows.

Equating the expressions in Equations~(\ref{eq:Eminus}) and (\ref{eq:Eplus}), we obtain the gas depletion time
\begin{equation}
    t_{\rm dep} \equiv \left(\dfrac{4\pi}{3}\,\rho_{\rm gas}\,R^{3}\right)/ \dot{M}_{\ast} 
    = \dfrac{1}{4\,G}\,\left(\dfrac{P_{\ast}}{m_{\ast}}\right)\,\dfrac{1}{\rho_{\rm gas}\,R} .  
\end{equation}
Assuming $\Sigma_{\rm gas} \simeq \rho_{\rm gas}\,R$, we obtain
\begin{equation}
    t_{\rm dep} \simeq 0.5\,\Gyr\,\left(\dfrac{f}{0.3}\right)\, \left( \dfrac{P_{\ast}/m_{\ast}}{3000\kms} \right)\,\left( \dfrac{\Sigma_{\rm gas}}{100\msun\,{\rm pc}^{-2}}\right)^{-1},
    \label{eq:tdep}3
\end{equation}
where $f$ is an order-unity fudge factor we introduce to encapsulate order-unity constants in the derivation and the unaccounted geometrical effects. In \thesanzoom, $P_{\ast}/m_{\ast} = (P_{\ast}/m_{\ast})_{\rm SNe} + (P_{\ast}/m_{\ast})_{\rm ESF}$, where the two terms stand for momentum injection from SNe feedback and the additional ESF we impose. Other sources of feedback like radiation pressure are likely not important at $\Sigma_{\rm gas}\lesssim 10^{4}\msun\,{\rm pc}^{-2}$~\citep{OstrikerShetty2011}. $(P_{\ast}/m_{\ast})_{\rm ESF} = 5000 \kms$ as configured. We take $(P_{\ast}/m_{\ast})_{\rm SNe}$ to be the typical value $3000\kms$~\citep[e.g.][]{OstrikerShetty2011, Kim2015, Martizzi2015, Marinacci2019}. This is computed based on the terminal momentum of a SNe blast wave, $p_{\ast} \simeq 2-5\times 10^{5}\msun \kms$~\citep[e.g.][]{Cioffi1988} albeit its weak dependence on gas density and stellar metallicity, and the total mass of stars per high mass star (that will explode as a SNe) formed, $m_{\ast}\simeq 100\msun$. For the fudge factor, in practice, we find $f\simeq 0.3$ can nicely fit the neutral gas KS relations in simulations. The $t_{\rm dep}\propto \Sigma^{-1}_{\rm gas}$ and $\Sigma_{\rm SFR}\propto \Sigma^{2}_{\rm gas}$ scalings are in good agreement with the simulations results and the pure dependence on feedback explains the universality of the KS relation found in \thesanzoom. 

One can obtain a similar scaling in a disk configuration using the Toomre criterion for a marginally stable disk as shown in many previous works~\citep[e.g.][]{OstrikerShetty2011, FG2013, Torrey2017, Dib2017, Orr2018, Ostriker2022}. At lower redshifts, when geometrically thin stellar disks emerge, the local total surface density will be dominated by stars and one can obtain $\Sigma_{\rm SFR}\propto \Sigma_{\rm gas}$~\citep[e.g.][]{Orr2018,Ostriker2022} with the same equilibrium argument in a thin disk. This agrees better with the shallower slopes of the KS relation found in low-redshift observations (see e.g. the observational data points shown in Figure~\ref{fig:kslaw-yso} at $\Sigma_{\rm HI+H_2}\lesssim 100\msun \pc^{-2}$). Meanwhile, in the extremely dense regime where $\Sigma_{\rm gas}\gtrsim 10^{4}\msun \pc^{-2}$, the dominant feedback mechanism for momentum injection is the pressure from dust-reprocessed IR radiation~\citep[e.g.][]{Thompson2005,OstrikerShetty2011} with $(P_{\ast}/m_{\ast}) \propto \kappa_{\rm IR}\,\Sigma_{\rm gas}$, where $\kappa_{\rm IR}$ is IR opacity. In such a regime, one can also obtain $\Sigma_{\rm sfr}\propto \Sigma_{\rm gas}$ assuming a constant $\kappa_{\rm IR}$, although we cannot directly test this scenario limited by the maximum gas surface density reached in \thesanzoom.

\begin{figure}
    \centering
    \includegraphics[width=\linewidth]{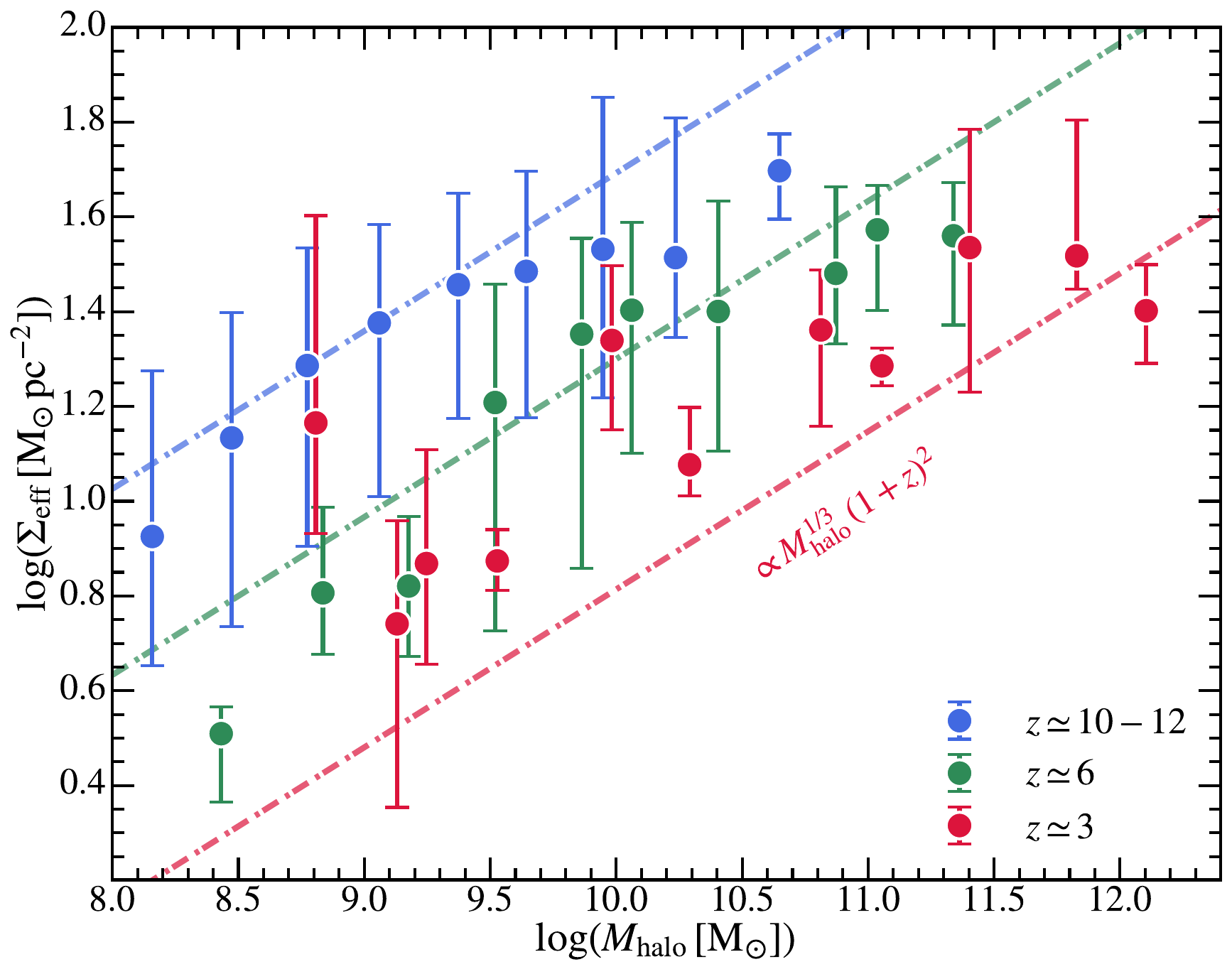}
    \caption{Effective surface density ($\Sigma_{\rm eff}$) of the star-forming regions of galaxies versus halo mass at three redshifts. We measure $\Sigma_{\rm eff}$ for individual galaxies and plot the median and $1\sigma$ scatter in each halo mass bin. The dashed lines show the analytical estimations, $\Sigma_{\rm gas}\propto M^{1/3}_{\rm halo}\,(1+z)^{2}$, in Equation~(\ref{eq:surface-density}). Despite a slight underprediction in $\lesssim 10^{11}\msun$ haloes at $z\simeq 3$, the analytical model fits the simulation results.}
    \label{fig:surface-density}
\end{figure}

Moving back to Equation~(\ref{eq:epshalo-decompose}), we can now finish the loop and connect the galaxy-scale SFE (gas depletion time) with the halo-scale SFE. The key point is that while the KS relation appears nearly universal in \thesanzoom, variations in the surface density distribution of galaxies with different masses at different redshifts can result in distinct aggregated SFEs at the halo scale. Acknowledging the $\Sigma_{\rm SFR}\propto \Sigma^{2}_{\rm gas}$ dependence, it is trivial to show that the effective surface density that determines the overall depletion time of star-forming gas in a galaxy should be $\Sigma_{\rm eff} \equiv \langle \Sigma^{2}_{\rm gas} \rangle/\langle \Sigma_{\rm gas} \rangle$, where the average is taken over all pixels with star-formation in that galaxy. This quantity effectively captures the clumping factor on kpc scales with respect to the mean gas surface density. In Figure~\ref{fig:surface-density}, we show the effective density of \thesanzoom galaxies as a function of host halo mass at $z\simeq 3$, 6, and $10-12$. The effective density is higher in more massive galaxies and at higher redshifts. 

At high redshifts, the star-forming region is fueled by cold gas streams in connection with the filamentary structures in the cosmic web~\citep[e.g.][]{Birnboim2003, Keres2005, Keres2009c, Dekel2009}. The characteristic density $\rho_{\rm gas}$ should be some clumping factor times the mean baryon density of the streams, which scales with the critical density of the Universe $\rho_{\rm crit}(z)$. Meanwhile, the turbulent velocity should roughly scale with the inflow velocity of the cold streams, which scales with $V_{\rm vir}$. Combining these with Equation~(\ref{eq:sigma-rho-R}), we obtain
\begin{equation}
    \Sigma_{\rm gas} \propto \rho_{\rm gas}\,R \propto \rho_{\rm gas}^{1/2}\,\sigma \propto M^{1/3}_{\rm halo}\,(1+z)^{2}.
    \label{eq:surface-density}
\end{equation}
This is shown in Figure~\ref{fig:surface-density} as reference lines, which agree with \thesanzoom predictions. Large scatters exist at fixed halo mass and could reflect the ``compaction-depletion-replenishment'' cycles around the main sequence~\citep[e.g.][]{Tacchella2016}. In fact, at $z\simeq 3$, our measurements of $\Sigma_{\rm eff}$ in some low halo mass bins could be biased towards the ``compaction'' (and star-forming) phase due to limited statistics.

Taking the Equation above and the decomposition of $\epsilon^{\ast}_{\rm halo}$ in Equation~(\ref{eq:epshalo-decompose}), when the mass fraction of the gas reservoir remains constant, we have
\begin{equation}
    \epsilon^{\ast}_{\rm halo} \propto \dfrac{t_{\rm H}(z)}{t_{\rm dep}} \,(1+z)^{-\gamma} \sim  M^{1/3}_{\rm halo}\,(1+z)^{1/2 - \gamma}.
    \label{eq:epshalo-analytical}
\end{equation}
The halo mass dependence is the same as we derived in Section~\ref{sec:result-epsgal} based on the mass-loading factor of feedback-driven winds in the momentum-conserving regime, and is consistent with the halo-scale SFE scaling in \thesanzoom. Since $\gamma \simeq 0.5$ at $z\lesssim 8$ (see the discussions above Equation~\ref{eq:epshalo-decompose}), no redshift dependence is expected from this calculation and is consistent with the weak/no redshift dependence of $\epsilon^{\ast}_{\rm halo}$ found in Section~\ref{sec:result-epshalo}. This is due to the cancellation between the redshift dependence of $t_{\rm H}(z)$ and the gas surface density. In low-mass haloes, the equilibrium described above may not be realized in the first place as superbubbles generated by stellar feedback break out and feedback energy freely escape the ISM. It results in the steeper dependence of $\epsilon^{\ast}_{\rm halo}$ on halo mass in the low-mass regime found in Section~\ref{sec:result-epsgal}.

There are several interesting implications from this redshift-independent power-law scaling of $\epsilon^{\ast}_{\rm halo}$. If we assume $\epsilon^{\ast}_{\rm halo} = \epsilon_{0}\,M^{\alpha}_{\rm halo}$, the integral in Equation~(\ref{eq:epshalo-integral}) gives $M_{\ast} \simeq f_{\rm b}\,\epsilon_{0}\,M^{1+\alpha}_{\rm halo} / (1+\alpha)$. From the analysis in Section~\ref{sec:result-epshalo}, we find $\alpha = 1/3$ in relatively massive galaxies ($M_{\rm halo}\gtrsim 10^{9}\msun$) likely due to the momentum-driven nature of gas outflows. This implies a log-log slope of $4/3$ in the $M_{\ast}-M_{\rm halo}$ relation, which is in good agreement with observational constraints at $z\gtrsim 3$ (see e.g. \citealt{Behroozi2019}). However, the slope is expected to transition to $5/3$ in lower-mass galaxies or at higher redshifts, motivated by our findings in Section~\ref{sec:result-epshalo}. If we combine this with the definition of $\epsilon^{\ast}_{\rm halo}$ (i.e. ${\rm SFR}/f_{\rm b}\,\dot{M}_{\rm halo}$) and the halo accretion rates in Equation~\ref{eq:acc}, we obtain the specific SFR (sSFR) as
\begin{equation}
    {\rm sSFR} \equiv {\rm SFR}/M_{\ast}  \simeq 2\,\Gyr^{-1}\,\left(\dfrac{1+z}{5}\right)^{2},
\end{equation}
for $\gamma = 0.5$, regardless of the value of $\alpha$. The sSFR predicted here is formally independent of halo mass and has a steep redshift dependence. Such scaling is in good agreement with the evolution of the star-forming main sequence at $z\gtrsim 3$ in \thesanzoom, which is studied in detail in McClymont et al. (2025).

\subsection{Galaxy-scale SFE in simulations with model variations}
\label{sec:result-epsgal-var}

\begin{figure*}
    \centering
    \includegraphics[width=0.33\linewidth]{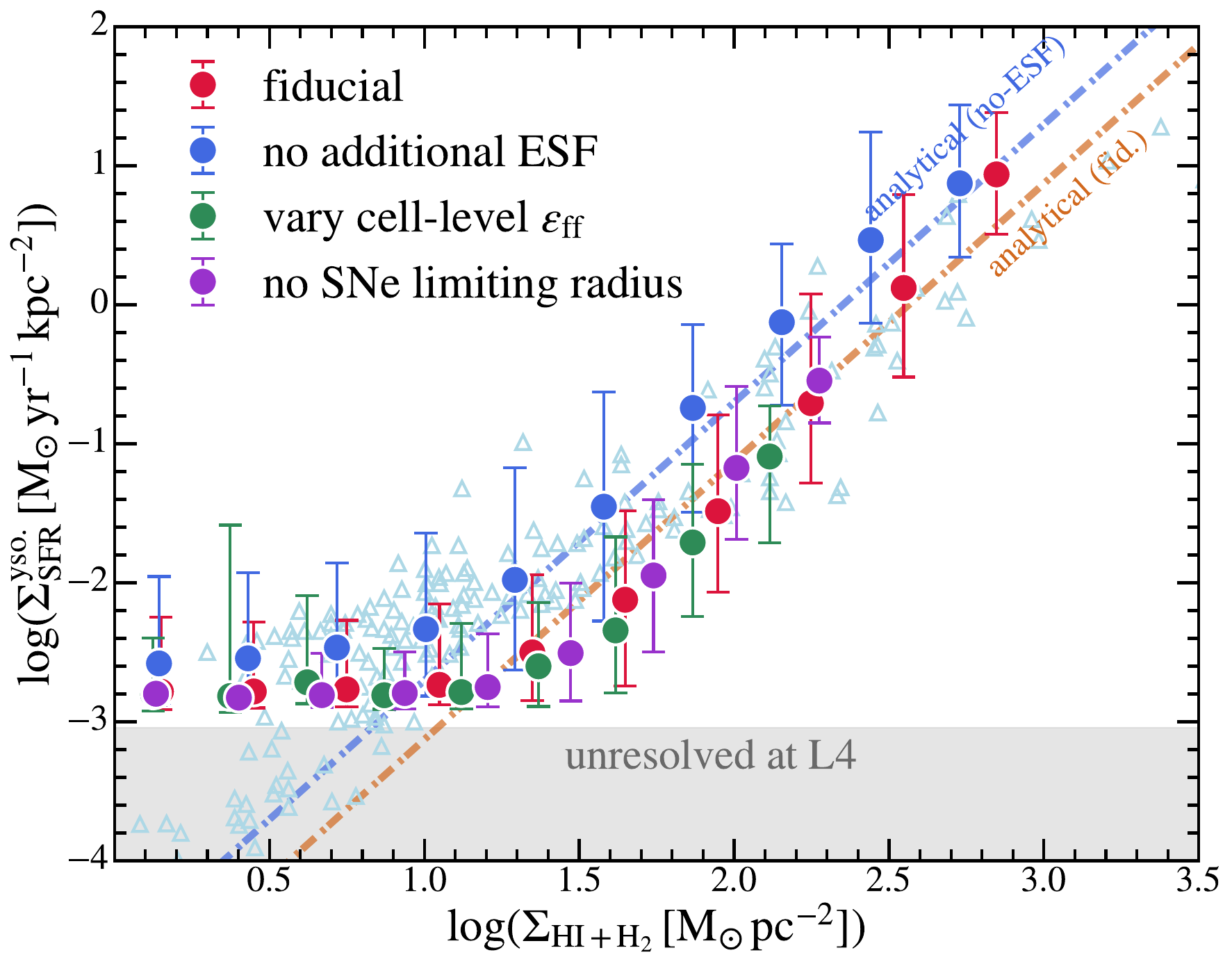}
    \includegraphics[width=0.33\linewidth]{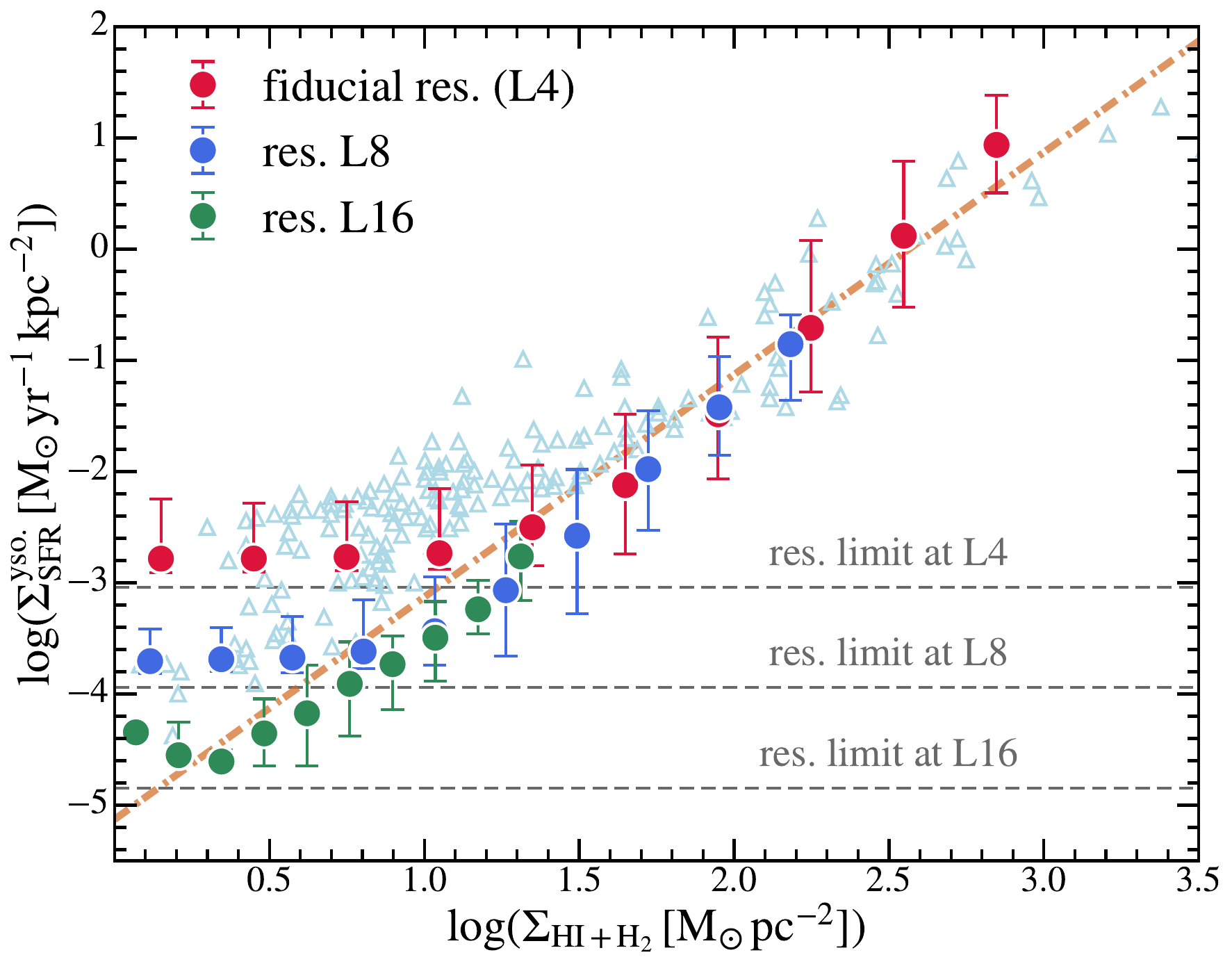}
    \includegraphics[width=0.33\linewidth]{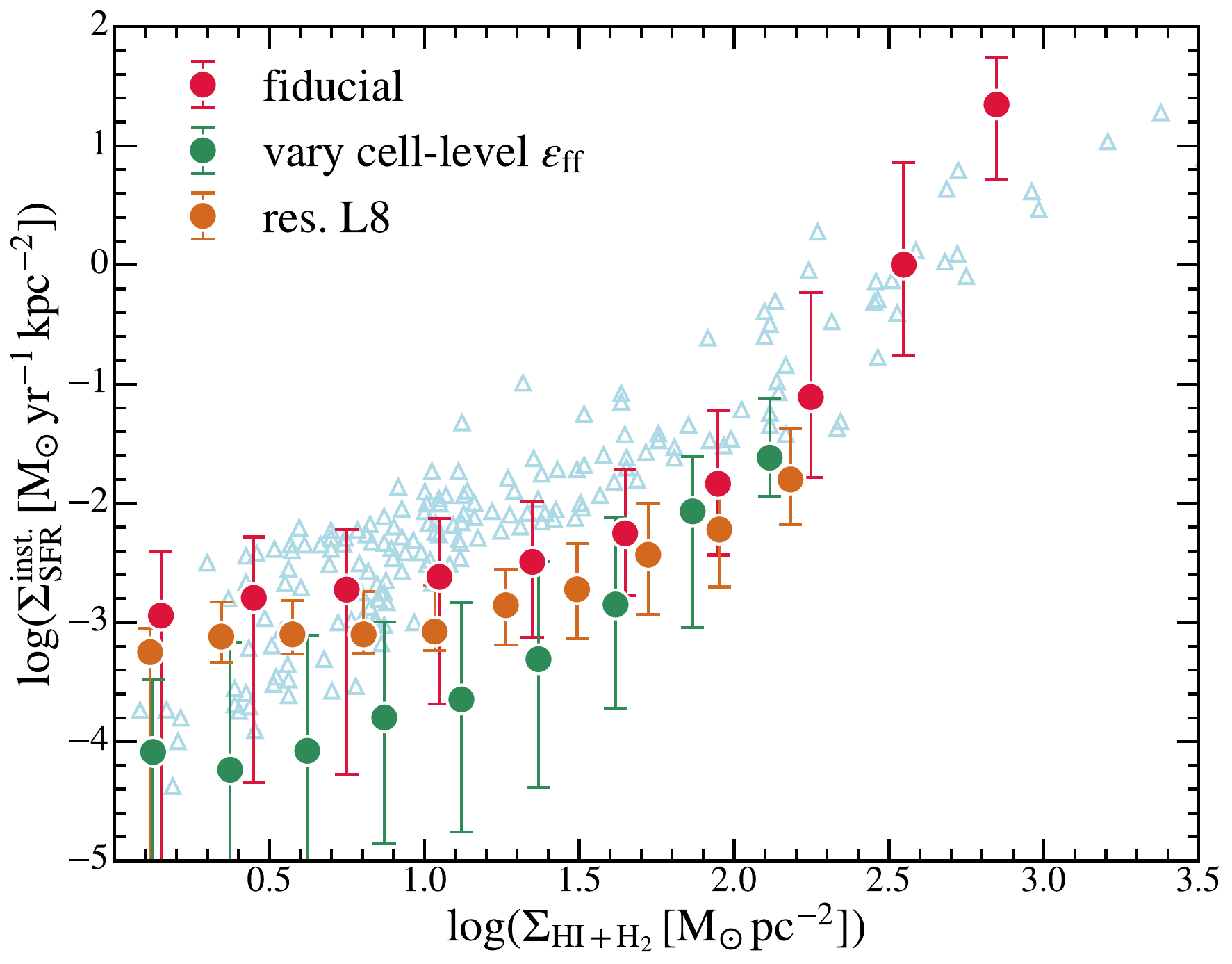} 
    \caption{KS relations of neutral gas at $z\simeq 3$ in \thesanzoom simulations with different physics variations (left), numerical resolutions (middle), and definitions (right). The background observational data points are the same as in Figure~\ref{fig:kslaw-yso}. Varying cell-level SFE or removing the SNe feedback limiting radius does not affect the KS relation. This demonstrates that the integrated SFR, even on relatively short timescales, is already insensitive to the details in star-formation and feedback recipes due to self-regulation. Removing the additional ESF, on the other hand, increases the normalization of the KS relation. The orange and blue dashed lines show the predictions from our analytical model assuming the feedback momentum injection rates $P_{\ast}/m_{\ast}$ in the fiducial and ``no ESF'' runs. In the middle panel, as expected, the numerical resolution affects the minimum SFR surface density that can be resolved, but the KS relation above the limit is converged. In the right panel, the SFR is traced by instantaneous rates in gas cells. In this case, we find that the KS relation is less sensitive to numerical resolution but depends on the cell-level SFE. This is most apparent in the low-surface density regime where low-density gas can form stars with $\epsilon_{\rm ff}$ as low as $1\%$.}
    \label{fig:kslaw-compare}
\end{figure*}

In this section, we evaluate the KS relations of neutral gas in runs with model variations and different numerical resolutions. These are shown in the left and middle panels of Figure~\ref{fig:kslaw-compare}. Additionally, in the right panel, we show the KS relations derived using the instantaneous SFR of gas cells in different simulation configurations. 

First, the KS relation is sensitive to the inclusion of additional ESF. This can be understood through the dependence of gas depletion time on $P_{\ast}/m_{\ast}$ in Equation~(\ref{eq:tdep}). Removing the ESF contribution roughly decreases $P_{\ast}/m_{\ast}$ by a factor of $8/3$ and should consequentially decrease the depletion time by the same factor. This agrees with the increasing factor of $\Sigma_{\rm SFR}$ in the ``no ESF'' runs shown in Figure~\ref{fig:kslaw-compare}. At $\Sigma_{\rm HI+H_2}\lesssim 100 \msun \pc^{-2}$, the increasing factor can be slightly larger than implied in the change of $P_{\ast}/m_{\ast}$, indicating the non-linear effect of aggregating different sources of feedback. For example, the additional ESF we impose can reduce the ambient gas density of ISM and change its geometry, which enhances the effectiveness of SNe feedback later on~\citep[e.g.][]{Krumholz2019,Kannan2020feedback}. However, despite the sensitivity to the total feedback momentum injection, the limiting radius of SNe feedback does not affect the KS relation.

In terms of numerical resolution, it is not surprising that the minimum $\Sigma^{\rm yso.}_{\rm SFR}$ scales linearly with the baryonic mass resolution, $m_{\rm b}/(10\Myr)/(1\kpc)^{2}$. However, the KS relation measured based on young stars converges above this threshold. As confirmed in high-resolution runs, the $\Sigma_{\rm SFR}\propto \Sigma^{2}_{\rm gas}$ scaling extends to the low-surface density end ($\Sigma_{\rm HI+H_2}\lesssim 10 \msun \pc^{-2}$), which deviates from local observations. As discussed above, this could be due to the more prominent stellar contributions to total matter surface densities in low-redshift mature disks, leading to different scalings between $\Sigma_{\rm SFR}$ and $\Sigma_{\rm gas}$ using similar equilibrium arguments. While a floor in $\Sigma^{\rm yso.}_{\rm SFR}$ exists, this does not necessarily imply an overestimation of the total SFR in this regime. Rather, it reflects an artificial discreteness in star-formation, where smooth star-formation along the KS relation is replaced by a Bernoulli distribution, with some gas cells forming stars at the minimum $\Sigma^{\rm yso.}_{\rm SFR}$ and some not forming stars at all. The total stellar mass formed is nevertheless converged, as indicated by our analysis of the $M_{\ast}-M_{\rm halo}$ relation and halo-scale SFE. On the other hand, $\Sigma^{\rm inst.}_{\rm SFR}$ is slightly smaller in the high-resolution runs, but the change is less significant than the change in baryonic mass resolution. In high-resolution runs, dense gas is better resolved, leading to a lower $t_{\rm ff}(\rho_{\rm cell})$ and increased SFR at fixed $m_{\rm b}$.

Finally, in terms of the cell-level SFE, $\Sigma^{\rm inst.}_{\rm SFR}$ decreases in the ``varying-$\epsilon_{\rm ff}$'' runs in low-density environments. This is anticipated as $\epsilon_{\rm ff}$ starts from low values in low-density regions and ramps back to unity in dense gas in these runs. When $\epsilon_{\rm ff}$ is small, star-formation at the cell level proceeds more slowly, increasing the likelihood of observing a patch in a star-forming state with a lower SFR. In contrast, $\Sigma^{\rm yso.}_{\rm SFR}$ is insensitive to the choice of $\epsilon_{\rm ff}$, at least up to the resolution limit. This suggests that the integrated star-formation, even on relatively short timescales (e.g. the 10 Myr we used to define young stars), is already insensitive to cell-level SFE likely due to self-regulation. Two feedback loops contribute to this. When $\epsilon_{\rm ff}$ is chosen smaller, gas cells are allowed to fragment and collapse to higher densities before being converted to stellar particles, which decreases $t_{\rm ff}(\rho_{\rm cell})$ of star-forming gas. At the same time, star-forming clouds can persist longer before dissociation from stellar feedback, which increases the fraction of gas in the star-forming phase. Both of these effects help stabilize the depletion time measured on a larger physical scale to an equilibrium value. 

The experiments here are reassuring that the predictions on halo-scale and galaxy-scale SFE are mostly insensitive to the uncertainties in the star-formation and feedback recipes.

\section{Discussion}
\label{sec:discussion}

\subsection{In-situ versus ex-situ star-formation}
\label{sec:finsitu}

\begin{figure}
    \centering
    \includegraphics[width=\linewidth]{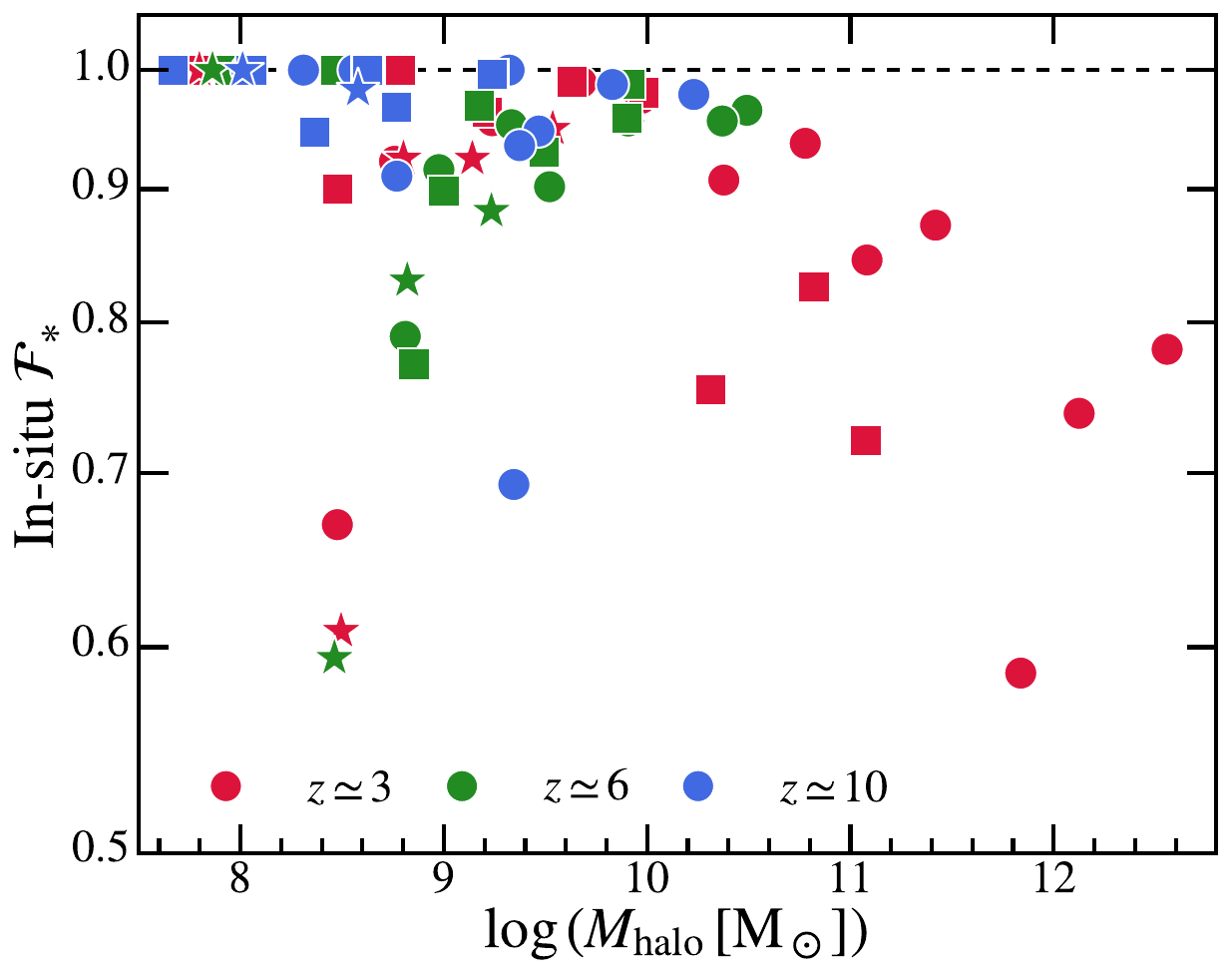}
    \caption{Fraction of stellar mass formed in-situ versus halo mass at $z\simeq 3$, 6, and 10. Results from the runs at resolution levels L4, L8, and L16 are shown in circles, squares, and stars, respectively. At $M_{\rm halo}\lesssim 10^{10}\msun$, most of the stellar mass forms in-situ (within $R_{\rm vir}$ of the host halo), regardless of redshift. In more massive galaxies at $z\simeq 3$, the in-situ mass fraction decreases likely due to late-time mergers.}
    \label{fig:insitu}
\end{figure}

A key underlying assumption in the above discussions of the connection between halo-scale and galaxy-scale resolved star formation is that most stellar mass must be built in-situ rather than through mergers. If it is the latter case, stars obtained by the halo are originally formed in lower-mass haloes at higher redshifts, resulting in different properties of star-forming regions and hence distinct gas depletion times. Therefore, it is important to understand the primary mode of stellar mass growth at high redshifts. In Figure~\ref{fig:insitu}, we show the fraction of stellar mass formed in-situ versus halo mass at $z\simeq 3$, 6, and 10. Stellar particles formed within $R_{\rm vir}$ of the main progenitor halo are classified as in-situ. $\mathcal{F}_{\ast}$ is the ratio between the total birth mass of stellar particles formed in situ and that of all stellar particles within $R_{\rm vir}$ of the current halo. For most of the galaxies with $M_{\rm halo}\lesssim 10^{10}\msun$, $\mathcal{F}_{\ast}$ remains close to unity and is insensitive to redshift or numerical resolution, suggesting that in-situ star-formation dominates. But external perturbations during e.g. mergers can still play a role in driving in-situ starbursts. At $M_{\rm halo}\gtrsim 10^{10}\msun$ at lower redshifts, stellar masses formed ex-situ and brought by galaxy mergers start to be important. Nevertheless, $\mathcal{F}_{\ast}$ maintains larger than $\sim 60\%$ in all cases. These findings are in good agreement with recent \textit{JWST} observations based on galaxy close pair fractions~\citep[e.g.][]{Puskas2025}.

\subsection{The further connection to cloud-scale SFE}

The galaxy-scale star-formation in kpc-scale patches of the ISM represents the aggregated star-formation of an ensemble of GMCs. In theoretical studies of GMCs, the SFE is often defined as the fraction of an initial gas reservoir converted into stars over the lifetime of a GMC ($\epsilon_{\rm int}$; e.g. \citealt{McKee2007}). The global SFE is connected to the microscopic SFE of GMCs as~\citep[e.g.,][]{FG2013}
\begin{equation}
    t_{\rm dep} = t^{\rm GMC}_{\rm life}/ \left(f_{\rm GMC}\, \epsilon_{\rm int}\right), 
\end{equation}
where $t^{\rm GMC}_{\rm life}$ is the GMC lifetime and $f_{\rm GMC}$ is the fraction of gas mass in GMCs. The global depletion time is therefore determined by the competition between $\epsilon_{\rm int}$, $t^{\rm GMC}_{\rm life}$, and $f_{\rm GMC}$, and the underlying equilibrium of GMC formation from gravitational collapse and dissociation by feedback~\citep[e.g.][]{McKee2007,FG2013,Krumholz2014,Semenov2017,Semenov2018,Kruijssen2019,Chevance2020,Tacchella2020}. In a follow-up study (Wang et al. in prep.), we will analyze the properties of GMCs (or in general self-gravitating star-forming complexes) in \thesanzoom and understand the connection between cloud and galaxy-scale SFE.

\subsection{Implications for UV-bright galaxies at $z\gtrsim 12$}

\textit{JWST} observations revealed a large abundance of UV-bright galaxies at $z\gtrsim 12$~\citep{Finkelstein2023,Finkelstein2024,Harikane2023} that is challenging for canonical models of galaxy formation. In Section~\ref{sec:result-epshalo}, we find that halo-scale SFE in \thesanzoom is, in general, higher than what has been assumed in empirical galaxy formation models in low-mass haloes and it features a mild increase with redshift in the high-mass end. Obviously, these trends have the potential to explain the abundance of bright galaxies at cosmic dawn. To investigate this, we pair the $\epsilon^{\ast}_{\rm halo}(M_{\rm halo})$ in \thesanzoom with the empirical framework developed in \citet{Shen2023,Shen2024}. This framework connects the halo mass function to the UV luminosity function, accounting for the variability/scatter of UV luminosity at fixed halo mass ($\sigma_{\rm UV}$) due to e.g. bursty star-formation.

We adopt the analytical form $\epsilon^{\ast}_{\rm halo}$ fitted in Section~\ref{sec:result-epshalo} at $z \simeq 12$ (the $z \simeq 14$ is essentially identical to the $z \simeq 12$ fitting). Since the dynamical range covered by \thesanzoom ends around $M_{\rm halo} = 10^{10}\msun$, we truncate the double power-law function there and assume $\epsilon^{\ast}_{\rm halo}$ is capped constant at $M_{\rm halo} \geq 10^{10}\msun$. This provides a conservative estimate of the SFE in massive haloes outside our simulation coverage, since $\epsilon^{\ast}_{\rm halo}$ generally increases with halo mass in most models without AGN feedback. We assume the \citet{Chabrier2003} stellar initial mass function to be consistent with the simulations. For UV variability, following \citet{Gelli2024,Shen2024}, we adopt the scaling $\sigma_{\rm UV}\propto M^{-1/3}_{\rm halo}$, the normalization $\sigma_{\rm UV}(10^{10.5}\msun)=0.8 \mmag$, and a minimum value $\sigma_{\rm UV}^{\rm min}=0.4 \mmag$. We adopt an analytical form of $\sigma_{\rm UV}$ here. The overall value is consistent with the analysis of main-sequence scatters in \thesanzoom from McClymont et al. (2025). But a more detailed analysis directly based on star-formation histories of \thesanzoom galaxies is expected in Shen et al. (in prep.). In Figure~\ref{appfig:var} of the Appendix, we show that this level of $\sigma_{\rm UV}$ is consistent with predictions from cosmological simulations~\citep[e.g.][]{Katz2023,Pallottini2023,Feldmann2024}. The same $\sigma_{\rm UV}(M_{\rm halo})$ also leads to model predictions that are consistent with the UV luminosity function constraints at low redshifts when using the fitted $\epsilon^{\ast}_{\rm halo}(M_{\rm halo})$ from \thesanzoom. An example at $z\simeq 6$ is shown in Figure~\ref{appfig:var} as well. 

\begin{figure}
    \centering
    \includegraphics[width=\linewidth]{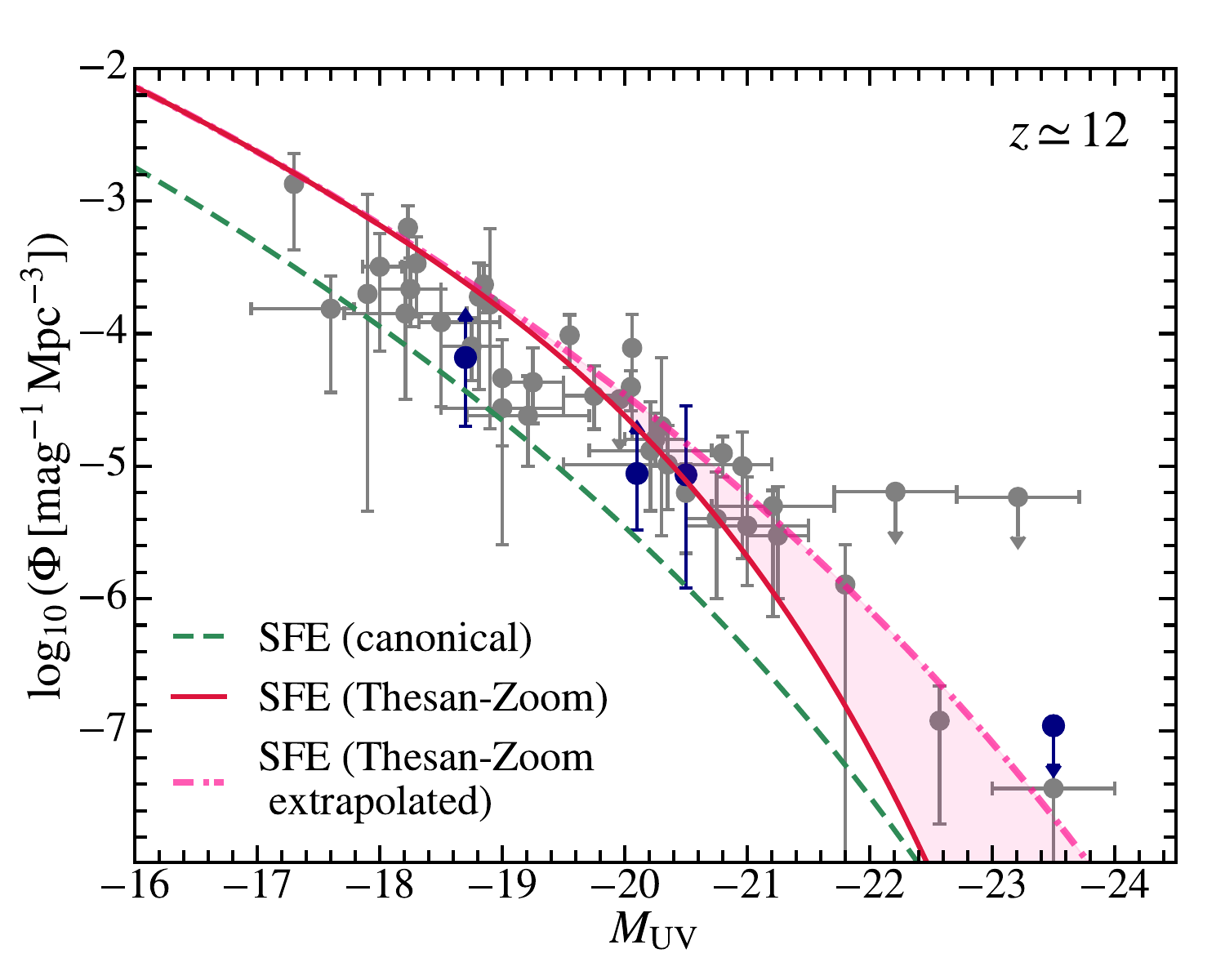}
    \includegraphics[width=\linewidth]{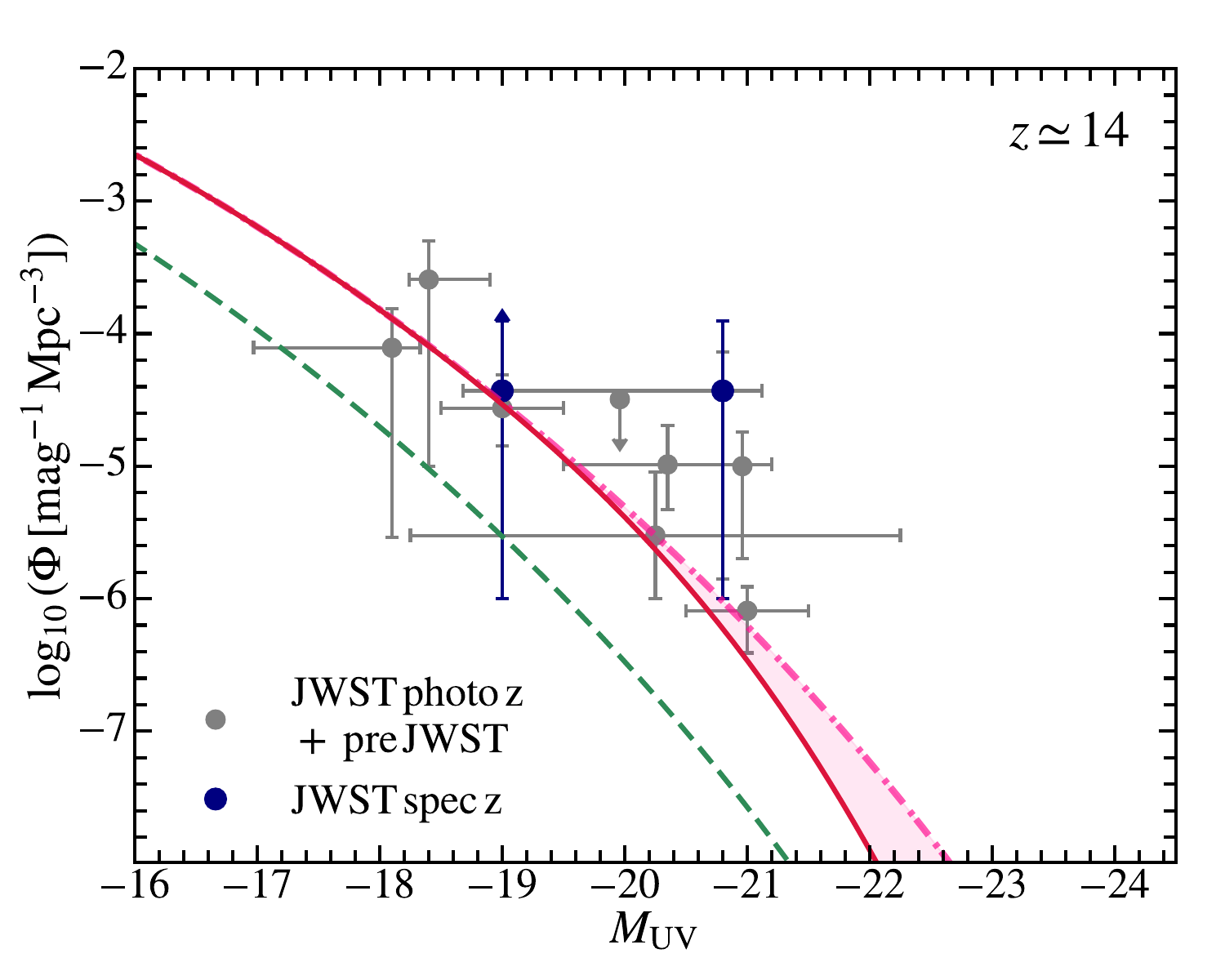}
    \caption{Galaxy UV luminosity function at $z\simeq 12$ and $z\simeq 14$ predicted using the halo-scale SFE found in \thesanzoom. The red solid line shows predictions using SFE measured in \thesanzoom, which is terminated at $M_{\rm halo} = 10^{10}\msun$ and capped there. The pink dot-dashed line shows the prediction extrapolating \thesanzoom SFE to $10^{12}\msun$. The green dashed line shows a ``canonical'' model prediction using halo-scale SFE from previous theoretical and observational works. In all cases, we assume a reasonable value of UV variability depending on halo mass~\citep[see the main text for details;][]{Shen2024}. We compare them to the observational constraints compiled in \citet{Shen2023,Shen2024}. While the ``canonical'' model underpredicts the UV luminosity function by more than half a dex, the \thesanzoom predictions are in good agreement with observations to $M_{\rm UV} \sim -21$. Matching the bright-end UVLF at $z\simeq 12$ requires extrapolating the halo-scale SFE to larger halo masses.}
    \label{fig:uvlf}
\end{figure}

In Figure~\ref{fig:uvlf}, we show the predictions combining \thesanzoom with the analytical model at $z\simeq 12$ and $14$. We show the prediction of a ``canonical'' model in \citet{Shen2024} using the median $\epsilon^{\ast}_{\rm halo}$ in pre-\textit{JWST} theoretical works with the dashed line. The solid line shows the prediction using $\epsilon^{\ast}_{\rm halo}$ derived from \thesanzoom as discussed above. The pink dot-dashed line shows the prediction when further extrapolating \thesanzoom SFE to $M_{\rm halo}=10^{12}\msun$. We compare them to the observational constraints compiled in \citet{Shen2023,Shen2024}\footnote{This includes the HST observations compiled in \citet{Vogelsberger2020} and additional ones from \citet{McLeod2016,Oesch2018,Morishita2018,Stefanon2019,Bowler2020,Bouwens2021}. The JWST constraints are taken from \citet{Castellano2022, Finkelstein2022, Naidu2022, Adams2023, Bouwens2023a, Bouwens2023b, Donnan2023, Harikane2023, Leetho2023, Morishita2023, Perez2023, Robertson2024, McLeod2024, Donnan2024, Casey2024}.}. In particular, the spectroscopic constraints are taken from \citet{Harikane2024b,Harikane2024a}. While the ``canonical'' model underpredicts the abundance of bright galaxies, the \thesanzoom model agrees quite well with observations due to the slightly higher $\epsilon^{\ast}_{\rm halo}$ in low-mass haloes. A moderately low value of $\epsilon^{\ast}_{\rm halo}$ ($\lesssim 10\%$ at $10^{10}\msun$) is adequate to explain the UV luminosity function at these redshifts until reaching the brightest end ($M_{\rm UV}\sim -21$). The bright-end UV luminosity function constraints at $z\simeq 12$ would require extrapolating $\epsilon^{\ast}_{\rm halo}$ to order unity in massive haloes. \citet{Feldmann2024} found similar results from binned estimations of the UV luminosity function in FIREBox to $M_{\rm UV}\sim -19$, which is also attributed to the elevated halo-scale SFE in low-mass haloes. As discussed there, considering low-luminosity galaxies dominate the total UV luminosity density of the Universe at these redshifts, \textit{JWST} constraints on UV luminosity density at $z\gtrsim 10$ will not be in tension with theoretical models considering the enhanced SFE we found in \thesanzoom.

\section{Conclusions}
\label{sec:conclusion}

In this work, we have investigated the efficiency of star-formation on both halo and galactic scales using the \thesanzoom simulations, a state-of-the-art zoom-in radiation-hydrodynamic simulation campaign designed to study high-redshift galaxies ($z\gtrsim 3$). \thesanzoom utilizes an on-the-fly radiation-hydrodynamic solver with self-consistent boundary conditions taken from the parent \thesanone simulation, and a non-equilibrium thermochemistry module to model the interactions between radiation, gas, and dust. \thesanzoom incorporates an advanced galaxy formation model with star-formation and multiple channels of stellar feedback in a resolved multiphase ISM. It provides a robust framework for understanding the interplay between star formation, feedback processes, and the large-scale reionization environment. Our key findings are summarized below.

\begin{itemize}

    \item The \textbf{Halo-scale SFE} quantifies the fraction of baryons accreted by the DM halo that is converted into stars. As a measure of integrated star-formation, the stellar-to-halo-mass ratios of galaxies in the fiducial \thesanzoom runs align well with observational constraints up to $z\simeq 8$, with the additional ESF playing a critical role in preventing the overproduction of stellar mass at low redshifts. Meanwhile, the instantaneous halo-scale SFE, $\epsilon^{\ast}_{\rm halo}$, exhibits a clear dependence on halo mass following approximately a double power-law function. The break halo mass is $M_{\rm halo}\sim 10^{9}\msun$, and the characteristic slope of the power-law is $1/3$ and $2/3$ in the high and low-mass regimes, respectively. The change of the slope can be understood as a transition of gas outflows from a momentum-driven to an energy-driven regime. At $z\gtrsim 8$, this relationship simplifies to a single power-law in the mass range covered by \thesanzoom. This results in a mild increase of $\epsilon^{\ast}_{\rm halo}$ at $M_{\rm halo}\gtrsim 10^{9}\msun$ at $z\gtrsim 8$. The $\epsilon^{\ast}_{\rm halo}$ we find in \thesanzoom is systematically higher than what is typically assumed in canonical empirical/semi-analytical models for galaxy formation at $M_{\rm halo}\lesssim 10^{11}\msun$.

    \item \textbf{Supply of cold gas for star-formation:} We decompose the halo-scale SFE into a ``supply'' term (proportional to the cold, neutral gas fraction in the halo) and a ``consumption'' term (related to the depletion time of gas). For the ``supply'' term, the total gas fraction within the central star-forming region ($<0.3\,R_{\rm vir}$) increases at lower redshifts, while the fraction of cold, neutral gas in the star-forming region stabilizes at approximately 20\%, independent of redshift or halo mass, likely as a consequence of the radiative feedback from local young stars regulating the ionized fractions of gas. This suggests that the halo-scale SFE is primarily governed by the ``consumption'' rate of neutral gas in the central star-forming region.
    
    \item \textbf{The KS relation and galaxy-scale SFE:} Regarding the ``consumption'' term, we find that the KS relation of neutral gas on kpc-scales is independent of halo mass and redshift, featuring roughly a $\Sigma_{\rm SFR}\propto \Sigma^{2}_{\rm HI+H_2}$ scaling. Such a universality of the KS relation and the power-law scaling are expected from the feedback-regulated star-formation when the turbulent energy dissipation in the ISM is balanced by the injection from stellar feedback. None of the numerical parameters affect the KS relation. The only factor we find that affects the KS relation is the additional ESF, which reduces the normalization of the relation when included. This is consistent with the analytical picture that the total feedback momentum injection dictates the gas depletion time in equilibrium. Combining the findings for the neutral gas fraction in haloes and the depletion time learned from the KS relations, one can reproduce the halo mass and redshift dependence of the halo-scale SFE. The increase of $\epsilon^{\ast}_{\rm halo}$ with halo mass is mainly driven by the increased gas surface densities in massive haloes.

    \item \textbf{Implications for UV luminosity functions at $z\gtrsim 12$:} While \thesanzoom does not yet explore the regime of extremely massive haloes and feedback-free/failure star formation scenarios, the mild increase in $\epsilon^{\ast}_{\rm halo}$ with redshift is sufficient to explain the observed abundance of UV-bright galaxies at $z\gtrsim 12$. We show this by pairing the $\epsilon^{\ast}_{\rm halo}$ found in \thesanzoom with an empirical model of galaxy formation. We assume a reasonable level of UV variability of galaxies ($\sigma_{\rm UV}\propto M^{-1/3}_{\rm halo}$ and $\sigma_{\rm UV}(10^{10.5}\msun)=0.8\mmag$) that is consistent with predictions from cosmological simulations. However, we note that explaining the UV luminosity function at bright-end ($M_{\rm UV}\lesssim -21$) requires extrapolations of the $\epsilon^{\ast}_{\rm halo}$ in \thesanzoom to more massive haloes.
    
\end{itemize}

In summary, we find that star-formation in high-redshift galaxies exhibits self-regulated, quasi-universal behavior at both the galaxy and halo scales. The halo-scale SFE is linked to kpc-scale gas depletion times self-regulated by stellar feedback. In the halo mass range covered by \thesanzoom, no out-of-equilibrium star-formation is found in high-redshift galaxies, likely due to the relatively low gas surface densities compared to the critical thresholds for unregulated star-formation proposed in many theories. We will conduct a direct study of these more massive and extreme systems in follow-up simulations using the same RHD method and ISM models but incorporating the physics of SMBH seeding, growth, and feedback, which are likely also important in this regime.

\section*{Acknowledgements}

The authors gratefully acknowledge the Gauss Centre for Supercomputing e.V. (\url{www.gauss-centre.eu}) for funding this project by providing computing time on the GCS Supercomputer SuperMUC-NG at Leibniz Supercomputing Centre (\url{www.lrz.de}), under project pn29we. XS acknowledges support from the National Aeronautics and Space Administration (NASA) theory grant JWST-AR-04814. RK acknowledges the support of the Natural Sciences and Engineering Research Council of Canada (NSERC) through a Discovery Grant and a Discovery Launch Supplement (funding reference numbers RGPIN-2024-06222 and DGECR-2024-00144) and York University's Global Research Excellence Initiative. Support for OZ was provided by Harvard University through the Institute for Theory and Computation Fellowship. EG is grateful to the Canon Foundation Europe and the Osaka University for their support through the Canon Fellowship. WM thanks the Science and Technology Facilities Council (STFC) Center for Doctoral Training (CDT) in Data intensive Science at the University of Cambridge (STFC grant number 2742968) for a PhD studentship. LH acknowledges support from the Simons Collaboration on ``Learning the Universe''.

\section*{Data Availability}
All simulation data, including snapshots, group catalogs, and merger trees will be made publicly available in the near future. Data will be distributed via \url{www.thesan-project.com}. Before the public data release, the data underlying this article can be shared upon reasonable request to the corresponding author(s).




\begin{thebibliography}{}
\makeatletter
\relax
\def\mn@urlcharsother{\let\do\@makeother \do\$\do\&\do\#\do\^\do\_\do\%\do\~}
\def\mn@doi{\begingroup\mn@urlcharsother \@ifnextchar [ {\mn@doi@}
  {\mn@doi@[]}}
\def\mn@doi@[#1]#2{\def\@tempa{#1}\ifx\@tempa\@empty \href
  {http://dx.doi.org/#2} {doi:#2}\else \href {http://dx.doi.org/#2} {#1}\fi
  \endgroup}
\def\mn@eprint#1#2{\mn@eprint@#1:#2::\@nil}
\def\mn@eprint@arXiv#1{\href {http://arxiv.org/abs/#1} {{\tt arXiv:#1}}}
\def\mn@eprint@dblp#1{\href {http://dblp.uni-trier.de/rec/bibtex/#1.xml}
  {dblp:#1}}
\def\mn@eprint@#1:#2:#3:#4\@nil{\def\@tempa {#1}\def\@tempb {#2}\def\@tempc
  {#3}\ifx \@tempc \@empty \let \@tempc \@tempb \let \@tempb \@tempa \fi \ifx
  \@tempb \@empty \def\@tempb {arXiv}\fi \@ifundefined
  {mn@eprint@\@tempb}{\@tempb:\@tempc}{\expandafter \expandafter \csname
  mn@eprint@\@tempb\endcsname \expandafter{\@tempc}}}

\bibitem[\protect\citeauthoryear{{Aarseth}}{{Aarseth}}{2003}]{Aarseth2003}
{Aarseth} S.~J.,  2003, {Gravitational N-Body Simulations}.
{Cambridge University Press}

\bibitem[\protect\citeauthoryear{{Adams} et~al.,}{{Adams}
  et~al.}{2023}]{Adams2023}
{Adams} N.~J.,  et~al., 2023, \mn@doi [\mnras] {10.1093/mnras/stac3347}, \href
  {https://ui.adsabs.harvard.edu/abs/2023MNRAS.518.4755A} {518, 4755}

\bibitem[\protect\citeauthoryear{{Agertz}, {Kravtsov}, {Leitner}  \&
  {Gnedin}}{{Agertz} et~al.}{2013}]{Agertz2013}
{Agertz} O.,  {Kravtsov} A.~V.,  {Leitner} S.~N.,   {Gnedin} N.~Y.,  2013,
  \mn@doi [\apj] {10.1088/0004-637X/770/1/25}, \href
  {https://ui.adsabs.harvard.edu/abs/2013ApJ...770...25A} {770, 25}

\bibitem[\protect\citeauthoryear{{Agertz} et~al.,}{{Agertz}
  et~al.}{2021}]{Agertz2021}
{Agertz} O.,  et~al., 2021, \mn@doi [\mnras] {10.1093/mnras/stab322}, \href
  {https://ui.adsabs.harvard.edu/abs/2021MNRAS.503.5826A} {503, 5826}

\bibitem[\protect\citeauthoryear{{Aguirre}, {Hernquist}, {Schaye}, {Weinberg},
  {Katz}  \& {Gardner}}{{Aguirre} et~al.}{2001}]{Aguirre2001}
{Aguirre} A.,  {Hernquist} L.,  {Schaye} J.,  {Weinberg} D.~H.,  {Katz} N.,
  {Gardner} J.,  2001, \mn@doi [\apj] {10.1086/323070}, \href
  {https://ui.adsabs.harvard.edu/abs/2001ApJ...560..599A} {560, 599}

\bibitem[\protect\citeauthoryear{{Angl{\'e}s-Alc{\'a}zar},
  {Faucher-Gigu{\`e}re}, {Kere{\v{s}}}, {Hopkins}, {Quataert}  \&
  {Murray}}{{Angl{\'e}s-Alc{\'a}zar} et~al.}{2017}]{Angles2017}
{Angl{\'e}s-Alc{\'a}zar} D.,  {Faucher-Gigu{\`e}re} C.-A.,  {Kere{\v{s}}} D.,
  {Hopkins} P.~F.,  {Quataert} E.,   {Murray} N.,  2017, \mn@doi [\mnras]
  {10.1093/mnras/stx1517}, \href
  {https://ui.adsabs.harvard.edu/abs/2017MNRAS.470.4698A} {470, 4698}

\bibitem[\protect\citeauthoryear{{Barnes} \& {Hut}}{{Barnes} \&
  {Hut}}{1986}]{Barnes1986}
{Barnes} J.,  {Hut} P.,  1986, \mn@doi [\nat] {10.1038/324446a0}, \href
  {https://ui.adsabs.harvard.edu/abs/1986Natur.324..446B} {324, 446}

\bibitem[\protect\citeauthoryear{{Bate}}{{Bate}}{2012}]{Bate2012}
{Bate} M.~R.,  2012, \mn@doi [\mnras] {10.1111/j.1365-2966.2011.19955.x}, \href
  {https://ui.adsabs.harvard.edu/abs/2012MNRAS.419.3115B} {419, 3115}

\bibitem[\protect\citeauthoryear{{Behroozi} \& {Silk}}{{Behroozi} \&
  {Silk}}{2015}]{Behroozi2015}
{Behroozi} P.~S.,  {Silk} J.,  2015, \mn@doi [\apj]
  {10.1088/0004-637X/799/1/32}, \href
  {https://ui.adsabs.harvard.edu/abs/2015ApJ...799...32B} {799, 32}

\bibitem[\protect\citeauthoryear{{Behroozi}, {Wechsler}  \&
  {Conroy}}{{Behroozi} et~al.}{2013}]{Behroozi2013}
{Behroozi} P.~S.,  {Wechsler} R.~H.,   {Conroy} C.,  2013, \mn@doi [\apj]
  {10.1088/0004-637X/770/1/57}, \href
  {https://ui.adsabs.harvard.edu/abs/2013ApJ...770...57B} {770, 57}

\bibitem[\protect\citeauthoryear{{Behroozi}, {Wechsler}, {Hearin}  \&
  {Conroy}}{{Behroozi} et~al.}{2019}]{Behroozi2019}
{Behroozi} P.,  {Wechsler} R.~H.,  {Hearin} A.~P.,   {Conroy} C.,  2019,
  \mn@doi [\mnras] {10.1093/mnras/stz1182}, \href
  {https://ui.adsabs.harvard.edu/abs/2019MNRAS.488.3143B} {488, 3143}

\bibitem[\protect\citeauthoryear{{Behroozi} et~al.,}{{Behroozi}
  et~al.}{2020}]{Behroozi2020}
{Behroozi} P.,  et~al., 2020, \mn@doi [\mnras] {10.1093/mnras/staa3164}, \href
  {https://ui.adsabs.harvard.edu/abs/2020MNRAS.499.5702B} {499, 5702}

\bibitem[\protect\citeauthoryear{{Birnboim} \& {Dekel}}{{Birnboim} \&
  {Dekel}}{2003}]{Birnboim2003}
{Birnboim} Y.,  {Dekel} A.,  2003, \mn@doi [\mnras]
  {10.1046/j.1365-8711.2003.06955.x}, \href
  {https://ui.adsabs.harvard.edu/abs/2003MNRAS.345..349B} {345, 349}

\bibitem[\protect\citeauthoryear{{Blumenthal}, {Faber}, {Primack}  \&
  {Rees}}{{Blumenthal} et~al.}{1984}]{Blumenthal1984}
{Blumenthal} G.~R.,  {Faber} S.~M.,  {Primack} J.~R.,   {Rees} M.~J.,  1984,
  \mn@doi [\nat] {10.1038/311517a0}, \href
  {https://ui.adsabs.harvard.edu/abs/1984Natur.311..517B} {311, 517}

\bibitem[\protect\citeauthoryear{{Borrow} \& {Borrisov}}{{Borrow} \&
  {Borrisov}}{2020}]{Borrow2020}
{Borrow} J.,  {Borrisov} A.,  2020, \mn@doi [The Journal of Open Source
  Software] {10.21105/joss.02430}, \href
  {https://ui.adsabs.harvard.edu/abs/2020JOSS....5.2430B} {5, 2430}

\bibitem[\protect\citeauthoryear{{Borrow} \& {Kelly}}{{Borrow} \&
  {Kelly}}{2021}]{Borrow2021}
{Borrow} J.,  {Kelly} A.~J.,  2021, \mn@doi [arXiv e-prints]
  {10.48550/arXiv.2106.05281}, \href
  {https://ui.adsabs.harvard.edu/abs/2021arXiv210605281B} {p. arXiv:2106.05281}

\bibitem[\protect\citeauthoryear{{Borrow}, {Kannan}, {Garaldi}, {Smith},
  {Vogelsberger}, {Pakmor}, {Springel}  \& {Hernquist}}{{Borrow}
  et~al.}{2023}]{Borrow2023}
{Borrow} J.,  {Kannan} R.,  {Garaldi} E.,  {Smith} A.,  {Vogelsberger} M.,
  {Pakmor} R.,  {Springel} V.,   {Hernquist} L.,  2023, \mn@doi [\mnras]
  {10.1093/mnras/stad2523}, \href
  {https://ui.adsabs.harvard.edu/abs/2023MNRAS.525.5932B} {525, 5932}

\bibitem[\protect\citeauthoryear{{Bouch{\'e}} et~al.,}{{Bouch{\'e}}
  et~al.}{2010}]{Bouche2010}
{Bouch{\'e}} N.,  et~al., 2010, \mn@doi [\apj] {10.1088/0004-637X/718/2/1001},
  \href {https://ui.adsabs.harvard.edu/abs/2010ApJ...718.1001B} {718, 1001}

\bibitem[\protect\citeauthoryear{{Bournaud}, {Elmegreen}  \&
  {Elmegreen}}{{Bournaud} et~al.}{2007}]{Bournaud2007}
{Bournaud} F.,  {Elmegreen} B.~G.,   {Elmegreen} D.~M.,  2007, \mn@doi [\apj]
  {10.1086/522077}, \href
  {https://ui.adsabs.harvard.edu/abs/2007ApJ...670..237B} {670, 237}

\bibitem[\protect\citeauthoryear{{Bouwens} et~al.,}{{Bouwens}
  et~al.}{2021}]{Bouwens2021}
{Bouwens} R.~J.,  et~al., 2021, \mn@doi [\aj] {10.3847/1538-3881/abf83e}, \href
  {https://ui.adsabs.harvard.edu/abs/2021AJ....162...47B} {162, 47}

\bibitem[\protect\citeauthoryear{{Bouwens}, {Illingworth}, {Oesch}, {Stefanon},
  {Naidu}, {van Leeuwen}  \& {Magee}}{{Bouwens} et~al.}{2023a}]{Bouwens2023a}
{Bouwens} R.,  {Illingworth} G.,  {Oesch} P.,  {Stefanon} M.,  {Naidu} R.,
  {van Leeuwen} I.,   {Magee} D.,  2023a, \mn@doi [\mnras]
  {10.1093/mnras/stad1014}, \href
  {https://ui.adsabs.harvard.edu/abs/2023MNRAS.523.1009B} {523, 1009}

\bibitem[\protect\citeauthoryear{{Bouwens} et~al.,}{{Bouwens}
  et~al.}{2023b}]{Bouwens2023b}
{Bouwens} R.~J.,  et~al., 2023b, \mn@doi [\mnras] {10.1093/mnras/stad1145},
  \href {https://ui.adsabs.harvard.edu/abs/2023MNRAS.523.1036B} {523, 1036}

\bibitem[\protect\citeauthoryear{{Bowler}, {Jarvis}, {Dunlop}, {McLure},
  {McLeod}, {Adams}, {Milvang-Jensen}  \& {McCracken}}{{Bowler}
  et~al.}{2020}]{Bowler2020}
{Bowler} R.~A.~A.,  {Jarvis} M.~J.,  {Dunlop} J.~S.,  {McLure} R.~J.,  {McLeod}
  D.~J.,  {Adams} N.~J.,  {Milvang-Jensen} B.,   {McCracken} H.~J.,  2020,
  \mn@doi [\mnras] {10.1093/mnras/staa313}, \href
  {https://ui.adsabs.harvard.edu/abs/2020MNRAS.493.2059B} {493, 2059}

\bibitem[\protect\citeauthoryear{{Boylan-Kolchin}}{{Boylan-Kolchin}}{2024}]{BK2024}
{Boylan-Kolchin} M.,  2024, \mn@doi [arXiv e-prints]
  {10.48550/arXiv.2407.10900}, \href
  {https://ui.adsabs.harvard.edu/abs/2024arXiv240710900B} {p. arXiv:2407.10900}

\bibitem[\protect\citeauthoryear{{Buck}, {Pfrommer}, {Pakmor}, {Grand}  \&
  {Springel}}{{Buck} et~al.}{2020}]{Buck2020}
{Buck} T.,  {Pfrommer} C.,  {Pakmor} R.,  {Grand} R. J.~J.,   {Springel} V.,
  2020, \mn@doi [\mnras] {10.1093/mnras/staa1960}, \href
  {https://ui.adsabs.harvard.edu/abs/2020MNRAS.497.1712B} {497, 1712}

\bibitem[\protect\citeauthoryear{{Bullock}, {Kravtsov}  \&
  {Weinberg}}{{Bullock} et~al.}{2000}]{Bullock2000}
{Bullock} J.~S.,  {Kravtsov} A.~V.,   {Weinberg} D.~H.,  2000, \mn@doi [\apj]
  {10.1086/309279}, \href
  {https://ui.adsabs.harvard.edu/abs/2000ApJ...539..517B} {539, 517}

\bibitem[\protect\citeauthoryear{{Casey} et~al.,}{{Casey}
  et~al.}{2024}]{Casey2024}
{Casey} C.~M.,  et~al., 2024, \mn@doi [\apj] {10.3847/1538-4357/ad2075}, \href
  {https://ui.adsabs.harvard.edu/abs/2024ApJ...965...98C} {965, 98}

\bibitem[\protect\citeauthoryear{{Castellano} et~al.,}{{Castellano}
  et~al.}{2022}]{Castellano2022}
{Castellano} M.,  et~al., 2022, \mn@doi [\apjl] {10.3847/2041-8213/ac94d0},
  \href {https://ui.adsabs.harvard.edu/abs/2022ApJ...938L..15C} {938, L15}

\bibitem[\protect\citeauthoryear{{Ceverino}, {Glover}  \& {Klessen}}{{Ceverino}
  et~al.}{2017}]{Ceverino2017}
{Ceverino} D.,  {Glover} S. C.~O.,   {Klessen} R.~S.,  2017, \mn@doi [\mnras]
  {10.1093/mnras/stx1386}, \href
  {https://ui.adsabs.harvard.edu/abs/2017MNRAS.470.2791C} {470, 2791}

\bibitem[\protect\citeauthoryear{{Ceverino}, {Nakazato}, {Yoshida}, {Klessen}
  \& {Glover}}{{Ceverino} et~al.}{2024}]{Ceverino2024}
{Ceverino} D.,  {Nakazato} Y.,  {Yoshida} N.,  {Klessen} R.~S.,   {Glover}
  S.~C.~O.,  2024, \mn@doi [\aap] {10.1051/0004-6361/202450224}, \href
  {https://ui.adsabs.harvard.edu/abs/2024A&A...689A.244C} {689, A244}

\bibitem[\protect\citeauthoryear{{Chabrier}}{{Chabrier}}{2003}]{Chabrier2003}
{Chabrier} G.,  2003, \mn@doi [\pasp] {10.1086/376392}, \href
  {https://ui.adsabs.harvard.edu/abs/2003PASP..115..763C} {115, 763}

\bibitem[\protect\citeauthoryear{{Chevalier} \& {Clegg}}{{Chevalier} \&
  {Clegg}}{1985}]{Chevalier1985-wind}
{Chevalier} R.~A.,  {Clegg} A.~W.,  1985, \mn@doi [\nat] {10.1038/317044a0},
  \href {https://ui.adsabs.harvard.edu/abs/1985Natur.317...44C} {317, 44}

\bibitem[\protect\citeauthoryear{{Chevance} et~al.,}{{Chevance}
  et~al.}{2020}]{Chevance2020}
{Chevance} M.,  et~al., 2020, \mn@doi [\mnras] {10.1093/mnras/stz3525}, \href
  {https://ui.adsabs.harvard.edu/abs/2020MNRAS.493.2872C} {493, 2872}

\bibitem[\protect\citeauthoryear{{Cioffi}, {McKee}  \& {Bertschinger}}{{Cioffi}
  et~al.}{1988}]{Cioffi1988}
{Cioffi} D.~F.,  {McKee} C.~F.,   {Bertschinger} E.,  1988, \mn@doi [\apj]
  {10.1086/166834}, \href
  {https://ui.adsabs.harvard.edu/abs/1988ApJ...334..252C} {334, 252}

\bibitem[\protect\citeauthoryear{{Coil}, {Weiner}, {Holz}, {Cooper}, {Yan}  \&
  {Aird}}{{Coil} et~al.}{2011}]{Coil2011}
{Coil} A.~L.,  {Weiner} B.~J.,  {Holz} D.~E.,  {Cooper} M.~C.,  {Yan} R.,
  {Aird} J.,  2011, \mn@doi [\apj] {10.1088/0004-637X/743/1/46}, \href
  {https://ui.adsabs.harvard.edu/abs/2011ApJ...743...46C} {743, 46}

\bibitem[\protect\citeauthoryear{{Cole}, {Lacey}, {Baugh}  \& {Frenk}}{{Cole}
  et~al.}{2000}]{Cole2000}
{Cole} S.,  {Lacey} C.~G.,  {Baugh} C.~M.,   {Frenk} C.~S.,  2000, \mn@doi
  [\mnras] {10.1046/j.1365-8711.2000.03879.x}, \href
  {https://ui.adsabs.harvard.edu/abs/2000MNRAS.319..168C} {319, 168}

\bibitem[\protect\citeauthoryear{{Conroy}, {Wechsler}  \& {Kravtsov}}{{Conroy}
  et~al.}{2006}]{Conroy2006}
{Conroy} C.,  {Wechsler} R.~H.,   {Kravtsov} A.~V.,  2006, \mn@doi [\apj]
  {10.1086/503602}, \href
  {https://ui.adsabs.harvard.edu/abs/2006ApJ...647..201C} {647, 201}

\bibitem[\protect\citeauthoryear{{Daddi} et~al.,}{{Daddi}
  et~al.}{2010}]{Daddi2010}
{Daddi} E.,  et~al., 2010, \mn@doi [\apjl] {10.1088/2041-8205/714/1/L118},
  \href {https://ui.adsabs.harvard.edu/abs/2010ApJ...714L.118D} {714, L118}

\bibitem[\protect\citeauthoryear{{Dav{\'e}}, {Finlator}  \&
  {Oppenheimer}}{{Dav{\'e}} et~al.}{2012}]{Dave2012}
{Dav{\'e}} R.,  {Finlator} K.,   {Oppenheimer} B.~D.,  2012, \mn@doi [\mnras]
  {10.1111/j.1365-2966.2011.20148.x}, \href
  {https://ui.adsabs.harvard.edu/abs/2012MNRAS.421...98D} {421, 98}

\bibitem[\protect\citeauthoryear{{Davis}, {Efstathiou}, {Frenk}  \&
  {White}}{{Davis} et~al.}{1985}]{Davis1985}
{Davis} M.,  {Efstathiou} G.,  {Frenk} C.~S.,   {White} S.~D.~M.,  1985,
  \mn@doi [\apj] {10.1086/163168}, \href
  {https://ui.adsabs.harvard.edu/abs/1985ApJ...292..371D} {292, 371}

\bibitem[\protect\citeauthoryear{{Dehnen} \& {Aly}}{{Dehnen} \&
  {Aly}}{2012}]{Dehnen2012}
{Dehnen} W.,  {Aly} H.,  2012, \mn@doi [\mnras]
  {10.1111/j.1365-2966.2012.21439.x}, \href
  {https://ui.adsabs.harvard.edu/abs/2012MNRAS.425.1068D} {425, 1068}

\bibitem[\protect\citeauthoryear{{Dekel}, {Sari}  \& {Ceverino}}{{Dekel}
  et~al.}{2009}]{Dekel2009}
{Dekel} A.,  {Sari} R.,   {Ceverino} D.,  2009, \mn@doi [\apj]
  {10.1088/0004-637X/703/1/785}, \href
  {https://ui.adsabs.harvard.edu/abs/2009ApJ...703..785D} {703, 785}

\bibitem[\protect\citeauthoryear{{Dekel}, {Sarkar}, {Birnboim}, {Mandelker}  \&
  {Li}}{{Dekel} et~al.}{2023}]{Dekel2023}
{Dekel} A.,  {Sarkar} K.~C.,  {Birnboim} Y.,  {Mandelker} N.,   {Li} Z.,  2023,
  \mn@doi [\mnras] {10.1093/mnras/stad1557}, \href
  {https://ui.adsabs.harvard.edu/abs/2023MNRAS.523.3201D} {523, 3201}

\bibitem[\protect\citeauthoryear{{Deng}, {Li}, {Kannan}, {Smith},
  {Vogelsberger}  \& {Bryan}}{{Deng} et~al.}{2024}]{Deng2024}
{Deng} Y.,  {Li} H.,  {Kannan} R.,  {Smith} A.,  {Vogelsberger} M.,   {Bryan}
  G.~L.,  2024, \mn@doi [\mnras] {10.1093/mnras/stad3202}, \href
  {https://ui.adsabs.harvard.edu/abs/2024MNRAS.527..478D} {527, 478}

\bibitem[\protect\citeauthoryear{{Dib}, {Hony}  \& {Blanc}}{{Dib}
  et~al.}{2017}]{Dib2017}
{Dib} S.,  {Hony} S.,   {Blanc} G.,  2017, \mn@doi [\mnras]
  {10.1093/mnras/stx934}, \href
  {https://ui.adsabs.harvard.edu/abs/2017MNRAS.469.1521D} {469, 1521}

\bibitem[\protect\citeauthoryear{{Dome}, {Tacchella}, {Fialkov}, {Ceverino},
  {Dekel}, {Ginzburg}, {Lapiner}  \& {Looser}}{{Dome} et~al.}{2024}]{Dome2024}
{Dome} T.,  {Tacchella} S.,  {Fialkov} A.,  {Ceverino} D.,  {Dekel} A.,
  {Ginzburg} O.,  {Lapiner} S.,   {Looser} T.~J.,  2024, \mn@doi [\mnras]
  {10.1093/mnras/stad3239}, \href
  {https://ui.adsabs.harvard.edu/abs/2024MNRAS.527.2139D} {527, 2139}

\bibitem[\protect\citeauthoryear{{Donnan} et~al.,}{{Donnan}
  et~al.}{2023}]{Donnan2023}
{Donnan} C.~T.,  et~al., 2023, \mn@doi [\mnras] {10.1093/mnras/stac3472}, \href
  {https://ui.adsabs.harvard.edu/abs/2023MNRAS.518.6011D} {518, 6011}

\bibitem[\protect\citeauthoryear{{Donnan} et~al.,}{{Donnan}
  et~al.}{2024}]{Donnan2024}
{Donnan} C.~T.,  et~al., 2024, \mn@doi [arXiv e-prints]
  {10.48550/arXiv.2403.03171}, \href
  {https://ui.adsabs.harvard.edu/abs/2024arXiv240303171D} {p. arXiv:2403.03171}

\bibitem[\protect\citeauthoryear{{Draine}}{{Draine}}{2011}]{Draine2011}
{Draine} B.~T.,  2011, {Physics of the Interstellar and Intergalactic Medium}.
Princeton University Press

\bibitem[\protect\citeauthoryear{Dubroca \& Feugeas}{Dubroca \&
  Feugeas}{1999}]{Dubroca1999}
Dubroca B.,  Feugeas J.-L.,  1999, \mn@doi [Comptes Rendus de l'Académie des
  Sciences - Series I - Mathematics]
  {https://doi.org/10.1016/S0764-4442(00)87499-6}, 329, 915

\bibitem[\protect\citeauthoryear{{Eldridge}, {Stanway}, {Xiao}, {McClelland},
  {Taylor}, {Ng}, {Greis}  \& {Bray}}{{Eldridge} et~al.}{2017}]{Eldridge2017}
{Eldridge} J.~J.,  {Stanway} E.~R.,  {Xiao} L.,  {McClelland} L.~A.~S.,
  {Taylor} G.,  {Ng} M.,  {Greis} S.~M.~L.,   {Bray} J.~C.,  2017, \mn@doi
  [\pasa] {10.1017/pasa.2017.51}, \href
  {https://ui.adsabs.harvard.edu/abs/2017PASA...34...58E} {34, e058}

\bibitem[\protect\citeauthoryear{{Elmegreen}, {Elmegreen}, {Fernandez}  \&
  {Lemonias}}{{Elmegreen} et~al.}{2009}]{Elmegreen2009}
{Elmegreen} B.~G.,  {Elmegreen} D.~M.,  {Fernandez} M.~X.,   {Lemonias} J.~J.,
  2009, \mn@doi [\apj] {10.1088/0004-637X/692/1/12}, \href
  {https://ui.adsabs.harvard.edu/abs/2009ApJ...692...12E} {692, 12}

\bibitem[\protect\citeauthoryear{{Evans}}{{Evans}}{1999}]{Evans1999}
{Evans} II N.~J.,  1999, \mn@doi [\araa] {10.1146/annurev.astro.37.1.311},
  \href {https://ui.adsabs.harvard.edu/abs/1999ARA&A..37..311E} {37, 311}

\bibitem[\protect\citeauthoryear{{Evans} II et~al.,}{{Evans}
  et~al.}{2009}]{Evans2009}
{Evans} II N.~J.,  et~al., 2009, \mn@doi [\apjs] {10.1088/0067-0049/181/2/321},
  \href {https://ui.adsabs.harvard.edu/abs/2009ApJS..181..321E} {181, 321}

\bibitem[\protect\citeauthoryear{{Fakhouri}, {Ma}  \&
  {Boylan-Kolchin}}{{Fakhouri} et~al.}{2010}]{Fakhouri2010}
{Fakhouri} O.,  {Ma} C.-P.,   {Boylan-Kolchin} M.,  2010, \mn@doi [\mnras]
  {10.1111/j.1365-2966.2010.16859.x}, \href
  {https://ui.adsabs.harvard.edu/abs/2010MNRAS.406.2267F} {406, 2267}

\bibitem[\protect\citeauthoryear{{Faucher-Gigu{\`e}re}, {Lidz}, {Zaldarriaga}
  \& {Hernquist}}{{Faucher-Gigu{\`e}re} et~al.}{2009}]{FG2009}
{Faucher-Gigu{\`e}re} C.-A.,  {Lidz} A.,  {Zaldarriaga} M.,   {Hernquist} L.,
  2009, \mn@doi [\apj] {10.1088/0004-637X/703/2/1416}, \href
  {https://ui.adsabs.harvard.edu/abs/2009ApJ...703.1416F} {703, 1416}

\bibitem[\protect\citeauthoryear{{Faucher-Gigu{\`e}re}, {Quataert}  \&
  {Hopkins}}{{Faucher-Gigu{\`e}re} et~al.}{2013}]{FG2013}
{Faucher-Gigu{\`e}re} C.-A.,  {Quataert} E.,   {Hopkins} P.~F.,  2013, \mn@doi
  [\mnras] {10.1093/mnras/stt866}, \href
  {https://ui.adsabs.harvard.edu/abs/2013MNRAS.433.1970F} {433, 1970}

\bibitem[\protect\citeauthoryear{{Federrath} \& {Klessen}}{{Federrath} \&
  {Klessen}}{2012}]{Federrath2012}
{Federrath} C.,  {Klessen} R.~S.,  2012, \mn@doi [\apj]
  {10.1088/0004-637X/761/2/156}, \href
  {https://ui.adsabs.harvard.edu/abs/2012ApJ...761..156F} {761, 156}

\bibitem[\protect\citeauthoryear{{Feldmann} et~al.,}{{Feldmann}
  et~al.}{2024}]{Feldmann2024}
{Feldmann} R.,  et~al., 2024, \mn@doi [\mnras] {10.1093/mnras/stae2633}, \href
  {https://ui.adsabs.harvard.edu/abs/2024MNRAS.tmp.2560F} {}

\bibitem[\protect\citeauthoryear{{Ferreira} et~al.,}{{Ferreira}
  et~al.}{2022}]{Ferreira2022}
{Ferreira} L.,  et~al., 2022, \mn@doi [\apjl] {10.3847/2041-8213/ac947c}, \href
  {https://ui.adsabs.harvard.edu/abs/2022ApJ...938L...2F} {938, L2}

\bibitem[\protect\citeauthoryear{{Ferreira} et~al.,}{{Ferreira}
  et~al.}{2023}]{Ferreira2023}
{Ferreira} L.,  et~al., 2023, \mn@doi [\apj] {10.3847/1538-4357/acec76}, \href
  {https://ui.adsabs.harvard.edu/abs/2023ApJ...955...94F} {955, 94}

\bibitem[\protect\citeauthoryear{{Fielding}, {Quataert}  \&
  {Martizzi}}{{Fielding} et~al.}{2018}]{Fielding2018-wind}
{Fielding} D.,  {Quataert} E.,   {Martizzi} D.,  2018, \mn@doi [\mnras]
  {10.1093/mnras/sty2466}, \href
  {https://ui.adsabs.harvard.edu/abs/2018MNRAS.481.3325F} {481, 3325}

\bibitem[\protect\citeauthoryear{{Finkelstein} et~al.,}{{Finkelstein}
  et~al.}{2022}]{Finkelstein2022}
{Finkelstein} S.~L.,  et~al., 2022, \mn@doi [\apjl] {10.3847/2041-8213/ac966e},
  \href {https://ui.adsabs.harvard.edu/abs/2022ApJ...940L..55F} {940, L55}

\bibitem[\protect\citeauthoryear{{Finkelstein} et~al.,}{{Finkelstein}
  et~al.}{2023}]{Finkelstein2023}
{Finkelstein} S.~L.,  et~al., 2023, \mn@doi [\apjl] {10.3847/2041-8213/acade4},
  \href {https://ui.adsabs.harvard.edu/abs/2023ApJ...946L..13F} {946, L13}

\bibitem[\protect\citeauthoryear{{Finkelstein} et~al.,}{{Finkelstein}
  et~al.}{2024}]{Finkelstein2024}
{Finkelstein} S.~L.,  et~al., 2024, \mn@doi [\apjl] {10.3847/2041-8213/ad4495},
  \href {https://ui.adsabs.harvard.edu/abs/2024ApJ...969L...2F} {969, L2}

\bibitem[\protect\citeauthoryear{{Fitts} et~al.,}{{Fitts}
  et~al.}{2017}]{Fitts2017}
{Fitts} A.,  et~al., 2017, \mn@doi [\mnras] {10.1093/mnras/stx1757}, \href
  {https://ui.adsabs.harvard.edu/abs/2017MNRAS.471.3547F} {471, 3547}

\bibitem[\protect\citeauthoryear{{F{\"o}rster Schreiber}, {Shapley}, {Erb},
  {Genzel}, {Steidel}, {Bouch{\'e}}, {Cresci}  \& {Davies}}{{F{\"o}rster
  Schreiber} et~al.}{2011}]{Forster2011}
{F{\"o}rster Schreiber} N.~M.,  {Shapley} A.~E.,  {Erb} D.~K.,  {Genzel} R.,
  {Steidel} C.~C.,  {Bouch{\'e}} N.,  {Cresci} G.,   {Davies} R.,  2011,
  \mn@doi [\apj] {10.1088/0004-637X/731/1/65}, \href
  {https://ui.adsabs.harvard.edu/abs/2011ApJ...731...65F} {731, 65}

\bibitem[\protect\citeauthoryear{{Garaldi}, {Kannan}, {Smith}, {Springel},
  {Pakmor}, {Vogelsberger}  \& {Hernquist}}{{Garaldi}
  et~al.}{2022}]{Garaldi2022}
{Garaldi} E.,  {Kannan} R.,  {Smith} A.,  {Springel} V.,  {Pakmor} R.,
  {Vogelsberger} M.,   {Hernquist} L.,  2022, \mn@doi [\mnras]
  {10.1093/mnras/stac257}, \href
  {https://ui.adsabs.harvard.edu/abs/2022MNRAS.512.4909G} {512, 4909}

\bibitem[\protect\citeauthoryear{{Garaldi} et~al.,}{{Garaldi}
  et~al.}{2024}]{Garaldi2024}
{Garaldi} E.,  et~al., 2024, \mn@doi [\mnras] {10.1093/mnras/stae839}, \href
  {https://ui.adsabs.harvard.edu/abs/2024MNRAS.530.3765G} {530, 3765}

\bibitem[\protect\citeauthoryear{{Garrison-Kimmel} et~al.,}{{Garrison-Kimmel}
  et~al.}{2019}]{GK2019}
{Garrison-Kimmel} S.,  et~al., 2019, \mn@doi [\mnras] {10.1093/mnras/stz1317},
  \href {https://ui.adsabs.harvard.edu/abs/2019MNRAS.487.1380G} {487, 1380}

\bibitem[\protect\citeauthoryear{{Gelli}, {Mason}  \& {Hayward}}{{Gelli}
  et~al.}{2024}]{Gelli2024}
{Gelli} V.,  {Mason} C.,   {Hayward} C.~C.,  2024, \mn@doi [\apj]
  {10.3847/1538-4357/ad7b36}, \href
  {https://ui.adsabs.harvard.edu/abs/2024ApJ...975..192G} {975, 192}

\bibitem[\protect\citeauthoryear{{Gnedin} \& {Kaurov}}{{Gnedin} \&
  {Kaurov}}{2014}]{Gnedin2014}
{Gnedin} N.~Y.,  {Kaurov} A.~A.,  2014, \mn@doi [\apj]
  {10.1088/0004-637X/793/1/30}, \href
  {https://ui.adsabs.harvard.edu/abs/2014ApJ...793...30G} {793, 30}

\bibitem[\protect\citeauthoryear{{Grudi{\'c}}, {Hopkins},
  {Faucher-Gigu{\`e}re}, {Quataert}, {Murray}  \& {Kere{\v{s}}}}{{Grudi{\'c}}
  et~al.}{2018}]{Grudic2018}
{Grudi{\'c}} M.~Y.,  {Hopkins} P.~F.,  {Faucher-Gigu{\`e}re} C.-A.,  {Quataert}
  E.,  {Murray} N.,   {Kere{\v{s}}} D.,  2018, \mn@doi [\mnras]
  {10.1093/mnras/sty035}, \href
  {https://ui.adsabs.harvard.edu/abs/2018MNRAS.475.3511G} {475, 3511}

\bibitem[\protect\citeauthoryear{{Haardt} \& {Madau}}{{Haardt} \&
  {Madau}}{2012}]{Haardt2012}
{Haardt} F.,  {Madau} P.,  2012, \mn@doi [\apj] {10.1088/0004-637X/746/2/125},
  \href {https://ui.adsabs.harvard.edu/abs/2012ApJ...746..125H} {746, 125}

\bibitem[\protect\citeauthoryear{{Harikane} et~al.,}{{Harikane}
  et~al.}{2016}]{Harikane2016}
{Harikane} Y.,  et~al., 2016, \mn@doi [\apj] {10.3847/0004-637X/821/2/123},
  \href {https://ui.adsabs.harvard.edu/abs/2016ApJ...821..123H} {821, 123}

\bibitem[\protect\citeauthoryear{{Harikane} et~al.,}{{Harikane}
  et~al.}{2018}]{Harikane2018}
{Harikane} Y.,  et~al., 2018, \mn@doi [\pasj] {10.1093/pasj/psx097}, \href
  {https://ui.adsabs.harvard.edu/abs/2018PASJ...70S..11H} {70, S11}

\bibitem[\protect\citeauthoryear{{Harikane} et~al.,}{{Harikane}
  et~al.}{2022}]{Harikane2022}
{Harikane} Y.,  et~al., 2022, \mn@doi [\apjs] {10.3847/1538-4365/ac3dfc}, \href
  {https://ui.adsabs.harvard.edu/abs/2022ApJS..259...20H} {259, 20}

\bibitem[\protect\citeauthoryear{{Harikane} et~al.,}{{Harikane}
  et~al.}{2023}]{Harikane2023}
{Harikane} Y.,  et~al., 2023, \mn@doi [\apjs] {10.3847/1538-4365/acaaa9}, \href
  {https://ui.adsabs.harvard.edu/abs/2023ApJS..265....5H} {265, 5}

\bibitem[\protect\citeauthoryear{{Harikane} et~al.,}{{Harikane}
  et~al.}{2024a}]{Harikane2024b}
{Harikane} Y.,  et~al., 2024a, \mn@doi [arXiv e-prints]
  {10.48550/arXiv.2406.18352}, \href
  {https://ui.adsabs.harvard.edu/abs/2024arXiv240618352H} {p. arXiv:2406.18352}

\bibitem[\protect\citeauthoryear{{Harikane}, {Nakajima}, {Ouchi}, {Umeda},
  {Isobe}, {Ono}, {Xu}  \& {Zhang}}{{Harikane} et~al.}{2024b}]{Harikane2024a}
{Harikane} Y.,  {Nakajima} K.,  {Ouchi} M.,  {Umeda} H.,  {Isobe} Y.,  {Ono}
  Y.,  {Xu} Y.,   {Zhang} Y.,  2024b, \mn@doi [\apj]
  {10.3847/1538-4357/ad0b7e}, \href
  {https://ui.adsabs.harvard.edu/abs/2024ApJ...960...56H} {960, 56}

\bibitem[\protect\citeauthoryear{{Heckman}, {Lehnert}, {Strickland}  \&
  {Armus}}{{Heckman} et~al.}{2000}]{Heckman2000}
{Heckman} T.~M.,  {Lehnert} M.~D.,  {Strickland} D.~K.,   {Armus} L.,  2000,
  \mn@doi [\apjs] {10.1086/313421}, \href
  {https://ui.adsabs.harvard.edu/abs/2000ApJS..129..493H} {129, 493}

\bibitem[\protect\citeauthoryear{{Hennebelle} \& {Chabrier}}{{Hennebelle} \&
  {Chabrier}}{2011}]{Hennebelle2011}
{Hennebelle} P.,  {Chabrier} G.,  2011, \mn@doi [\apjl]
  {10.1088/2041-8205/743/2/L29}, \href
  {https://ui.adsabs.harvard.edu/abs/2011ApJ...743L..29H} {743, L29}

\bibitem[\protect\citeauthoryear{{Hirschi}}{{Hirschi}}{2007}]{Hirschi2007}
{Hirschi} R.,  2007, \mn@doi [\aap] {10.1051/0004-6361:20065356}, \href
  {https://ui.adsabs.harvard.edu/abs/2007A&A...461..571H} {461, 571}

\bibitem[\protect\citeauthoryear{{Hopkins}}{{Hopkins}}{2024}]{Hopkins2024}
{Hopkins} P.~F.,  2024, \mn@doi [arXiv e-prints] {10.48550/arXiv.2404.16987},
  \href {https://ui.adsabs.harvard.edu/abs/2024arXiv240416987H} {p.
  arXiv:2404.16987}

\bibitem[\protect\citeauthoryear{{Hopkins}, {Quataert}  \& {Murray}}{{Hopkins}
  et~al.}{2012}]{Hopkins2012}
{Hopkins} P.~F.,  {Quataert} E.,   {Murray} N.,  2012, \mn@doi [\mnras]
  {10.1111/j.1365-2966.2012.20593.x}, \href
  {https://ui.adsabs.harvard.edu/abs/2012MNRAS.421.3522H} {421, 3522}

\bibitem[\protect\citeauthoryear{{Hopkins}, {Narayanan}  \& {Murray}}{{Hopkins}
  et~al.}{2013}]{Hopkins2013}
{Hopkins} P.~F.,  {Narayanan} D.,   {Murray} N.,  2013, \mn@doi [\mnras]
  {10.1093/mnras/stt723}, \href
  {https://ui.adsabs.harvard.edu/abs/2013MNRAS.432.2647H} {432, 2647}

\bibitem[\protect\citeauthoryear{{Hopkins}, {Kere{\v{s}}}, {O{\~n}orbe},
  {Faucher-Gigu{\`e}re}, {Quataert}, {Murray}  \& {Bullock}}{{Hopkins}
  et~al.}{2014}]{Hopkins2014}
{Hopkins} P.~F.,  {Kere{\v{s}}} D.,  {O{\~n}orbe} J.,  {Faucher-Gigu{\`e}re}
  C.-A.,  {Quataert} E.,  {Murray} N.,   {Bullock} J.~S.,  2014, \mn@doi
  [\mnras] {10.1093/mnras/stu1738}, \href
  {https://ui.adsabs.harvard.edu/abs/2014MNRAS.445..581H} {445, 581}

\bibitem[\protect\citeauthoryear{{Hopkins} et~al.,}{{Hopkins}
  et~al.}{2018a}]{Hopkins2018feedback}
{Hopkins} P.~F.,  et~al., 2018a, \mn@doi [\mnras] {10.1093/mnras/sty674}, \href
  {https://ui.adsabs.harvard.edu/abs/2018MNRAS.477.1578H} {477, 1578}

\bibitem[\protect\citeauthoryear{{Hopkins} et~al.,}{{Hopkins}
  et~al.}{2018b}]{Hopkins2018}
{Hopkins} P.~F.,  et~al., 2018b, \mn@doi [\mnras] {10.1093/mnras/sty1690},
  \href {https://ui.adsabs.harvard.edu/abs/2018MNRAS.480..800H} {480, 800}

\bibitem[\protect\citeauthoryear{{Hopkins}, {Grudi{\'c}}, {Wetzel},
  {Kere{\v{s}}}, {Faucher-Gigu{\`e}re}, {Ma}, {Murray}  \& {Butcher}}{{Hopkins}
  et~al.}{2020a}]{Hopkins2020-rad}
{Hopkins} P.~F.,  {Grudi{\'c}} M.~Y.,  {Wetzel} A.,  {Kere{\v{s}}} D.,
  {Faucher-Gigu{\`e}re} C.-A.,  {Ma} X.,  {Murray} N.,   {Butcher} N.,  2020a,
  \mn@doi [\mnras] {10.1093/mnras/stz3129}, \href
  {https://ui.adsabs.harvard.edu/abs/2020MNRAS.491.3702H} {491, 3702}

\bibitem[\protect\citeauthoryear{{Hopkins} et~al.,}{{Hopkins}
  et~al.}{2020b}]{Hopkins2020-cr}
{Hopkins} P.~F.,  et~al., 2020b, \mn@doi [\mnras] {10.1093/mnras/stz3321},
  \href {https://ui.adsabs.harvard.edu/abs/2020MNRAS.492.3465H} {492, 3465}

\bibitem[\protect\citeauthoryear{{Hopkins}, {Wellons},
  {Angl{\'e}s-Alc{\'a}zar}, {Faucher-Gigu{\`e}re}  \& {Grudi{\'c}}}{{Hopkins}
  et~al.}{2022}]{Hopkins2022}
{Hopkins} P.~F.,  {Wellons} S.,  {Angl{\'e}s-Alc{\'a}zar} D.,
  {Faucher-Gigu{\`e}re} C.-A.,   {Grudi{\'c}} M.~Y.,  2022, \mn@doi [\mnras]
  {10.1093/mnras/stab3458}, \href
  {https://ui.adsabs.harvard.edu/abs/2022MNRAS.510..630H} {510, 630}

\bibitem[\protect\citeauthoryear{{Hopkins} et~al.,}{{Hopkins}
  et~al.}{2023}]{Hopkins2023fire3}
{Hopkins} P.~F.,  et~al., 2023, \mn@doi [\mnras] {10.1093/mnras/stac3489},
  \href {https://ui.adsabs.harvard.edu/abs/2023MNRAS.519.3154H} {519, 3154}

\bibitem[\protect\citeauthoryear{{Iliev}, {Shapiro}  \& {Raga}}{{Iliev}
  et~al.}{2005}]{Iliev2005}
{Iliev} I.~T.,  {Shapiro} P.~R.,   {Raga} A.~C.,  2005, \mn@doi [\mnras]
  {10.1111/j.1365-2966.2005.09155.x}, \href
  {https://ui.adsabs.harvard.edu/abs/2005MNRAS.361..405I} {361, 405}

\bibitem[\protect\citeauthoryear{{Kannan}, {Vogelsberger}, {Marinacci},
  {McKinnon}, {Pakmor}  \& {Springel}}{{Kannan} et~al.}{2019}]{Kannan2019}
{Kannan} R.,  {Vogelsberger} M.,  {Marinacci} F.,  {McKinnon} R.,  {Pakmor} R.,
    {Springel} V.,  2019, \mn@doi [\mnras] {10.1093/mnras/stz287}, \href
  {https://ui.adsabs.harvard.edu/abs/2019MNRAS.485..117K} {485, 117}

\bibitem[\protect\citeauthoryear{{Kannan}, {Marinacci}, {Simpson}, {Glover}  \&
  {Hernquist}}{{Kannan} et~al.}{2020a}]{Kannan2020feedback}
{Kannan} R.,  {Marinacci} F.,  {Simpson} C.~M.,  {Glover} S. C.~O.,
  {Hernquist} L.,  2020a, \mn@doi [\mnras] {10.1093/mnras/stz3078}, \href
  {https://ui.adsabs.harvard.edu/abs/2020MNRAS.491.2088K} {491, 2088}

\bibitem[\protect\citeauthoryear{{Kannan}, {Marinacci}, {Vogelsberger},
  {Sales}, {Torrey}, {Springel}  \& {Hernquist}}{{Kannan}
  et~al.}{2020b}]{Kannan2020smuggle}
{Kannan} R.,  {Marinacci} F.,  {Vogelsberger} M.,  {Sales} L.~V.,  {Torrey} P.,
   {Springel} V.,   {Hernquist} L.,  2020b, \mn@doi [\mnras]
  {10.1093/mnras/staa3249}, \href
  {https://ui.adsabs.harvard.edu/abs/2020MNRAS.499.5732K} {499, 5732}

\bibitem[\protect\citeauthoryear{{Kannan}, {Marinacci}, {Vogelsberger},
  {Sales}, {Torrey}, {Springel}  \& {Hernquist}}{{Kannan}
  et~al.}{2020c}]{Kannan2020}
{Kannan} R.,  {Marinacci} F.,  {Vogelsberger} M.,  {Sales} L.~V.,  {Torrey} P.,
   {Springel} V.,   {Hernquist} L.,  2020c, \mn@doi [\mnras]
  {10.1093/mnras/staa3249}, \href
  {https://ui.adsabs.harvard.edu/abs/2020MNRAS.499.5732K} {499, 5732}

\bibitem[\protect\citeauthoryear{{Kannan}, {Vogelsberger}, {Marinacci},
  {Sales}, {Torrey}  \& {Hernquist}}{{Kannan} et~al.}{2021}]{Kannan2021}
{Kannan} R.,  {Vogelsberger} M.,  {Marinacci} F.,  {Sales} L.~V.,  {Torrey} P.,
    {Hernquist} L.,  2021, \mn@doi [\mnras] {10.1093/mnras/stab416}, \href
  {https://ui.adsabs.harvard.edu/abs/2021MNRAS.503..336K} {503, 336}

\bibitem[\protect\citeauthoryear{{Kannan}, {Garaldi}, {Smith}, {Pakmor},
  {Springel}, {Vogelsberger}  \& {Hernquist}}{{Kannan}
  et~al.}{2022}]{Kannan2022thesan}
{Kannan} R.,  {Garaldi} E.,  {Smith} A.,  {Pakmor} R.,  {Springel} V.,
  {Vogelsberger} M.,   {Hernquist} L.,  2022, \mn@doi [\mnras]
  {10.1093/mnras/stab3710}, \href
  {https://ui.adsabs.harvard.edu/abs/2022MNRAS.511.4005K} {511, 4005}

\bibitem[\protect\citeauthoryear{{Kannan} et~al.,}{{Kannan}
  et~al.}{2023}]{Kannan2023}
{Kannan} R.,  et~al., 2023, \mn@doi [\mnras] {10.1093/mnras/stac3743}, \href
  {https://ui.adsabs.harvard.edu/abs/2023MNRAS.524.2594K} {524, 2594}

\bibitem[\protect\citeauthoryear{{Kannan} et~al.,}{{Kannan}
  et~al.}{2025}]{Kannan2025}
{Kannan} R.,  et~al., 2025, arXiv e-prints, \href
  {https://ui.adsabs.harvard.edu/abs/2025arXiv250220437K} {p. arXiv:2502.20437}

\bibitem[\protect\citeauthoryear{{Katz}, {Weinberg}  \& {Hernquist}}{{Katz}
  et~al.}{1996}]{Katz1996}
{Katz} N.,  {Weinberg} D.~H.,   {Hernquist} L.,  1996, \mn@doi [\apjs]
  {10.1086/192305}, \href
  {https://ui.adsabs.harvard.edu/abs/1996ApJS..105...19K} {105, 19}

\bibitem[\protect\citeauthoryear{{Katz} et~al.,}{{Katz}
  et~al.}{2020}]{Katz2020}
{Katz} H.,  et~al., 2020, \mn@doi [\mnras] {10.1093/mnras/staa639}, \href
  {https://ui.adsabs.harvard.edu/abs/2020MNRAS.494.2200K} {494, 2200}

\bibitem[\protect\citeauthoryear{{Katz} et~al.,}{{Katz}
  et~al.}{2023}]{Katz2023}
{Katz} H.,  et~al., 2023, \mn@doi [The Open Journal of Astrophysics]
  {10.21105/astro.2309.03269}, \href
  {https://ui.adsabs.harvard.edu/abs/2023OJAp....6E..44K} {6, 44}

\bibitem[\protect\citeauthoryear{{Kennicutt}}{{Kennicutt}}{1998}]{Kennicutt1998}
{Kennicutt} Robert~C. J.,  1998, \mn@doi [\araa]
  {10.1146/annurev.astro.36.1.189}, \href
  {https://ui.adsabs.harvard.edu/abs/1998ARA&A..36..189K} {36, 189}

\bibitem[\protect\citeauthoryear{{Kennicutt} \& {De Los Reyes}}{{Kennicutt} \&
  {De Los Reyes}}{2021}]{Kennicutt2021}
{Kennicutt} Robert~C. J.,  {De Los Reyes} M. A.~C.,  2021, \mn@doi [\apj]
  {10.3847/1538-4357/abd3a2}, \href
  {https://ui.adsabs.harvard.edu/abs/2021ApJ...908...61K} {908, 61}

\bibitem[\protect\citeauthoryear{{Kennicutt} \& {Evans}}{{Kennicutt} \&
  {Evans}}{2012}]{Kennicutt2012}
{Kennicutt} R.~C.,  {Evans} N.~J.,  2012, \mn@doi [\araa]
  {10.1146/annurev-astro-081811-125610}, \href
  {https://ui.adsabs.harvard.edu/abs/2012ARA&A..50..531K} {50, 531}

\bibitem[\protect\citeauthoryear{{Kere{\v{s}}}, {Katz}, {Weinberg}  \&
  {Dav{\'e}}}{{Kere{\v{s}}} et~al.}{2005}]{Keres2005}
{Kere{\v{s}}} D.,  {Katz} N.,  {Weinberg} D.~H.,   {Dav{\'e}} R.,  2005,
  \mn@doi [\mnras] {10.1111/j.1365-2966.2005.09451.x}, \href
  {https://ui.adsabs.harvard.edu/abs/2005MNRAS.363....2K} {363, 2}

\bibitem[\protect\citeauthoryear{{Kere{\v{s}}}, {Katz}, {Fardal}, {Dav{\'e}}
  \& {Weinberg}}{{Kere{\v{s}}} et~al.}{2009a}]{Keres2009c}
{Kere{\v{s}}} D.,  {Katz} N.,  {Fardal} M.,  {Dav{\'e}} R.,   {Weinberg} D.~H.,
   2009a, \mn@doi [\mnras] {10.1111/j.1365-2966.2009.14541.x}, \href
  {https://ui.adsabs.harvard.edu/abs/2009MNRAS.395..160K} {395, 160}

\bibitem[\protect\citeauthoryear{{Kere{\v{s}}}, {Katz}, {Dav{\'e}}, {Fardal}
  \& {Weinberg}}{{Kere{\v{s}}} et~al.}{2009b}]{Keres2009b}
{Kere{\v{s}}} D.,  {Katz} N.,  {Dav{\'e}} R.,  {Fardal} M.,   {Weinberg} D.~H.,
   2009b, \mn@doi [\mnras] {10.1111/j.1365-2966.2009.14924.x}, \href
  {https://ui.adsabs.harvard.edu/abs/2009MNRAS.396.2332K} {396, 2332}

\bibitem[\protect\citeauthoryear{{Kim} \& {Ostriker}}{{Kim} \&
  {Ostriker}}{2015}]{Kim2015}
{Kim} C.-G.,  {Ostriker} E.~C.,  2015, \mn@doi [\apj]
  {10.1088/0004-637X/802/2/99}, \href
  {https://ui.adsabs.harvard.edu/abs/2015ApJ...802...99K} {802, 99}

\bibitem[\protect\citeauthoryear{{Kim} \& {Ostriker}}{{Kim} \&
  {Ostriker}}{2017}]{Kim2017}
{Kim} C.-G.,  {Ostriker} E.~C.,  2017, \mn@doi [\apj]
  {10.3847/1538-4357/aa8599}, \href
  {https://ui.adsabs.harvard.edu/abs/2017ApJ...846..133K} {846, 133}

\bibitem[\protect\citeauthoryear{{Kim}, {Ostriker}  \& {Raileanu}}{{Kim}
  et~al.}{2017}]{Kim2017-wind}
{Kim} C.-G.,  {Ostriker} E.~C.,   {Raileanu} R.,  2017, \mn@doi [\apj]
  {10.3847/1538-4357/834/1/25}, \href
  {https://ui.adsabs.harvard.edu/abs/2017ApJ...834...25K} {834, 25}

\bibitem[\protect\citeauthoryear{{Kimm}, {Haehnelt}, {Blaizot}, {Katz},
  {Michel-Dansac}, {Garel}, {Rosdahl}  \& {Teyssier}}{{Kimm}
  et~al.}{2018}]{Kimm2018}
{Kimm} T.,  {Haehnelt} M.,  {Blaizot} J.,  {Katz} H.,  {Michel-Dansac} L.,
  {Garel} T.,  {Rosdahl} J.,   {Teyssier} R.,  2018, \mn@doi [\mnras]
  {10.1093/mnras/sty126}, \href
  {https://ui.adsabs.harvard.edu/abs/2018MNRAS.475.4617K} {475, 4617}

\bibitem[\protect\citeauthoryear{{Kimmig}, {Remus}, {Seidel}, {Valenzuela},
  {Dolag}  \& {Burkert}}{{Kimmig} et~al.}{2025}]{Kimmig2025-agn}
{Kimmig} L.~C.,  {Remus} R.-S.,  {Seidel} B.,  {Valenzuela} L.~M.,  {Dolag} K.,
    {Burkert} A.,  2025, \mn@doi [\apj] {10.3847/1538-4357/ad9472}, \href
  {https://ui.adsabs.harvard.edu/abs/2025ApJ...979...15K} {979, 15}

\bibitem[\protect\citeauthoryear{{Kokorev} et~al.,}{{Kokorev}
  et~al.}{2024}]{Kokorev2024-agn}
{Kokorev} V.,  et~al., 2024, \mn@doi [\apj] {10.3847/1538-4357/ad7d03}, \href
  {https://ui.adsabs.harvard.edu/abs/2024ApJ...975..178K} {975, 178}

\bibitem[\protect\citeauthoryear{{Kruijssen} et~al.,}{{Kruijssen}
  et~al.}{2019}]{Kruijssen2019}
{Kruijssen} J.~M.~D.,  et~al., 2019, \mn@doi [\nat]
  {10.1038/s41586-019-1194-3}, \href
  {https://ui.adsabs.harvard.edu/abs/2019Natur.569..519K} {569, 519}

\bibitem[\protect\citeauthoryear{{Krumholz}}{{Krumholz}}{2014}]{Krumholz2014}
{Krumholz} M.~R.,  2014, \mn@doi [\physrep] {10.1016/j.physrep.2014.02.001},
  \href {https://ui.adsabs.harvard.edu/abs/2014PhR...539...49K} {539, 49}

\bibitem[\protect\citeauthoryear{{Krumholz} \& {McKee}}{{Krumholz} \&
  {McKee}}{2005}]{Krumholz2005}
{Krumholz} M.~R.,  {McKee} C.~F.,  2005, \mn@doi [\apj] {10.1086/431734}, \href
  {https://ui.adsabs.harvard.edu/abs/2005ApJ...630..250K} {630, 250}

\bibitem[\protect\citeauthoryear{{Krumholz} \& {Tan}}{{Krumholz} \&
  {Tan}}{2007}]{Krumholz2007}
{Krumholz} M.~R.,  {Tan} J.~C.,  2007, \mn@doi [\apj] {10.1086/509101}, \href
  {https://ui.adsabs.harvard.edu/abs/2007ApJ...654..304K} {654, 304}

\bibitem[\protect\citeauthoryear{{Krumholz}, {Dekel}  \& {McKee}}{{Krumholz}
  et~al.}{2012}]{Krumholz2012}
{Krumholz} M.~R.,  {Dekel} A.,   {McKee} C.~F.,  2012, \mn@doi [\apj]
  {10.1088/0004-637X/745/1/69}, \href
  {https://ui.adsabs.harvard.edu/abs/2012ApJ...745...69K} {745, 69}

\bibitem[\protect\citeauthoryear{{Krumholz}, {McKee}  \&
  {Bland-Hawthorn}}{{Krumholz} et~al.}{2019}]{Krumholz2019}
{Krumholz} M.~R.,  {McKee} C.~F.,   {Bland-Hawthorn} J.,  2019, \mn@doi [\araa]
  {10.1146/annurev-astro-091918-104430}, \href
  {https://ui.adsabs.harvard.edu/abs/2019ARA&A..57..227K} {57, 227}

\bibitem[\protect\citeauthoryear{{Labb{\'e}} et~al.,}{{Labb{\'e}}
  et~al.}{2023}]{Labbe2023}
{Labb{\'e}} I.,  et~al., 2023, \mn@doi [\nat] {10.1038/s41586-023-05786-2},
  \href {https://ui.adsabs.harvard.edu/abs/2023Natur.616..266L} {616, 266}

\bibitem[\protect\citeauthoryear{{Leethochawalit} et~al.,}{{Leethochawalit}
  et~al.}{2023}]{Leetho2023}
{Leethochawalit} N.,  et~al., 2023, \mn@doi [\apjl] {10.3847/2041-8213/ac959b},
  \href {https://ui.adsabs.harvard.edu/abs/2023ApJ...942L..26L} {942, L26}

\bibitem[\protect\citeauthoryear{{Leitherer} et~al.,}{{Leitherer}
  et~al.}{1999}]{Leitherer1999}
{Leitherer} C.,  et~al., 1999, \mn@doi [\apjs] {10.1086/313233}, \href
  {https://ui.adsabs.harvard.edu/abs/1999ApJS..123....3L} {123, 3}

\bibitem[\protect\citeauthoryear{{Levermore}}{{Levermore}}{1984}]{Levermore1984}
{Levermore} C.~D.,  1984, \mn@doi [\jqsrt] {10.1016/0022-4073(84)90112-2},
  \href {https://ui.adsabs.harvard.edu/abs/1984JQSRT..31..149L} {31, 149}

\bibitem[\protect\citeauthoryear{{Li}, {Dekel}, {Sarkar}, {Aung}, {Giavalisco},
  {Mandelker}  \& {Tacchella}}{{Li} et~al.}{2024}]{Li2024}
{Li} Z.,  {Dekel} A.,  {Sarkar} K.~C.,  {Aung} H.,  {Giavalisco} M.,
  {Mandelker} N.,   {Tacchella} S.,  2024, \mn@doi [\aap]
  {10.1051/0004-6361/202348727}, \href
  {https://ui.adsabs.harvard.edu/abs/2024A&A...690A.108L} {690, A108}

\bibitem[\protect\citeauthoryear{{Liu}, {Gao}  \& {Greve}}{{Liu}
  et~al.}{2015}]{Liu2015}
{Liu} L.,  {Gao} Y.,   {Greve} T.~R.,  2015, \mn@doi [\apj]
  {10.1088/0004-637X/805/1/31}, \href
  {https://ui.adsabs.harvard.edu/abs/2015ApJ...805...31L} {805, 31}

\bibitem[\protect\citeauthoryear{{Ma} et~al.,}{{Ma} et~al.}{2018}]{Ma2018}
{Ma} X.,  et~al., 2018, \mn@doi [\mnras] {10.1093/mnras/sty1024}, \href
  {https://ui.adsabs.harvard.edu/abs/2018MNRAS.478.1694M} {478, 1694}

\bibitem[\protect\citeauthoryear{{Madau}, {Weisz}  \& {Conroy}}{{Madau}
  et~al.}{2014}]{Madau2014}
{Madau} P.,  {Weisz} D.~R.,   {Conroy} C.,  2014, \mn@doi [\apjl]
  {10.1088/2041-8205/790/2/L17}, \href
  {https://ui.adsabs.harvard.edu/abs/2014ApJ...790L..17M} {790, L17}

\bibitem[\protect\citeauthoryear{{Maiolino} et~al.,}{{Maiolino}
  et~al.}{2024}]{Maiolino2024-agn}
{Maiolino} R.,  et~al., 2024, \mn@doi [arXiv e-prints]
  {10.48550/arXiv.2405.00504}, \href
  {https://ui.adsabs.harvard.edu/abs/2024arXiv240500504M} {p. arXiv:2405.00504}

\bibitem[\protect\citeauthoryear{{Marinacci} \& {Vogelsberger}}{{Marinacci} \&
  {Vogelsberger}}{2016}]{Marinacci2016}
{Marinacci} F.,  {Vogelsberger} M.,  2016, \mn@doi [\mnras]
  {10.1093/mnrasl/slv176}, \href
  {https://ui.adsabs.harvard.edu/abs/2016MNRAS.456L..69M} {456, L69}

\bibitem[\protect\citeauthoryear{{Marinacci}, {Sales}, {Vogelsberger}, {Torrey}
   \& {Springel}}{{Marinacci} et~al.}{2019}]{Marinacci2019}
{Marinacci} F.,  {Sales} L.~V.,  {Vogelsberger} M.,  {Torrey} P.,   {Springel}
  V.,  2019, \mn@doi [\mnras] {10.1093/mnras/stz2391}, \href
  {https://ui.adsabs.harvard.edu/abs/2019MNRAS.489.4233M} {489, 4233}

\bibitem[\protect\citeauthoryear{{Martin}}{{Martin}}{1999}]{Martin1999}
{Martin} C.~L.,  1999, \mn@doi [\apj] {10.1086/306863}, \href
  {https://ui.adsabs.harvard.edu/abs/1999ApJ...513..156M} {513, 156}

\bibitem[\protect\citeauthoryear{{Martin}, {Scannapieco}, {Ellison}, {Hennawi},
  {Djorgovski}  \& {Fournier}}{{Martin} et~al.}{2010}]{Martin2010}
{Martin} C.~L.,  {Scannapieco} E.,  {Ellison} S.~L.,  {Hennawi} J.~F.,
  {Djorgovski} S.~G.,   {Fournier} A.~P.,  2010, \mn@doi [\apj]
  {10.1088/0004-637X/721/1/174}, \href
  {https://ui.adsabs.harvard.edu/abs/2010ApJ...721..174M} {721, 174}

\bibitem[\protect\citeauthoryear{{Martizzi}, {Faucher-Gigu{\`e}re}  \&
  {Quataert}}{{Martizzi} et~al.}{2015}]{Martizzi2015}
{Martizzi} D.,  {Faucher-Gigu{\`e}re} C.-A.,   {Quataert} E.,  2015, \mn@doi
  [\mnras] {10.1093/mnras/stv562}, \href
  {https://ui.adsabs.harvard.edu/abs/2015MNRAS.450..504M} {450, 504}

\bibitem[\protect\citeauthoryear{{Mason}, {Trenti}  \& {Treu}}{{Mason}
  et~al.}{2015}]{Mason2015}
{Mason} C.~A.,  {Trenti} M.,   {Treu} T.,  2015, \mn@doi [\apj]
  {10.1088/0004-637X/813/1/21}, \href
  {https://ui.adsabs.harvard.edu/abs/2015ApJ...813...21M} {813, 21}

\bibitem[\protect\citeauthoryear{{Mason}, {Trenti}  \& {Treu}}{{Mason}
  et~al.}{2023}]{Mason2023}
{Mason} C.~A.,  {Trenti} M.,   {Treu} T.,  2023, \mn@doi [\mnras]
  {10.1093/mnras/stad035}, \href
  {https://ui.adsabs.harvard.edu/abs/2023MNRAS.521..497M} {521, 497}

\bibitem[\protect\citeauthoryear{{McKee} \& {Ostriker}}{{McKee} \&
  {Ostriker}}{2007}]{McKee2007}
{McKee} C.~F.,  {Ostriker} E.~C.,  2007, \mn@doi [\araa]
  {10.1146/annurev.astro.45.051806.110602}, \href
  {https://ui.adsabs.harvard.edu/abs/2007ARA&A..45..565M} {45, 565}

\bibitem[\protect\citeauthoryear{{McKinnon}, {Torrey}  \&
  {Vogelsberger}}{{McKinnon} et~al.}{2016}]{McKinnon2016}
{McKinnon} R.,  {Torrey} P.,   {Vogelsberger} M.,  2016, \mn@doi [\mnras]
  {10.1093/mnras/stw253}, \href
  {https://ui.adsabs.harvard.edu/abs/2016MNRAS.457.3775M} {457, 3775}

\bibitem[\protect\citeauthoryear{{McKinnon}, {Torrey}, {Vogelsberger},
  {Hayward}  \& {Marinacci}}{{McKinnon} et~al.}{2017}]{McKinnon2017}
{McKinnon} R.,  {Torrey} P.,  {Vogelsberger} M.,  {Hayward} C.~C.,
  {Marinacci} F.,  2017, \mn@doi [\mnras] {10.1093/mnras/stx467}, \href
  {https://ui.adsabs.harvard.edu/abs/2017MNRAS.468.1505M} {468, 1505}

\bibitem[\protect\citeauthoryear{{McLeod}, {McLure}  \& {Dunlop}}{{McLeod}
  et~al.}{2016}]{McLeod2016}
{McLeod} D.~J.,  {McLure} R.~J.,   {Dunlop} J.~S.,  2016, \mn@doi [\mnras]
  {10.1093/mnras/stw904}, \href
  {https://ui.adsabs.harvard.edu/abs/2016MNRAS.459.3812M} {459, 3812}

\bibitem[\protect\citeauthoryear{{McLeod} et~al.,}{{McLeod}
  et~al.}{2024}]{McLeod2024}
{McLeod} D.~J.,  et~al., 2024, \mn@doi [\mnras] {10.1093/mnras/stad3471}, \href
  {https://ui.adsabs.harvard.edu/abs/2024MNRAS.527.5004M} {527, 5004}

\bibitem[\protect\citeauthoryear{{Menon}, {Lancaster}, {Burkhart},
  {Somerville}, {Dekel}  \& {Krumholz}}{{Menon} et~al.}{2024}]{Menon2024}
{Menon} S.~H.,  {Lancaster} L.,  {Burkhart} B.,  {Somerville} R.~S.,  {Dekel}
  A.,   {Krumholz} M.~R.,  2024, \mn@doi [\apjl] {10.3847/2041-8213/ad462d},
  \href {https://ui.adsabs.harvard.edu/abs/2024ApJ...967L..28M} {967, L28}

\bibitem[\protect\citeauthoryear{{Mirocha} \& {Furlanetto}}{{Mirocha} \&
  {Furlanetto}}{2023}]{Mirocha2023}
{Mirocha} J.,  {Furlanetto} S.~R.,  2023, \mn@doi [\mnras]
  {10.1093/mnras/stac3578}, \href
  {https://ui.adsabs.harvard.edu/abs/2023MNRAS.519..843M} {519, 843}

\bibitem[\protect\citeauthoryear{{Morishita} \& {Stiavelli}}{{Morishita} \&
  {Stiavelli}}{2023}]{Morishita2023}
{Morishita} T.,  {Stiavelli} M.,  2023, \mn@doi [\apjl]
  {10.3847/2041-8213/acbf50}, \href
  {https://ui.adsabs.harvard.edu/abs/2023ApJ...946L..35M} {946, L35}

\bibitem[\protect\citeauthoryear{{Morishita} et~al.,}{{Morishita}
  et~al.}{2018}]{Morishita2018}
{Morishita} T.,  et~al., 2018, \mn@doi [\apj] {10.3847/1538-4357/aae68c}, \href
  {https://ui.adsabs.harvard.edu/abs/2018ApJ...867..150M} {867, 150}

\bibitem[\protect\citeauthoryear{{Moster}, {Somerville}, {Maulbetsch}, {van den
  Bosch}, {Macci{\`o}}, {Naab}  \& {Oser}}{{Moster} et~al.}{2010}]{Moster2010}
{Moster} B.~P.,  {Somerville} R.~S.,  {Maulbetsch} C.,  {van den Bosch} F.~C.,
  {Macci{\`o}} A.~V.,  {Naab} T.,   {Oser} L.,  2010, \mn@doi [\apj]
  {10.1088/0004-637X/710/2/903}, \href
  {https://ui.adsabs.harvard.edu/abs/2010ApJ...710..903M} {710, 903}

\bibitem[\protect\citeauthoryear{{Muratov}, {Kere{\v{s}}},
  {Faucher-Gigu{\`e}re}, {Hopkins}, {Quataert}  \& {Murray}}{{Muratov}
  et~al.}{2015}]{Muratov2015}
{Muratov} A.~L.,  {Kere{\v{s}}} D.,  {Faucher-Gigu{\`e}re} C.-A.,  {Hopkins}
  P.~F.,  {Quataert} E.,   {Murray} N.,  2015, \mn@doi [\mnras]
  {10.1093/mnras/stv2126}, \href
  {https://ui.adsabs.harvard.edu/abs/2015MNRAS.454.2691M} {454, 2691}

\bibitem[\protect\citeauthoryear{{Murray}, {Quataert}  \& {Thompson}}{{Murray}
  et~al.}{2005}]{Murray2005}
{Murray} N.,  {Quataert} E.,   {Thompson} T.~A.,  2005, \mn@doi [\apj]
  {10.1086/426067}, \href
  {https://ui.adsabs.harvard.edu/abs/2005ApJ...618..569M} {618, 569}

\bibitem[\protect\citeauthoryear{{Naidu} et~al.,}{{Naidu}
  et~al.}{2022}]{Naidu2022}
{Naidu} R.~P.,  et~al., 2022, \mn@doi [arXiv e-prints]
  {10.48550/arXiv.2208.02794}, \href
  {https://ui.adsabs.harvard.edu/abs/2022arXiv220802794N} {p. arXiv:2208.02794}

\bibitem[\protect\citeauthoryear{{Nebrin}, {Smith}, {Lorinc}, {H{\"o}rnquist},
  {Larson}, {Mellema}  \& {Giri}}{{Nebrin} et~al.}{2025}]{Nebrin2025}
{Nebrin} O.,  {Smith} A.,  {Lorinc} K.,  {H{\"o}rnquist} J.,  {Larson}
  {\r{A}}.,  {Mellema} G.,   {Giri} S.~K.,  2025, \mn@doi [\mnras]
  {10.1093/mnras/staf038}, \href
  {https://ui.adsabs.harvard.edu/abs/2025MNRAS.tmp...39N} {}

\bibitem[\protect\citeauthoryear{{Newman} et~al.,}{{Newman}
  et~al.}{2012}]{Newman2012}
{Newman} S.~F.,  et~al., 2012, \mn@doi [\apj] {10.1088/0004-637X/761/1/43},
  \href {https://ui.adsabs.harvard.edu/abs/2012ApJ...761...43N} {761, 43}

\bibitem[\protect\citeauthoryear{{Oesch}, {Bouwens}, {Illingworth}, {Labb{\'e}}
   \& {Stefanon}}{{Oesch} et~al.}{2018}]{Oesch2018}
{Oesch} P.~A.,  {Bouwens} R.~J.,  {Illingworth} G.~D.,  {Labb{\'e}} I.,
  {Stefanon} M.,  2018, \mn@doi [\apj] {10.3847/1538-4357/aab03f}, \href
  {https://ui.adsabs.harvard.edu/abs/2018ApJ...855..105O} {855, 105}

\bibitem[\protect\citeauthoryear{{Offner}, {Klein}, {McKee}  \&
  {Krumholz}}{{Offner} et~al.}{2009}]{Offner2009}
{Offner} S. S.~R.,  {Klein} R.~I.,  {McKee} C.~F.,   {Krumholz} M.~R.,  2009,
  \mn@doi [\apj] {10.1088/0004-637X/703/1/131}, \href
  {https://ui.adsabs.harvard.edu/abs/2009ApJ...703..131O} {703, 131}

\bibitem[\protect\citeauthoryear{{Okamoto}, {Gao}  \& {Theuns}}{{Okamoto}
  et~al.}{2008}]{Okamoto2008}
{Okamoto} T.,  {Gao} L.,   {Theuns} T.,  2008, \mn@doi [\mnras]
  {10.1111/j.1365-2966.2008.13830.x}, \href
  {https://ui.adsabs.harvard.edu/abs/2008MNRAS.390..920O} {390, 920}

\bibitem[\protect\citeauthoryear{{Oppenheimer} \& {Dav{\'e}}}{{Oppenheimer} \&
  {Dav{\'e}}}{2006}]{Oppenheimer2006}
{Oppenheimer} B.~D.,  {Dav{\'e}} R.,  2006, \mn@doi [\mnras]
  {10.1111/j.1365-2966.2006.10989.x}, \href
  {https://ui.adsabs.harvard.edu/abs/2006MNRAS.373.1265O} {373, 1265}

\bibitem[\protect\citeauthoryear{{Orr} et~al.,}{{Orr} et~al.}{2018}]{Orr2018}
{Orr} M.~E.,  et~al., 2018, \mn@doi [\mnras] {10.1093/mnras/sty1241}, \href
  {https://ui.adsabs.harvard.edu/abs/2018MNRAS.478.3653O} {478, 3653}

\bibitem[\protect\citeauthoryear{{Ostriker} \& {Kim}}{{Ostriker} \&
  {Kim}}{2022}]{Ostriker2022}
{Ostriker} E.~C.,  {Kim} C.-G.,  2022, \mn@doi [\apj]
  {10.3847/1538-4357/ac7de2}, \href
  {https://ui.adsabs.harvard.edu/abs/2022ApJ...936..137O} {936, 137}

\bibitem[\protect\citeauthoryear{{Ostriker} \& {Shetty}}{{Ostriker} \&
  {Shetty}}{2011}]{OstrikerShetty2011}
{Ostriker} E.~C.,  {Shetty} R.,  2011, \mn@doi [\apj]
  {10.1088/0004-637X/731/1/41}, \href
  {https://ui.adsabs.harvard.edu/abs/2011ApJ...731...41O} {731, 41}

\bibitem[\protect\citeauthoryear{{Padoan} \& {Nordlund}}{{Padoan} \&
  {Nordlund}}{2011}]{Padoan2011}
{Padoan} P.,  {Nordlund} {\r{A}}.,  2011, \mn@doi [\apj]
  {10.1088/0004-637X/730/1/40}, \href
  {https://ui.adsabs.harvard.edu/abs/2011ApJ...730...40P} {730, 40}

\bibitem[\protect\citeauthoryear{{Pakmor}, {Springel}, {Bauer}, {Mocz},
  {Munoz}, {Ohlmann}, {Schaal}  \& {Zhu}}{{Pakmor} et~al.}{2016a}]{Pakmor2016}
{Pakmor} R.,  {Springel} V.,  {Bauer} A.,  {Mocz} P.,  {Munoz} D.~J.,
  {Ohlmann} S.~T.,  {Schaal} K.,   {Zhu} C.,  2016a, \mn@doi [\mnras]
  {10.1093/mnras/stv2380}, \href
  {https://ui.adsabs.harvard.edu/abs/2016MNRAS.455.1134P} {455, 1134}

\bibitem[\protect\citeauthoryear{{Pakmor}, {Pfrommer}, {Simpson}  \&
  {Springel}}{{Pakmor} et~al.}{2016b}]{Pakmor2016b}
{Pakmor} R.,  {Pfrommer} C.,  {Simpson} C.~M.,   {Springel} V.,  2016b, \mn@doi
  [\apjl] {10.3847/2041-8205/824/2/L30}, \href
  {https://ui.adsabs.harvard.edu/abs/2016ApJ...824L..30P} {824, L30}

\bibitem[\protect\citeauthoryear{{Pallottini} \& {Ferrara}}{{Pallottini} \&
  {Ferrara}}{2023}]{Pallottini2023}
{Pallottini} A.,  {Ferrara} A.,  2023, \mn@doi [\aap]
  {10.1051/0004-6361/202347384}, \href
  {https://ui.adsabs.harvard.edu/abs/2023A&A...677L...4P} {677, L4}

\bibitem[\protect\citeauthoryear{{Pallottini} et~al.,}{{Pallottini}
  et~al.}{2022}]{Pallottini2022}
{Pallottini} A.,  et~al., 2022, \mn@doi [\mnras] {10.1093/mnras/stac1281},
  \href {https://ui.adsabs.harvard.edu/abs/2022MNRAS.513.5621P} {513, 5621}

\bibitem[\protect\citeauthoryear{{Pawlik}, {Rahmati}, {Schaye}, {Jeon}  \&
  {Dalla Vecchia}}{{Pawlik} et~al.}{2017}]{Pawlik2017}
{Pawlik} A.~H.,  {Rahmati} A.,  {Schaye} J.,  {Jeon} M.,   {Dalla Vecchia} C.,
  2017, \mn@doi [\mnras] {10.1093/mnras/stw2869}, \href
  {https://ui.adsabs.harvard.edu/abs/2017MNRAS.466..960P} {466, 960}

\bibitem[\protect\citeauthoryear{{P{\'e}rez-Gonz{\'a}lez}
  et~al.,}{{P{\'e}rez-Gonz{\'a}lez} et~al.}{2023}]{Perez2023}
{P{\'e}rez-Gonz{\'a}lez} P.~G.,  et~al., 2023, \mn@doi [\apjl]
  {10.3847/2041-8213/acd9d0}, \href
  {https://ui.adsabs.harvard.edu/abs/2023ApJ...951L...1P} {951, L1}

\bibitem[\protect\citeauthoryear{{Planck Collaboration} et~al.,}{{Planck
  Collaboration} et~al.}{2016}]{Planck2016}
{Planck Collaboration} et~al., 2016, \mn@doi [\aap]
  {10.1051/0004-6361/201525830}, \href
  {https://ui.adsabs.harvard.edu/abs/2016A&A...594A..13P} {594, A13}

\bibitem[\protect\citeauthoryear{{Puchwein}, {Haardt}, {Haehnelt}  \&
  {Madau}}{{Puchwein} et~al.}{2019}]{Puchwein2019}
{Puchwein} E.,  {Haardt} F.,  {Haehnelt} M.~G.,   {Madau} P.,  2019, \mn@doi
  [\mnras] {10.1093/mnras/stz222}, \href
  {https://ui.adsabs.harvard.edu/abs/2019MNRAS.485...47P} {485, 47}

\bibitem[\protect\citeauthoryear{{Puchwein} et~al.,}{{Puchwein}
  et~al.}{2023}]{Puchwein2023}
{Puchwein} E.,  et~al., 2023, \mn@doi [\mnras] {10.1093/mnras/stac3761}, \href
  {https://ui.adsabs.harvard.edu/abs/2023MNRAS.519.6162P} {519, 6162}

\bibitem[\protect\citeauthoryear{{Pusk{\'a}s} et~al.,}{{Pusk{\'a}s}
  et~al.}{2025}]{Puskas2025}
{Pusk{\'a}s} D.,  et~al., 2025, arXiv e-prints, \href
  {https://ui.adsabs.harvard.edu/abs/2025arXiv250201721P} {p. arXiv:2502.01721}

\bibitem[\protect\citeauthoryear{{Rahmati}, {Pawlik}, {Rai{\v{c}}evi{\'c}}  \&
  {Schaye}}{{Rahmati} et~al.}{2013}]{Rahmati2013}
{Rahmati} A.,  {Pawlik} A.~H.,  {Rai{\v{c}}evi{\'c}} M.,   {Schaye} J.,  2013,
  \mn@doi [\mnras] {10.1093/mnras/stt066}, \href
  {https://ui.adsabs.harvard.edu/abs/2013MNRAS.430.2427R} {430, 2427}

\bibitem[\protect\citeauthoryear{{Rees}}{{Rees}}{1986}]{Rees1986}
{Rees} M.~J.,  1986, \mn@doi [\mnras] {10.1093/mnras/218.1.25P}, \href
  {https://ui.adsabs.harvard.edu/abs/1986MNRAS.218P..25R} {218, 25P}

\bibitem[\protect\citeauthoryear{{Rizzo}, {Vegetti}, {Powell}, {Fraternali},
  {McKean}, {Stacey}  \& {White}}{{Rizzo} et~al.}{2020}]{Rizzo2020}
{Rizzo} F.,  {Vegetti} S.,  {Powell} D.,  {Fraternali} F.,  {McKean} J.~P.,
  {Stacey} H.~R.,   {White} S.~D.~M.,  2020, \mn@doi [\nat]
  {10.1038/s41586-020-2572-6}, \href
  {https://ui.adsabs.harvard.edu/abs/2020Natur.584..201R} {584, 201}

\bibitem[\protect\citeauthoryear{{Robertson} et~al.,}{{Robertson}
  et~al.}{2023a}]{Robertson2024}
{Robertson} B.,  et~al., 2023a, \mn@doi [arXiv e-prints]
  {10.48550/arXiv.2312.10033}, \href
  {https://ui.adsabs.harvard.edu/abs/2023arXiv231210033R} {p. arXiv:2312.10033}

\bibitem[\protect\citeauthoryear{{Robertson} et~al.,}{{Robertson}
  et~al.}{2023b}]{Robertson2023}
{Robertson} B.~E.,  et~al., 2023b, \mn@doi [\apjl] {10.3847/2041-8213/aca086},
  \href {https://ui.adsabs.harvard.edu/abs/2023ApJ...942L..42R} {942, L42}

\bibitem[\protect\citeauthoryear{{Rodr{\'\i}guez-Puebla}, {Behroozi},
  {Primack}, {Klypin}, {Lee}  \& {Hellinger}}{{Rodr{\'\i}guez-Puebla}
  et~al.}{2016}]{RP2016}
{Rodr{\'\i}guez-Puebla} A.,  {Behroozi} P.,  {Primack} J.,  {Klypin} A.,  {Lee}
  C.,   {Hellinger} D.,  2016, \mn@doi [\mnras] {10.1093/mnras/stw1705}, \href
  {https://ui.adsabs.harvard.edu/abs/2016MNRAS.462..893R} {462, 893}

\bibitem[\protect\citeauthoryear{{Rodr{\'\i}guez-Puebla}, {Primack},
  {Avila-Reese}  \& {Faber}}{{Rodr{\'\i}guez-Puebla} et~al.}{2017}]{RP2017}
{Rodr{\'\i}guez-Puebla} A.,  {Primack} J.~R.,  {Avila-Reese} V.,   {Faber}
  S.~M.,  2017, \mn@doi [\mnras] {10.1093/mnras/stx1172}, \href
  {https://ui.adsabs.harvard.edu/abs/2017MNRAS.470..651R} {470, 651}

\bibitem[\protect\citeauthoryear{{Roman-Oliveira}, {Fraternali}  \&
  {Rizzo}}{{Roman-Oliveira} et~al.}{2023}]{RO2023}
{Roman-Oliveira} F.,  {Fraternali} F.,   {Rizzo} F.,  2023, \mn@doi [\mnras]
  {10.1093/mnras/stad530}, \href
  {https://ui.adsabs.harvard.edu/abs/2023MNRAS.521.1045R} {521, 1045}

\bibitem[\protect\citeauthoryear{{Rosdahl} et~al.,}{{Rosdahl}
  et~al.}{2018}]{Rosdahl2018}
{Rosdahl} J.,  et~al., 2018, \mn@doi [\mnras] {10.1093/mnras/sty1655}, \href
  {https://ui.adsabs.harvard.edu/abs/2018MNRAS.479..994R} {479, 994}

\bibitem[\protect\citeauthoryear{{Rowland} et~al.,}{{Rowland}
  et~al.}{2024}]{Rowland2024}
{Rowland} L.~E.,  et~al., 2024, \mn@doi [\mnras] {10.1093/mnras/stae2217},
  \href {https://ui.adsabs.harvard.edu/abs/2024MNRAS.535.2068R} {535, 2068}

\bibitem[\protect\citeauthoryear{{Schmidt}}{{Schmidt}}{1959}]{Schmidt1959}
{Schmidt} M.,  1959, \mn@doi [\apj] {10.1086/146614}, \href
  {https://ui.adsabs.harvard.edu/abs/1959ApJ...129..243S} {129, 243}

\bibitem[\protect\citeauthoryear{{Semenov}, {Kravtsov}  \& {Gnedin}}{{Semenov}
  et~al.}{2016}]{Semenov2016}
{Semenov} V.~A.,  {Kravtsov} A.~V.,   {Gnedin} N.~Y.,  2016, \mn@doi [\apj]
  {10.3847/0004-637X/826/2/200}, \href
  {https://ui.adsabs.harvard.edu/abs/2016ApJ...826..200S} {826, 200}

\bibitem[\protect\citeauthoryear{{Semenov}, {Kravtsov}  \& {Gnedin}}{{Semenov}
  et~al.}{2017}]{Semenov2017}
{Semenov} V.~A.,  {Kravtsov} A.~V.,   {Gnedin} N.~Y.,  2017, \mn@doi [\apj]
  {10.3847/1538-4357/aa8096}, \href
  {https://ui.adsabs.harvard.edu/abs/2017ApJ...845..133S} {845, 133}

\bibitem[\protect\citeauthoryear{{Semenov}, {Kravtsov}  \& {Gnedin}}{{Semenov}
  et~al.}{2018}]{Semenov2018}
{Semenov} V.~A.,  {Kravtsov} A.~V.,   {Gnedin} N.~Y.,  2018, \mn@doi [\apj]
  {10.3847/1538-4357/aac6eb}, \href
  {https://ui.adsabs.harvard.edu/abs/2018ApJ...861....4S} {861, 4}

\bibitem[\protect\citeauthoryear{{Semenov}, {Conroy}, {Smith}, {Puchwein}  \&
  {Hernquist}}{{Semenov} et~al.}{2024a}]{Semenov2024b}
{Semenov} V.~A.,  {Conroy} C.,  {Smith} A.,  {Puchwein} E.,   {Hernquist} L.,
  2024a, \mn@doi [arXiv e-prints] {10.48550/arXiv.2409.18173}, \href
  {https://ui.adsabs.harvard.edu/abs/2024arXiv240918173S} {p. arXiv:2409.18173}

\bibitem[\protect\citeauthoryear{{Semenov}, {Conroy}  \& {Hernquist}}{{Semenov}
  et~al.}{2024b}]{Semenov2024a}
{Semenov} V.~A.,  {Conroy} C.,   {Hernquist} L.,  2024b, \mn@doi [arXiv
  e-prints] {10.48550/arXiv.2410.09205}, \href
  {https://ui.adsabs.harvard.edu/abs/2024arXiv241009205S} {p. arXiv:2410.09205}

\bibitem[\protect\citeauthoryear{{Shapiro}, {Iliev}  \& {Raga}}{{Shapiro}
  et~al.}{2004}]{Shapiro2004}
{Shapiro} P.~R.,  {Iliev} I.~T.,   {Raga} A.~C.,  2004, \mn@doi [\mnras]
  {10.1111/j.1365-2966.2004.07364.x}, \href
  {https://ui.adsabs.harvard.edu/abs/2004MNRAS.348..753S} {348, 753}

\bibitem[\protect\citeauthoryear{{Shen}, {Vogelsberger}, {Boylan-Kolchin},
  {Tacchella}  \& {Kannan}}{{Shen} et~al.}{2023}]{Shen2023}
{Shen} X.,  {Vogelsberger} M.,  {Boylan-Kolchin} M.,  {Tacchella} S.,
  {Kannan} R.,  2023, \mn@doi [\mnras] {10.1093/mnras/stad2508}, \href
  {https://ui.adsabs.harvard.edu/abs/2023MNRAS.525.3254S} {525, 3254}

\bibitem[\protect\citeauthoryear{{Shen}, {Vogelsberger}, {Boylan-Kolchin},
  {Tacchella}  \& {Naidu}}{{Shen} et~al.}{2024a}]{Shen2024b-ede}
{Shen} X.,  {Vogelsberger} M.,  {Boylan-Kolchin} M.,  {Tacchella} S.,   {Naidu}
  R.~P.,  2024a, \mn@doi [\mnras] {10.1093/mnras/stae1932}, \href
  {https://ui.adsabs.harvard.edu/abs/2024MNRAS.533.3923S} {533, 3923}

\bibitem[\protect\citeauthoryear{{Shen}, {Vogelsberger}, {Boylan-Kolchin},
  {Tacchella}  \& {Naidu}}{{Shen} et~al.}{2024b}]{Shen2024}
{Shen} X.,  {Vogelsberger} M.,  {Boylan-Kolchin} M.,  {Tacchella} S.,   {Naidu}
  R.~P.,  2024b, \mn@doi [\mnras] {10.1093/mnras/stae1932}, \href
  {https://ui.adsabs.harvard.edu/abs/2024MNRAS.533.3923S} {533, 3923}

\bibitem[\protect\citeauthoryear{{Silk}, {Begelman}, {Norman}, {Nusser}  \&
  {Wyse}}{{Silk} et~al.}{2024}]{Silk2024-agn}
{Silk} J.,  {Begelman} M.~C.,  {Norman} C.,  {Nusser} A.,   {Wyse} R. F.~G.,
  2024, \mn@doi [\apjl] {10.3847/2041-8213/ad1bf0}, \href
  {https://ui.adsabs.harvard.edu/abs/2024ApJ...961L..39S} {961, L39}

\bibitem[\protect\citeauthoryear{{Smith}, {Bromm}  \& {Loeb}}{{Smith}
  et~al.}{2017}]{Smith2017}
{Smith} A.,  {Bromm} V.,   {Loeb} A.,  2017, \mn@doi [\mnras]
  {10.1093/mnras/stw2591}, \href
  {https://ui.adsabs.harvard.edu/abs/2017MNRAS.464.2963S} {464, 2963}

\bibitem[\protect\citeauthoryear{{Smith}, {Kannan}, {Garaldi}, {Vogelsberger},
  {Pakmor}, {Springel}  \& {Hernquist}}{{Smith} et~al.}{2022}]{Smith2022}
{Smith} A.,  {Kannan} R.,  {Garaldi} E.,  {Vogelsberger} M.,  {Pakmor} R.,
  {Springel} V.,   {Hernquist} L.,  2022, \mn@doi [\mnras]
  {10.1093/mnras/stac713}, \href
  {https://ui.adsabs.harvard.edu/abs/2022MNRAS.512.3243S} {512, 3243}

\bibitem[\protect\citeauthoryear{{Somerville} \& {Primack}}{{Somerville} \&
  {Primack}}{1999}]{Somerville1999}
{Somerville} R.~S.,  {Primack} J.~R.,  1999, \mn@doi [\mnras]
  {10.1046/j.1365-8711.1999.03032.x}, \href
  {https://ui.adsabs.harvard.edu/abs/1999MNRAS.310.1087S} {310, 1087}

\bibitem[\protect\citeauthoryear{{Springel}}{{Springel}}{2010}]{Springel2010}
{Springel} V.,  2010, \mn@doi [\mnras] {10.1111/j.1365-2966.2009.15715.x},
  \href {https://ui.adsabs.harvard.edu/abs/2010MNRAS.401..791S} {401, 791}

\bibitem[\protect\citeauthoryear{{Springel} \& {Hernquist}}{{Springel} \&
  {Hernquist}}{2003a}]{Springel2003a}
{Springel} V.,  {Hernquist} L.,  2003a, \mn@doi [\mnras]
  {10.1046/j.1365-8711.2003.06206.x}, \href
  {https://ui.adsabs.harvard.edu/abs/2003MNRAS.339..289S} {339, 289}

\bibitem[\protect\citeauthoryear{{Springel} \& {Hernquist}}{{Springel} \&
  {Hernquist}}{2003b}]{Springel2003b}
{Springel} V.,  {Hernquist} L.,  2003b, \mn@doi [\mnras]
  {10.1046/j.1365-8711.2003.06207.x}, \href
  {https://ui.adsabs.harvard.edu/abs/2003MNRAS.339..312S} {339, 312}

\bibitem[\protect\citeauthoryear{{Springel}, {Pakmor}, {Zier}  \&
  {Reinecke}}{{Springel} et~al.}{2021}]{Springel2021}
{Springel} V.,  {Pakmor} R.,  {Zier} O.,   {Reinecke} M.,  2021, \mn@doi
  [\mnras] {10.1093/mnras/stab1855}, \href
  {https://ui.adsabs.harvard.edu/abs/2021MNRAS.506.2871S} {506, 2871}

\bibitem[\protect\citeauthoryear{{Stefanon} et~al.,}{{Stefanon}
  et~al.}{2019}]{Stefanon2019}
{Stefanon} M.,  et~al., 2019, \mn@doi [\apj] {10.3847/1538-4357/ab3792}, \href
  {https://ui.adsabs.harvard.edu/abs/2019ApJ...883...99S} {883, 99}

\bibitem[\protect\citeauthoryear{{Steidel}, {Erb}, {Shapley}, {Pettini},
  {Reddy}, {Bogosavljevi{\'c}}, {Rudie}  \& {Rakic}}{{Steidel}
  et~al.}{2010}]{Steidel2010}
{Steidel} C.~C.,  {Erb} D.~K.,  {Shapley} A.~E.,  {Pettini} M.,  {Reddy} N.,
  {Bogosavljevi{\'c}} M.,  {Rudie} G.~C.,   {Rakic} O.,  2010, \mn@doi [\apj]
  {10.1088/0004-637X/717/1/289}, \href
  {https://ui.adsabs.harvard.edu/abs/2010ApJ...717..289S} {717, 289}

\bibitem[\protect\citeauthoryear{{Stone}, {Ostriker}  \& {Gammie}}{{Stone}
  et~al.}{1998}]{Stone1998}
{Stone} J.~M.,  {Ostriker} E.~C.,   {Gammie} C.~F.,  1998, \mn@doi [\apjl]
  {10.1086/311718}, \href
  {https://ui.adsabs.harvard.edu/abs/1998ApJ...508L..99S} {508, L99}

\bibitem[\protect\citeauthoryear{{Sun}, {Faucher-Gigu{\`e}re}, {Hayward}  \&
  {Shen}}{{Sun} et~al.}{2023a}]{Sun2023a}
{Sun} G.,  {Faucher-Gigu{\`e}re} C.-A.,  {Hayward} C.~C.,   {Shen} X.,  2023a,
  \mn@doi [\mnras] {10.1093/mnras/stad2902}, \href
  {https://ui.adsabs.harvard.edu/abs/2023MNRAS.526.2665S} {526, 2665}

\bibitem[\protect\citeauthoryear{{Sun} et~al.,}{{Sun}
  et~al.}{2023b}]{SunJY2023}
{Sun} J.,  et~al., 2023b, \mn@doi [\apjl] {10.3847/2041-8213/acbd9c}, \href
  {https://ui.adsabs.harvard.edu/abs/2023ApJ...945L..19S} {945, L19}

\bibitem[\protect\citeauthoryear{{Sun}, {Faucher-Gigu{\`e}re}, {Hayward},
  {Shen}, {Wetzel}  \& {Cochrane}}{{Sun} et~al.}{2023c}]{Sun2023b}
{Sun} G.,  {Faucher-Gigu{\`e}re} C.-A.,  {Hayward} C.~C.,  {Shen} X.,  {Wetzel}
  A.,   {Cochrane} R.~K.,  2023c, \mn@doi [\apjl] {10.3847/2041-8213/acf85a},
  \href {https://ui.adsabs.harvard.edu/abs/2023ApJ...955L..35S} {955, L35}

\bibitem[\protect\citeauthoryear{{Tacchella}, {Dekel}, {Carollo}, {Ceverino},
  {DeGraf}, {Lapiner}, {Mandelker}  \& {Primack Joel}}{{Tacchella}
  et~al.}{2016}]{Tacchella2016}
{Tacchella} S.,  {Dekel} A.,  {Carollo} C.~M.,  {Ceverino} D.,  {DeGraf} C.,
  {Lapiner} S.,  {Mandelker} N.,   {Primack Joel} R.,  2016, \mn@doi [\mnras]
  {10.1093/mnras/stw131}, \href
  {https://ui.adsabs.harvard.edu/abs/2016MNRAS.457.2790T} {457, 2790}

\bibitem[\protect\citeauthoryear{{Tacchella}, {Bose}, {Conroy}, {Eisenstein}
  \& {Johnson}}{{Tacchella} et~al.}{2018}]{Tacchella2018}
{Tacchella} S.,  {Bose} S.,  {Conroy} C.,  {Eisenstein} D.~J.,   {Johnson}
  B.~D.,  2018, \mn@doi [\apj] {10.3847/1538-4357/aae8e0}, \href
  {https://ui.adsabs.harvard.edu/abs/2018ApJ...868...92T} {868, 92}

\bibitem[\protect\citeauthoryear{{Tacchella}, {Forbes}  \&
  {Caplar}}{{Tacchella} et~al.}{2020}]{Tacchella2020}
{Tacchella} S.,  {Forbes} J.~C.,   {Caplar} N.,  2020, \mn@doi [\mnras]
  {10.1093/mnras/staa1838}, \href
  {https://ui.adsabs.harvard.edu/abs/2020MNRAS.497..698T} {497, 698}

\bibitem[\protect\citeauthoryear{{Thompson}, {Quataert}  \&
  {Murray}}{{Thompson} et~al.}{2005}]{Thompson2005}
{Thompson} T.~A.,  {Quataert} E.,   {Murray} N.,  2005, \mn@doi [\apj]
  {10.1086/431923}, \href
  {https://ui.adsabs.harvard.edu/abs/2005ApJ...630..167T} {630, 167}

\bibitem[\protect\citeauthoryear{{Torrey}, {Hopkins}, {Faucher-Gigu{\`e}re},
  {Vogelsberger}, {Quataert}, {Kere{\v{s}}}  \& {Murray}}{{Torrey}
  et~al.}{2017}]{Torrey2017}
{Torrey} P.,  {Hopkins} P.~F.,  {Faucher-Gigu{\`e}re} C.-A.,  {Vogelsberger}
  M.,  {Quataert} E.,  {Kere{\v{s}}} D.,   {Murray} N.,  2017, \mn@doi [\mnras]
  {10.1093/mnras/stx254}, \href
  {https://ui.adsabs.harvard.edu/abs/2017MNRAS.467.2301T} {467, 2301}

\bibitem[\protect\citeauthoryear{{Treu} et~al.,}{{Treu}
  et~al.}{2023}]{Treu2023}
{Treu} T.,  et~al., 2023, \mn@doi [\apjl] {10.3847/2041-8213/ac9283}, \href
  {https://ui.adsabs.harvard.edu/abs/2023ApJ...942L..28T} {942, L28}

\bibitem[\protect\citeauthoryear{{Truelove}, {Klein}, {McKee}, {Holliman},
  {Howell}  \& {Greenough}}{{Truelove} et~al.}{1997}]{Truelove1997}
{Truelove} J.~K.,  {Klein} R.~I.,  {McKee} C.~F.,  {Holliman} John~H. I.,
  {Howell} L.~H.,   {Greenough} J.~A.,  1997, \mn@doi [\apjl] {10.1086/310975},
  \href {https://ui.adsabs.harvard.edu/abs/1997ApJ...489L.179T} {489, L179}

\bibitem[\protect\citeauthoryear{{Tsukui} \& {Iguchi}}{{Tsukui} \&
  {Iguchi}}{2021}]{Tsukui2021}
{Tsukui} T.,  {Iguchi} S.,  2021, \mn@doi [Science] {10.1126/science.abe9680},
  \href {https://ui.adsabs.harvard.edu/abs/2021Sci...372.1201T} {372, 1201}

\bibitem[\protect\citeauthoryear{{Tumlinson}, {Peeples}  \& {Werk}}{{Tumlinson}
  et~al.}{2017}]{Tumlinson2017}
{Tumlinson} J.,  {Peeples} M.~S.,   {Werk} J.~K.,  2017, \mn@doi [\araa]
  {10.1146/annurev-astro-091916-055240}, \href
  {https://ui.adsabs.harvard.edu/abs/2017ARA&A..55..389T} {55, 389}

\bibitem[\protect\citeauthoryear{{Vallini} et~al.,}{{Vallini}
  et~al.}{2024}]{Vallini2024}
{Vallini} L.,  et~al., 2024, \mn@doi [\mnras] {10.1093/mnras/stad3150}, \href
  {https://ui.adsabs.harvard.edu/abs/2024MNRAS.527...10V} {527, 10}

\bibitem[\protect\citeauthoryear{{Vogelsberger} et~al.,}{{Vogelsberger}
  et~al.}{2020}]{Vogelsberger2020}
{Vogelsberger} M.,  et~al., 2020, \mn@doi [\mnras] {10.1093/mnras/staa137},
  \href {https://ui.adsabs.harvard.edu/abs/2020MNRAS.492.5167V} {492, 5167}

\bibitem[\protect\citeauthoryear{{Weinberger}, {Springel}  \&
  {Pakmor}}{{Weinberger} et~al.}{2020}]{Weinberger2020arepo}
{Weinberger} R.,  {Springel} V.,   {Pakmor} R.,  2020, \mn@doi [\apjs]
  {10.3847/1538-4365/ab908c}, \href
  {https://ui.adsabs.harvard.edu/abs/2020ApJS..248...32W} {248, 32}

\bibitem[\protect\citeauthoryear{{White} \& {Frenk}}{{White} \&
  {Frenk}}{1991}]{White1991}
{White} S. D.~M.,  {Frenk} C.~S.,  1991, \mn@doi [\apj] {10.1086/170483}, \href
  {https://ui.adsabs.harvard.edu/abs/1991ApJ...379...52W} {379, 52}

\bibitem[\protect\citeauthoryear{{Williams} \& {McKee}}{{Williams} \&
  {McKee}}{1997}]{Williams1997}
{Williams} J.~P.,  {McKee} C.~F.,  1997, \mn@doi [\apj] {10.1086/303588}, \href
  {https://ui.adsabs.harvard.edu/abs/1997ApJ...476..166W} {476, 166}

\bibitem[\protect\citeauthoryear{{Xiao} et~al.,}{{Xiao}
  et~al.}{2024}]{Xiao2024}
{Xiao} M.,  et~al., 2024, \mn@doi [\nat] {10.1038/s41586-024-08094-5}, \href
  {https://ui.adsabs.harvard.edu/abs/2024Natur.635..311X} {635, 311}

\bibitem[\protect\citeauthoryear{{Yung}, {Somerville}, {Finkelstein}, {Wilkins}
   \& {Gardner}}{{Yung} et~al.}{2024a}]{Yung2024b}
{Yung} L.~Y.~A.,  {Somerville} R.~S.,  {Finkelstein} S.~L.,  {Wilkins} S.~M.,
  {Gardner} J.~P.,  2024a, \mn@doi [\mnras] {10.1093/mnras/stad3484}, \href
  {https://ui.adsabs.harvard.edu/abs/2024MNRAS.527.5929Y} {527, 5929}

\bibitem[\protect\citeauthoryear{{Yung}, {Somerville}, {Nguyen}, {Behroozi},
  {Modi}  \& {Gardner}}{{Yung} et~al.}{2024b}]{Yung2024}
{Yung} L.~Y.~A.,  {Somerville} R.~S.,  {Nguyen} T.,  {Behroozi} P.,  {Modi} C.,
    {Gardner} J.~P.,  2024b, \mn@doi [\mnras] {10.1093/mnras/stae1188}, \href
  {https://ui.adsabs.harvard.edu/abs/2024MNRAS.530.4868Y} {530, 4868}

\bibitem[\protect\citeauthoryear{{Zhang} et~al.,}{{Zhang}
  et~al.}{2024}]{Zhang2024}
{Zhang} E.,  et~al., 2024, \mn@doi [\apj] {10.3847/1538-4357/ad7f57}, \href
  {https://ui.adsabs.harvard.edu/abs/2024ApJ...975..229Z} {975, 229}

\bibitem[\protect\citeauthoryear{{Zier}, {Kannan}, {Smith}, {Vogelsberger}  \&
  {Verbeek}}{{Zier} et~al.}{2024}]{Zier2024}
{Zier} O.,  {Kannan} R.,  {Smith} A.,  {Vogelsberger} M.,   {Verbeek} E.,
  2024, \mn@doi [\mnras] {10.1093/mnras/stae1837}, \href
  {https://ui.adsabs.harvard.edu/abs/2024MNRAS.533..268Z} {533, 268}

\bibitem[\protect\citeauthoryear{{Zuckerman} \& {Evans}}{{Zuckerman} \&
  {Evans}}{1974}]{Zuckerman1974}
{Zuckerman} B.,  {Evans} N.~J. I.,  1974, \mn@doi [\apjl] {10.1086/181613},
  \href {https://ui.adsabs.harvard.edu/abs/1974ApJ...192L.149Z} {192, L149}

\makeatother
\end{thebibliography}




\appendix

\section{KS relation with instantaneous SFR}

\begin{figure}
    \centering
    \includegraphics[width=\linewidth]{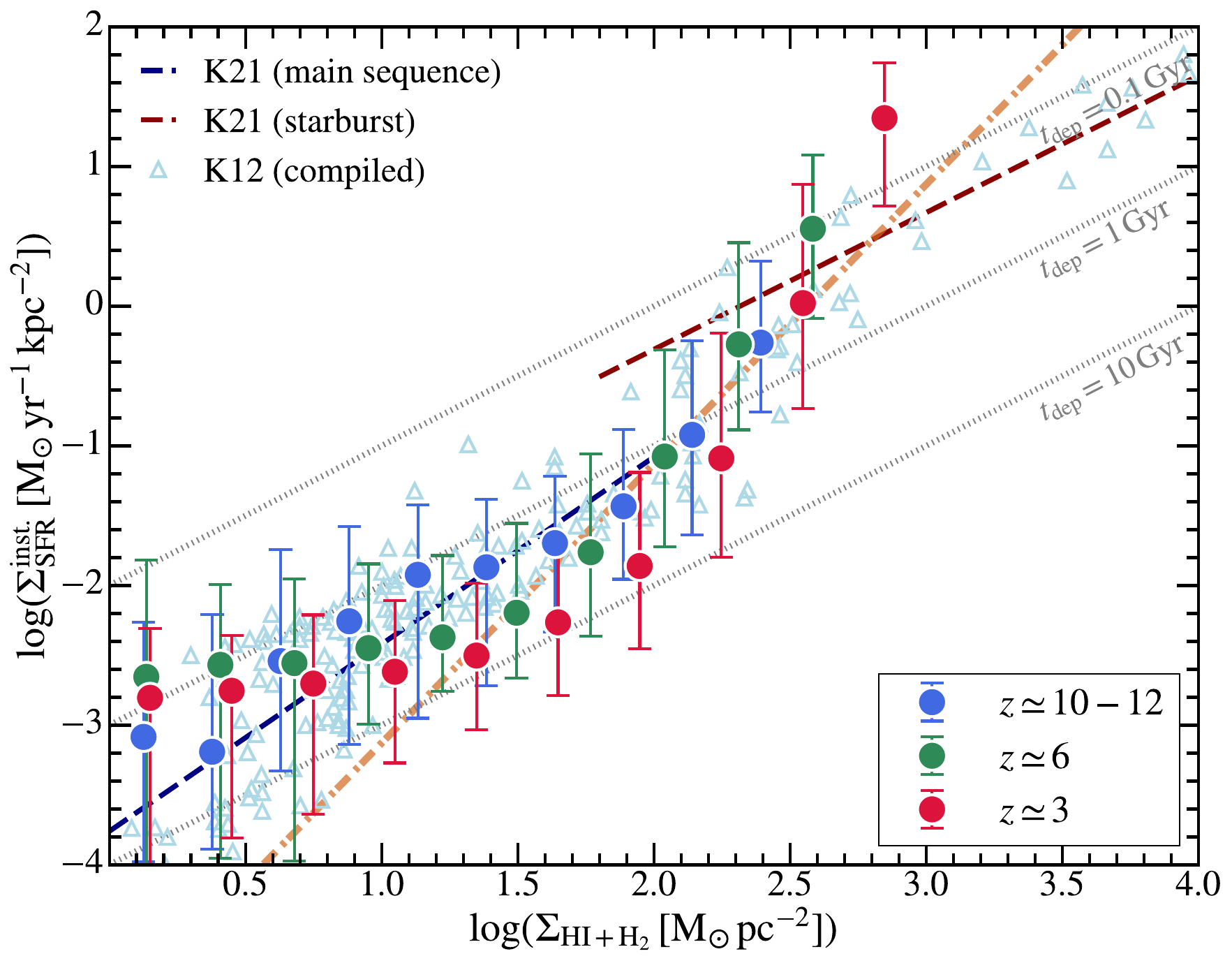} 
    \caption{KS relation of neutral gas at $1\,{\rm kpc}$ scale (same as Figure~\ref{fig:kslaw-yso}, but the SFR is calculated using the instantaneous SFR in gas cells).}
    \label{appfig:kslaw-inst}
\end{figure}

In the main text, when evaluating the KS relation, we use the total mass of young stellar particles with age $<10\Myr$ to compute SFR. This scheme is less affected by stochastic effects but is subject to a resolution limit, i.e. one young stellar particle per pixel. An alternative way to compute SFR is using the instantaneous SFR of star-forming gas cells and applying the same smoothing kernel as we did for the surface density. In Figure~\ref{appfig:kslaw-inst}, we show the KS relation of neutral gas with SFR measured using the instantaneous SFR in gas cells. Compared to the fiducial results in Figure~\ref{fig:kslaw-yso}, the slope of the KS relation and its independence of redshift are retained. However, the SFR surface density can extend to values below the resolution in this case. As has been demonstrated in Figure~\ref{fig:halo-sfe-compare-phy}, the KS relation defined in this way can be affected by the cell-level SFE in the low-surface density regime.

\section{UV variability and UV luminosity function at low redshifts}

\begin{figure}
    \centering
    \includegraphics[width=\linewidth]{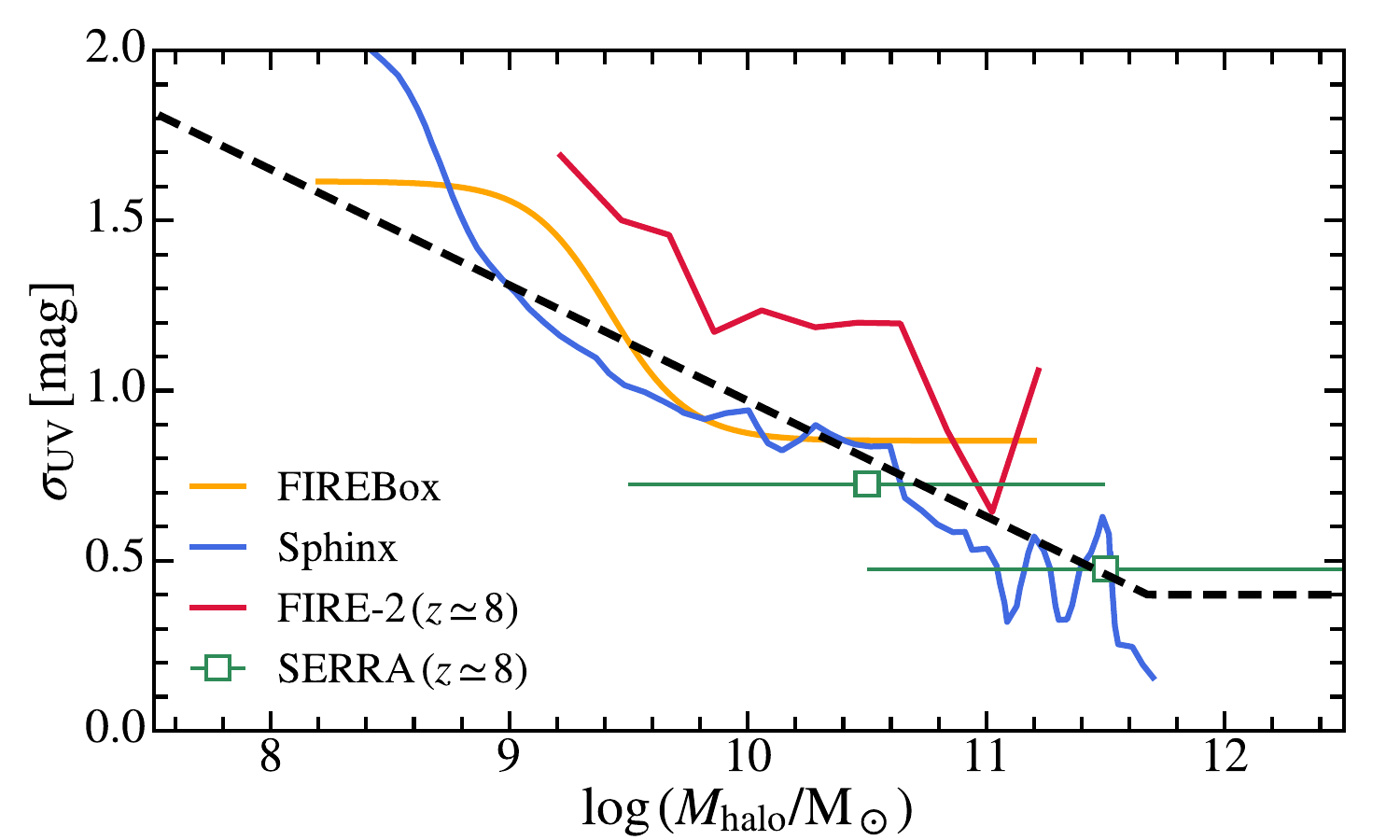}
    \includegraphics[width=\linewidth]{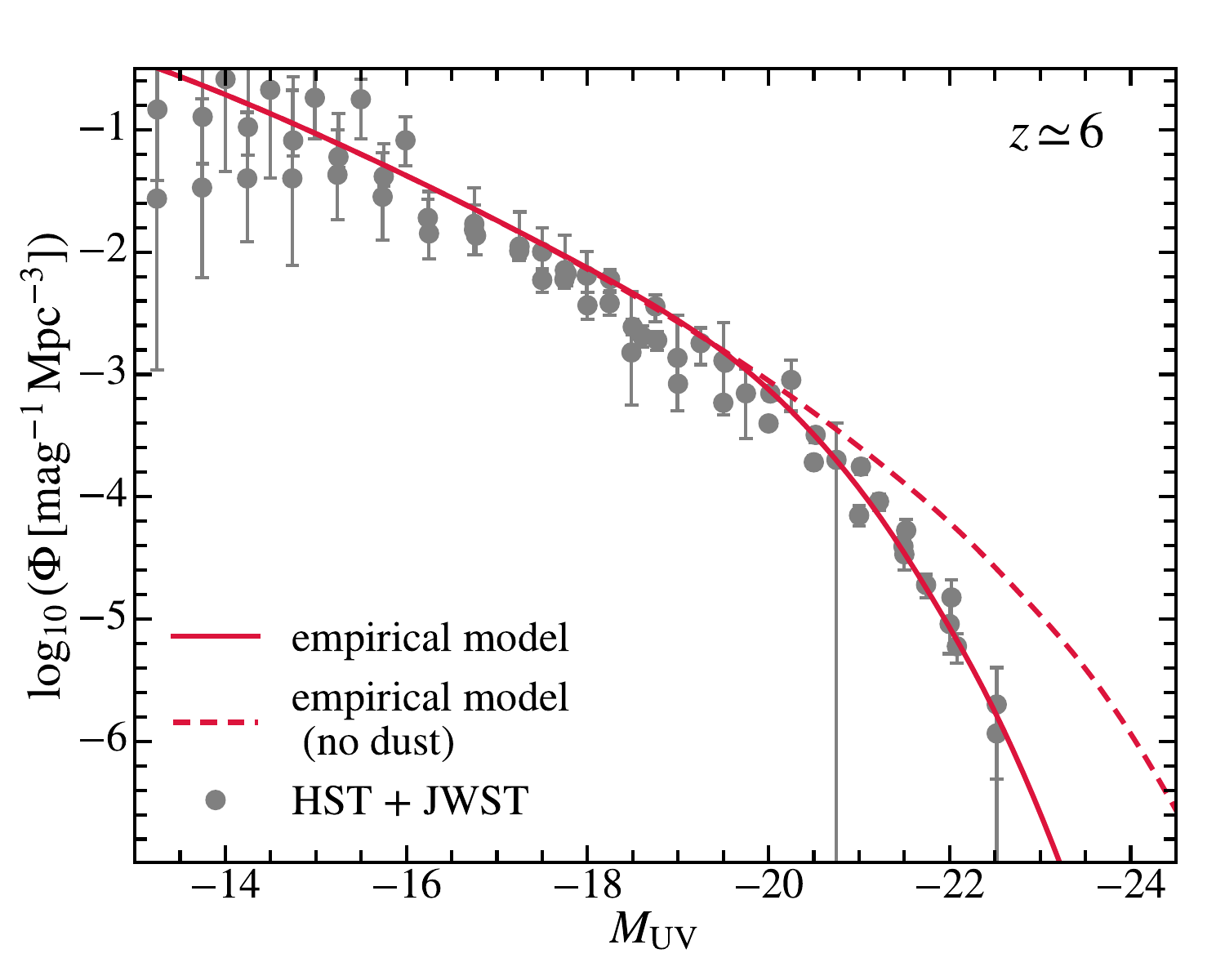}
    \caption{{\it Top:} UV variability $\sigma_{\rm UV}$ versus halo mass. We compare the assumed relation in our empirical model (black dashed line, no redshift dependence) with predictions from cosmological hydrodynamic simulations, including FIREBox~\citep{Feldmann2024}, Sphinx~\citep{Rosdahl2018,Katz2023}, FIRE-2~\citep{Ma2018,Sun2023b}, and SERRA~\citep{Pallottini2022,Pallottini2023}. The $\sigma_{\rm UV}(M_{\rm halo})$ we assume agrees with these simulation results in general while FIRE-2 predicts slightly higher $\sigma_{\rm UV}$ at $M_{\rm halo}\lesssim 10^{11}\msun$. {\it Bottom:} UV luminosity function predictions using the halo-scale SFE learned from \thesanzoom and the $\sigma_{\rm UV}(M_{\rm halo})$ shown above. The dust attenuation model is the same as discussed in \citet{Shen2023}. We find excellent agreement with observational constraints at $z\simeq 6$ compiled in \citet{Shen2023}.}
    \label{appfig:var}
\end{figure}

In Figure~\ref{appfig:var}, we illustrate both the assumed relation for UV variability, $\sigma_{\rm UV}$, as a function of halo mass (top panel) and the resulting UV luminosity function predictions at $z\simeq 6$ (bottom panel). In the top panel, we compare our empirical model (black dashed line) to predictions from various cosmological hydrodynamic simulations, including FIREBox \citep{Feldmann2024}, Sphinx \citep{Rosdahl2018, Katz2023}, FIRE-2 \citep{Ma2018,Sun2023b}, and SERRA \citep{Pallottini2022,Pallottini2023}. Overall, the empirically assumed UV variability lies within the range of simulated outcomes, although FIRE-2 yields slightly higher variability at lower halo masses $M_{\rm halo}\lesssim 10^{11}\msun$). In the bottom panel, we show the impact of this prescription on the UV luminosity function at $z\simeq 6$, adopting the halo-scale SFE derived from \thesanzoom and the same dust attenuation treatment described in \citet{Shen2023}. The resulting luminosity function compares favorably with the set of observational constraints compiled in \citet{Shen2023}.

\bsp	
\label{lastpage}
\end{document}